%% file: main.tex
\documentclass[pdflatex,sn-mathphys-num]{sn-jnl}

\usepackage[T1]{fontenc}
\usepackage[utf8]{inputenc}
\usepackage{graphicx}
\usepackage{multirow}
\usepackage{amsmath,amssymb,amsfonts}
\usepackage{amsthm}
\usepackage{mathrsfs}
\usepackage[title]{appendix}
\usepackage{xcolor}
\usepackage{textcomp}
\usepackage{booktabs}
\usepackage{tabularx}
\usepackage{algorithm}
\usepackage{enumitem}

\newcolumntype{L}{>{\raggedright\arraybackslash}X}
\providecommand{\tightlist}{\setlength{\itemsep}{0pt}\setlength{\parskip}{0pt}}
\floatname{algorithm}{Algorithm}

\newcommand{\snthmname}{}
\theoremstyle{thmstyleone}
\newtheorem*{thmx}{\snthmname}
\theoremstyle{thmstylethree}
\newtheorem*{defx}{\snthmname}

\makeatletter\patchcmd{\@maketitle}{Corresponding author(s). E-mail(s):}{Corresponding author. E-mail:}{}{}\makeatother
\begin{document}

\title[Equivalence of quantum circuit generators]{Parameterized equivalence checking of quantum circuit generators via weighted automata}

\author*[1]{\fnm{Pavel} \sur{Chuklanov}}\email{chuklanov.pavel@gmail.com}
\affil*[1]{\orgname{Moscow Institute of Physics and Technology}, \orgaddress{\street{Institutsky lane 9}, \city{Dolgoprudny, Moscow region}, \postcode{141700}, \country{Russia}}}

\abstract{A generator builds a quantum circuit from the number of qubits and a bit string. We check a generator against a reference for all parameter values under six equivalence relations. For ladder programs, we prove that above a computable threshold the circuit matrices form a uniform matrix product operator, a weighted automaton of dimension independent of the number of qubits. Equivalence reduces to the zeroness of a difference automaton, checked modulo primes. The answer is a counterexample or ``equal'', exact or with a computed error bound. Experimentally, the method found errors missed by fixed-size checks and produced no incorrect verdicts.}


\maketitle

\input{sections/01_intro}
\input{sections/02_preliminaries}
\input{sections/03_problem}
\input{sections/04_ladders}
\input{sections/05_method}
\input{sections/06_implementation}
\input{sections/07_evaluation}
\input{sections/08_related}
\input{sections/09_conclusion}

\backmatter

\bmhead{Supplementary information}

Online Resource 1, the electronic supplementary material of the journal version, is appended to this preprint after the references (Sects.~S1--S4, statements S.1--S.17, Tables S1, S2, Figures S1, S2). Numbers without the prefix S refer to the article.

\section*{Declarations}
\begin{itemize}
\item \textbf{Competing interests.} The author has no competing interests to declare that are relevant to the content of this article.
\item \textbf{Funding.} No funding was received for conducting this study.
\item \textbf{Author contributions.} P.~Chuklanov conceived the study, developed the method and its implementation, designed and ran the experiments, and wrote the manuscript.
\item \textbf{Data availability.} The benchmark of reference--solution pairs, the ground-truth answers, the raw outputs of all runs (the method, QCEC, the instance checks, AutoQ-Para, the third-party code of Section~7.5), our models for AutoQ-Para, and the scripts that build every table and figure are available in the Zenodo archive (version 1.0.0, \url{https://doi.org/10.5281/zenodo.24059428}). Third-party material in the archive (QuantumKatas tasks) is distributed under its original license.
\item \textbf{Code availability.} The implementation (the translator from Python with Qiskit to ladder programs and the verification kernel), version 1.0.0, is available in the same archive under the MIT license; repository: \url{https://github.com/alpha7en/ladder_ec}.
\item \textbf{Ethics approval and consent to participate.} Not applicable.
\item \textbf{Consent for publication.} Not applicable.
\item \textbf{Materials availability.} Not applicable.
\item \textbf{Use of generative AI.} The language model Claude Opus 5.5 (Anthropic) wrote the task solutions checked in Section~7.5 and assisted the author in developing the code, running the experiments, and editing and translating the manuscript. The author is responsible for the content of the article.
\end{itemize}

\begin{appendices}
\input{sections/A_proofs}
\end{appendices}

\clearpage
\renewcommand{\thesection}{S\arabic{section}}
\renewcommand{\thetable}{S\arabic{table}}
\renewcommand{\thefigure}{S\arabic{figure}}
\renewcommand{\theHsection}{S\arabic{section}}
\renewcommand{\theHtable}{S\arabic{table}}
\renewcommand{\theHfigure}{S\arabic{figure}}
\setcounter{section}{0}\setcounter{table}{0}\setcounter{figure}{0}
\begin{center}
{\Large\bfseries Online Resource 1}\\[1ex]
{\small Electronic supplementary material of the article; the references are in the list above.}
\end{center}
\noindent The proof of Theorem 1 of the article (Sects. S1.1--S1.5, statements S.1--S.6); statements S.7--S.14, on which the proof of Theorem 2 relies (Appendix A of the article); the proofs of Lemma 5 and Proposition 2 of the article (Sect. S1.8); the description of the Python+Qiskit translator with the proof of Theorem 3 (Section 6 of the article); a comparison of the exact and probabilistic plans on the benchmark (Figure S1, Section 7.4 of the article); and a check of the bound of Theorem 1 on the ladders of the benchmark (Figure S2, Section 7.2 of the article). Numbers without the prefix S are numbers of sections, statements, tables and formulas of the article.

\input{supplement/S1_lemmas}
\input{supplement/S2_translator}
\input{supplement/S3_plans}
\input{supplement/S4_bound}

\end{document}

%% file: sections/01_intro.tex
\section{Introduction}\label{sec:1}

Many quantum algorithms are defined as families of circuits parameterized by the number of qubits \(n\) and optionally a bit string \(a\in\{0,1\}^{n}\), including the Bernstein--Vazirani oracle with hidden string \(a\) \cite[Sect.~8.4]{Bernstein1997_QComplexity}, GHZ state preparation \cite{Greenberger1989_GHZ}, and the quantum Fourier transform (QFT) \cite{Coppersmith1994_ApproxQFT}. We call code that builds the circuit from \((n,a)\) a \emph{generator} (Definition 1 in Section 3). Whereas prior work on parameterized circuit verification treats continuous rotation angles as parameters \cite{Hong2024_ParamEC_TDD,Ross2025_Cutoff}, our parameters are the discrete system size \(n\) and the bit string \(a\). The circuits of MQT Bench \cite{Quetschlich2023_MQTBench} are built from templates (two of its 36 generators are in the class of ladder programs, Section 7.5), and each program of the QED-C suite \cite{Lubinski2023_QEDC} defines a scalable family of circuits \cite{Paltenghi2024_SurveyQSTesting}. Students also write such generators in programming assignments. In practice, such code is compared with a reference only on instances of fixed size: an autograder compares a submitted solution against a hidden reference implementation of one variant of the assignment, determined by a seed \cite{Kaltchenko2026_ChatGPTQiskitHW}, and the QuanBench benchmark of code written by language models computes the process fidelity with respect to a reference matrix at fixed \(n\) \cite{Guo2025_QuanBench}. Our goal is to check a generator against a reference for all \((n,a)\) at which the reference is defined.

Testing only finitely many instances can miss defects that emerge at larger scales. For example, suppose the reference \(R\) is the Bernstein--Vazirani oracle, while the candidate generator \(G\) (hereafter called the \emph{solution}) includes an extraneous loop over qubits \(i\ge20\) (Example 1, Section 3). The candidate agrees with the reference for all \(n\le20\), and for each \(n\ge21\) it diverges only on specific strings \(a\). Although synthetic, this example illustrates that no fixed bound on tested instance sizes can rule out bugs triggered only above a threshold \(n\).

Existing equivalence checking tools for circuits (for example, \cite{Burgholzer2020_AdvancedEC,Peham2022_ZX_EC}) compare two concrete circuits, so for a family they give only \emph{instance checking}, a check at finitely many pairs \((n,a)\). The survey \cite{Paltenghi2024_SurveyQSTesting} names the scalability of equivalence checking as an open problem. Some methods do work for all \(n\). The method of \cite{Abdulla2025_ParamVerification} checks the equivalence of parameterized circuits by representing the family with tree automata and transducers, built from a circuit with a given repetition pattern. For the Bernstein--Vazirani oracle, which does not fit this pattern, the model in \cite{Abdulla2025_ParamVerification} is built by hand and only for particular hidden strings \(a\). The verifier of the synthesis system QSynth \cite{Deng2023_QSynth} checks a program for all \(n\), but against a specification rather than a reference program. Section 8 compares these and other works in detail.

The main observation of this paper is that if a generator is given by a \emph{ladder program} (Section 4), then for all \(n\) at or above a computable threshold its matrix family \(\{U(n,a)\}\) can be represented as a uniform matrix product operator (MPO) whose local physical index at site \(j\) incorporates the classical bit \(a_j\) alongside the output \(x_j\) and input \(y_j\), with a bond dimension independent of \(n\) (Theorem 1, Section 5.1). A ladder program is a sequence of fixed segments (gates on the first and last qubits of the register) and loops over a counter \(i\) whose body acts on the qubits \(i+o\) with constant offsets \(o\) and on globally fixed qubits at the ends of the register.

A weighted automaton describes the states and operators of systems of all sizes by one function on words of any length. In the construction of Crosswhite and Bacon, the letters of an automaton for an operator are local operators \cite[Sect.~II.C]{Crosswhite2008_FiniteAutomataMPS}, and the values of the automaton on words of fixed length are the components of a uniform matrix product state (MPS) \cite{Li2022_WA_TN_RNN,Srinivasan2021_QTN_WA}. We take the triple \((a_j,x_j,y_j)\) as the letter of qubit \(j\); then a uniform MPO is a weighted automaton whose value on a word of length \(n+e\), where \(e\) is the number of ancillas, equals \(\langle x\rvert U(n,a)\lvert y\rangle\). The equality of two families for all \(n\) and \(a\) therefore reduces to the zeroness of the difference automaton of the two programs on words of sufficient length, which is decided by linear algebra \cite{Kiefer2020_WANotes}.

The entries of gate matrices lie in the cyclotomic field \(\mathbb{Q}(\zeta_m)\), \(\zeta_m=e^{2\pi\mathrm{i}/m}\). To compute with fixed-length integers, we check zeroness modulo primes \(p\), and we use the field norm to bound the number of primes modulo which a non-zero value vanishes.

\begin{figure}[tbp]
\centering
\includegraphics{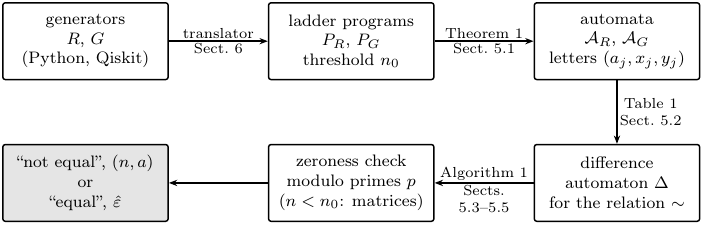}
\caption{Overview of the method. The translator turns the code of the reference \(R\) and the solution \(G\) into ladder programs and a threshold \(n_0\). For \(n\ge n_0\), the automata of the programs represent their matrix families, and the difference automaton of the relation is checked for zeroness modulo primes. The sizes \(n<n_0\) are checked by computing the matrices modulo primes}\label{fig:overview}
\end{figure}

Equivalence checking of weighted automata \cite{Kiefer2020_WANotes}, the description of uniform MPS and MPO by weighted automata \cite{Crosswhite2008_FiniteAutomataMPS,Li2022_WA_TN_RNN}, and sequential generation of MPS, in which an auxiliary system of finite dimension interacts with the qubits one at a time \cite{Schon2005_SequentialGeneration}, are known, and the automaton of a ladder also reads the qubits one at a time. Unlike these constructions, our automaton is built from the program, and its letter contains the bit \(a_j\), so one automaton represents the family for all \(a\), not one circuit; for the qubits \(n+c\), which come at the end of the word but are acted on by loop bodies, the automaton keeps guessed values and checks them at the end of the word (Theorem 1). Six relations, including equality up to global phase and equality with returned ancillas, reduce to a zeroness check (Lemma 2). The number of primes modulo which the check can fail is bounded for series whose power of \(\sqrt2\) in the denominator grows with the word length (Section 5.4). The code of a generator is translated into a ladder program with a certified threshold (Section 6). Figure~\ref{fig:overview} shows the steps of the method and the sections that describe them.

\textbf{Contributions.}

\begin{itemize}
\tightlist
\item
  We state the problem of checking a generator against a reference for all \((n,a)\) under six relations: equality of matrices, equality of prepared states, and equality of the action with ancillas returned to \(\lvert0\rangle\), each exact or up to global phase (Section 3). We define the class of ladder programs and build from a ladder program an automaton that represents its family of matrices, with an explicit dimension bound. The mirror layer \(\prod_{i<n/2}\mathrm{CX}(i,n-1-i)\) and the Fourier transform are not in the class (Section 4.2).
\item
  We give an algorithm that checks modulo primes (Section 5, Theorem 2). The answer ``not equal'' contains a counterexample \((n,a)\); the answer ``equal'' is either exact or comes with a computed bound \(\hat\varepsilon\) on the error probability over the random choice of primes. Sizes below the joint threshold \(n_0\) of the two programs are checked by direct computation.
\item
  We implement the method for generators in Python with Qiskit: the translator builds from the code a ladder program and a threshold such that for all \(n\) at or above it the code builds the same circuit up to replacing each floating-point gate angle \(\theta\) with the exact symbolic value \(q\pi\) of the corresponding angle expression, under assumptions on the semantics of Python and Qiskit (Section 6, Theorem 3).
\item
  On a benchmark of reference--solution pairs, the method produced no incorrect verdicts (a conclusive answer on 1876 of the 2200 pairs), confirmed equality for all \(n\) on 444 of the 483 equal pairs (median 0.29~s; enumeration by QCEC up to the cutoff \(n_{\rm cut}\), which gives the same guarantee, did not finish within 60~s on 176 of them), and found 139 of the 172 errors that appear only at \(n>20\) (Section 7).
\end{itemize}

%% file: sections/02_preliminaries.tex
\section{Preliminaries}\label{sec:2}

\subsection{Circuits}\label{sec:2.1}

The qubits of a register of \(N\) qubits are numbered \(0,\dots,N-1\). The basis vector \(\lvert x\rangle=\lvert x_0x_1\cdots x_{N-1}\rangle\), \(x\in\{0,1\}^{N}\), is the tensor product in which qubit 0 is the first factor. We write \(0^{N}\) for the string of \(N\) zeros.

We fix a gate set \(\mathcal{G}\) \cite[Sects.~4.2, 4.3]{Nielsen2010_QCQI}; each circuit and each program uses a finite subset of it. A gate \(g\in\mathcal{G}\) has an arity \(\operatorname{ar}(g)\) and a unitary matrix \(M_g\) of size \(2^{\operatorname{ar}(g)}\times2^{\operatorname{ar}(g)}\). An application \(g(q_1,\dots,q_{\operatorname{ar}(g)})\) to pairwise distinct qubits \(q_1,\dots,q_{\operatorname{ar}(g)}\) acts by \(M_g\) on these qubits and as the identity on the others. A \emph{circuit} on \(N\) qubits is a finite sequence of gate applications; its matrix \(U\) is the product of the matrices of the applications, with the first application rightmost. Matrices \(U\) and \(U'\) are \emph{equal up to phase} if \(U=\lambda U'\) for some \(\lambda\in\mathbb{C}\), \(\lvert\lambda\rvert=1\).

We take \(\mathcal{G}\) to consist of the gates H, X, Y, Z, S, S\(^{\dagger}\), T, T\(^{\dagger}\), CX, CY, CZ, CH, SWAP, CCX, CCZ, CSWAP and the rotations RX\((\theta)\), RY\((\theta)\), RZ\((\theta)\), P\((\theta)\), CP\((\theta)\), CRZ\((\theta)\), RZZ\((\theta)\) with angles \(\theta=q\pi\), \(q\in\mathbb{Q}\). Each rotation with a fixed rational multiple of \(\pi\) is a distinct symbol in \(\mathcal{G}\), meaning that rotation angles are fixed per gate rather than parameterized by loop counters or \(n\) (consequently, the exact QFT cannot be expressed in the language of Section 3; see Section 4.2).

\subsection{Weighted automata}\label{sec:2.2}

Weighted automata allow us to represent the matrix family \(U(n,a)\) for all \(n\) by a single finite object (Section 5). Let \(F\) be a field, \(\Sigma\) a finite alphabet, and \(\Sigma^{*}\) the set of words over \(\Sigma\). A \emph{series} is a map \(f:\Sigma^{*}\to F\). A \emph{weighted automaton} of dimension \(d\) is a triple \(\mathcal{A}=(\alpha,\{A(\sigma)\}_{\sigma\in\Sigma},\beta)\) with \(\alpha,\beta\in F^{d}\) and \(A(\sigma)\in F^{d\times d}\). It defines the series
\[f_{\mathcal{A}}(\sigma_0\cdots\sigma_{N-1})=\alpha^{\top}A(\sigma_0)\cdots A(\sigma_{N-1})\,\beta .\]
The values of a weighted automaton on words of length \(N\) are the components of a uniform (with the same tensor at every site) matrix product state (MPS) \cite[Sect.~5.1]{Orus2014_TNIntro}, \cite{Fannes1992_FCS,PerezGarcia2007_MPSRepresentations} with boundary vectors \cite[Sect.~3.1]{Li2022_WA_TN_RNN}: the letter is the physical index, the dimension \(d\) is the bond dimension, and \(\alpha\) and \(\beta\) are the boundary vectors. Conversely, any uniform MPS with boundary vectors directly specifies a weighted automaton. If the letter is a pair (output, input), we obtain a uniform matrix product operator (MPO; Figure 2a) \cite[Sect.~5]{Schollwock2011_DMRG_MPS}, \cite{Pirvu2010_MPORepresentations}. Crosswhite and Bacon build an MPO from a weighted automaton whose letters are local operators \cite[Sect.~II.C]{Crosswhite2008_FiniteAutomataMPS}. The form with the letter (output, input) is equivalent: substituting the matrix entries of the local operators gives the entry \(\langle x\rvert U\lvert y\rangle\).
An automaton is described by a graph: the states are the indices \(1,\dots,d\), the transition from \(q\) to \(q'\) on the letter \(\sigma\) has weight \(A(\sigma)_{qq'}\), and the initial and final weights of state \(q\) are \(\alpha_q\) and \(\beta_q\). Then \(f_{\mathcal{A}}(w)\) is the sum, over the paths that read \(w\), of the products of the weights. In the proof of Theorem 1 automata are built from a \emph{transition graph}: a finite set of states with initial and final weights and weighted transitions on letters, where two states may be joined by several transitions on the same letter. The automaton is obtained from the reachable states; \(A(\sigma)_{qq'}\) is the sum of the weights of the transitions \(q\to q'\) on \(\sigma\) (Online Resource 1, Sect. S1.1). Automata are \emph{equivalent} if they define the same series \cite[Sect.~1]{Kiefer2020_WANotes}.

A word \(w\) with \(f(w)\ne0\) is called a \emph{witness} that the series \(f\) is non-zero. We use four facts, all proved over an arbitrary field in \cite[Sects.~1--4]{Kiefer2020_WANotes}. Equivalence checking and minimization go back to Sch\"utzenberger \cite[Def.~\(1'\), Sect.~III.B.1]{Schutzenberger1961_Automata}, and Tzeng gave an explicit polynomial-time equivalence algorithm \cite[Thm.~3.1]{Tzeng1992_ProbAutomataEquiv}.

(F1) Given automata of dimensions \(d_1\) and \(d_2\), one can construct automata for the series \(f_1\pm f_2\) of dimension \(d_1+d_2\) and of the pointwise product \(f_1f_2\) of dimension \(d_1d_2\). The matrices of the product automaton are Kronecker products \cite[Prop.~1.1]{Kiefer2020_WANotes}.

(F2) If the series of an automaton of dimension \(d\) is non-zero, it has a witness of length at most \(d-1\). A basis of the span \(\operatorname{span}\{\alpha^{\top}A(w):w\in\Sigma^{*}\}\) and such a witness are found in \(O(\lvert\Sigma\rvert d^{3})\) operations in \(F\) \cite[Props.~2.1, 2.2]{Kiefer2020_WANotes}.

(F3) For any automaton, an equivalent automaton of least dimension is built in \(O(\lvert\Sigma\rvert d^{3})\) operations \cite[Thm.~3.6]{Kiefer2020_WANotes} (\emph{minimization}; for an MPS it corresponds to lossless compression).

(F4) The least dimension of an automaton for a series \(f\) equals the rank of its \emph{Hankel matrix} \(H[u,v]=f(uv)\), \(u,v\in\Sigma^{*}\) \cite[Thm.~4.3]{Kiefer2020_WANotes}. The rank of \(H\), and hence of any of its blocks, is at most the dimension of any automaton of the series \(f\) \cite[Prop.~3.3]{Kiefer2020_WANotes}. If \(f(w)=\langle x\rvert U\lvert y\rangle\) with the letter \((x_j,y_j)\) for qubit \(j\), then the block \(H[u,v]\), \(u\in\Sigma^{k}\), \(v\in\Sigma^{N-k}\), is the matricization of \(U\) along the cut of the register between qubits \(k-1\) and \(k\): the rows are indexed by the pairs \((x,y)\) on qubits \(0,\dots,k-1\), the columns by the pairs on qubits \(k,\dots,N-1\). The rank of this block is the operator Schmidt rank of \(U\) with respect to the cut. In these terms (F4) is a known property of MPOs: the bond dimension is at least the Schmidt rank across any cut (Figure 2a).

\begin{figure}[tbp]
\centering
\includegraphics{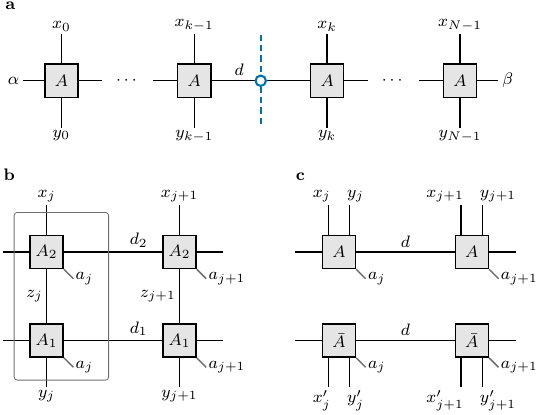}
\caption{An automaton as a uniform MPO. \textbf{a} The automaton \((\alpha,A,\beta)\): the lines (indices) \(x_j\), \(y_j\) are the output and input of qubit \(j\), the horizontal lines are bonds of dimension \(d\), the dashed line is the cut between qubits \(k-1\) and \(k\), and the circle is its intersection with a bond; \textbf{b} the product (Lemma 1): the indices \(z_j\) are contracted, the gray lines represent the bit \(a_j\), and the frame encloses the composite tensor of the product; \textbf{c} the doubled series (Section 5.2): the layers \(A\) and \(\bar A\) are not contracted}\label{fig:1}
\end{figure}

\subsection{Cyclotomic fields and computation modulo $p$}\label{sec:2.3}

Let \(m\ge1\), \(\zeta_m=e^{2\pi\mathrm{i}/m}\), and let \(\mathbb{K}=\mathbb{Q}(\zeta_m)\) be the cyclotomic field; its degree over \(\mathbb{Q}\) is \(\varphi(m)\), where \(\varphi\) is Euler's totient function. Let \(\mathcal{O}=\mathbb{Z}[\zeta_m]\). For a suitable \(m\), all gate matrix entries in \(\mathcal{G}\) lie in \(\mathbb{Z}[\zeta_m,\tfrac12]\) (Section 5.4, where \(\tfrac12\) enters via \(1/\sqrt2=\sqrt2/2\)), allowing exact symbolic evaluation over \(\mathbb{K}\). However, the bit-length of the coefficients grows with word length. Evaluating the automaton modulo a prime \(p\) restricts arithmetic to fixed-width machine integers, and we use the algebraic field norm to bound the number of primes modulo which a non-zero element can vanish. We need three standard facts.

(K1) The cyclotomic polynomial \(\Phi_m\) is irreducible over \(\mathbb{Q}\), and the automorphisms of the field \(\mathbb{K}\) are the maps \(\gamma_k:\zeta_m\mapsto\zeta_m^{k}\) for \(k\in(\mathbb{Z}/m\mathbb{Z})^{\times}\), giving \(\varphi(m)\) automorphisms in total \cite[Thm.~2.5]{Conrad_CyclotomicExtensions}. The automorphism \(\gamma_{-1}\) is complex conjugation; all \(\gamma_k\) commute with each other.

(K2) For \(\xi\in\mathcal{O}\) the norm \(\mathrm{Nm}(\xi)=\prod_{k}\gamma_k(\xi)\) is an integer, and \(\mathrm{Nm}(\xi)\ne0\) for \(\xi\ne0\).

(K3) Let \(p\) be an odd prime with \(p\equiv1\pmod m\). Then \(\mathbb{F}_p\) contains an element \(\omega_p\) of order \(m\), and the substitution \(\zeta_m\mapsto\omega_p\) defines a ring homomorphism \(\pi_p:\mathbb{Z}[\zeta_m,\tfrac12]\to\mathbb{F}_p\).

Facts (K2) and (K3) are standard (they follow from (K1), Appendix A).

The homomorphism \(\pi_p\) maps zero to zero. Hence \(\pi_p(\xi)\ne0\) implies \(\xi\ne0\); the converse is false. In Section 5.4 we bound the number of primes at which a non-zero \(\xi\) becomes zero by the norm \(\mathrm{Nm}(\xi)\).

For example, for \(m=8\) and \(p=17\) the element \(\omega_p=2\) has order 8 and \(\pi_p(\sqrt2)=\pi_p(\zeta_8+\zeta_8^{-1})=2+9=11\). So the non-zero element \(\xi=\sqrt2-11\) becomes zero modulo 17, and its norm \(\mathrm{Nm}(\xi)=\bigl((-11)^{2}-2\bigr)^{2}=119^{2}=7^{2}\cdot17^{2}\) is divisible by 17.

%% file: sections/03_problem.tex
\section{Generators and the problem}\label{sec:3}

\subsection{Generators and the language $\mathcal{L}$}\label{sec:3.1}

\renewcommand{\snthmname}{Definition 1}\begin{defx}[generator]

A generator \(P\) with \(e\ge0\) ancillas maps each pair \((n,a)\), where \(n\ge1\) and \(a=(a_0,\dots,a_{n-1})\in\{0,1\}^{n}\), to a circuit \(P(n,a)\) on \(N=n+e\) qubits, or to the undefined symbol \(\bot\). The qubits \(0,\dots,n-1\) are called \emph{working} qubits, and the qubits \(n,\dots,N-1\) \emph{ancillas}. The set of pairs with \(P(n,a)\ne\bot\) is denoted \(\mathrm{dom}(P)\). For \((n,a)\in\mathrm{dom}(P)\) the matrix of the circuit \(P(n,a)\) is denoted \(U_P(n,a)\).

\end{defx}

A generator without a bit parameter is the special case in which \(P(n,a)\) does not depend on \(a\).

We write generators in the language \(\mathcal{L}\). An \emph{index expression} is an expression \(\eta=c+k_0n+k_1i_1+\dots+k_ri_r\) with integers \(c,k_0,\dots,k_r\), where \(i_1,\dots,i_r\) are the counters of the enclosing loops. A \emph{program} of the language \(\mathcal{L}\) is a number \(e\ge0\) and a command \(P\) generated by the grammar
\[\begin{aligned}P::={}&g(\eta_1,\dots,\eta_{\operatorname{ar}(g)})\ \mid\ P;P\ \mid\ \mathbf{for}\ i:=\eta_1\ \mathbf{to}\ \eta_2\ \mathbf{step}\ s\ \mathbf{do}\ P\ \mathbf{od}\\ \mid{}&\mathbf{if}\ a[\eta]=0\ \mathbf{then}\ P\ \mathbf{fi}\ \mid\ \mathbf{if}\ a[\eta]=1\ \mathbf{then}\ P\ \mathbf{fi},\end{aligned}\]
where \(g\in\mathcal{G}\) and \(s\in\mathbb{Z}\setminus\{0\}\).

For given \(n\) and \(a\), a command computes a sequence of gate applications or \(\bot\):

\begin{itemize}
\tightlist
\item
  \(g(\eta_1,\dots,\eta_k)\): let \(q_1,\dots,q_k\) be the values of the expressions. If all \(q_j\) lie in \(\{0,\dots,N-1\}\) and are pairwise distinct, the result is the application \(g(q_1,\dots,q_k)\); otherwise it is \(\bot\).
\item
  \(P;Q\): the result of \(P\) followed by the result of \(Q\); if one of them is \(\bot\), the result is \(\bot\).
\item
  Loop: let \(v_1\) and \(v_2\) be the values of \(\eta_1\) and \(\eta_2\). The counter \(i\) sequentially takes the values \(v_1,v_1+s,v_1+2s,\dots\) that are at most \(v_2\) (for \(s>0\)) or at least \(v_2\) (for \(s<0\)). The resulting sequence is the concatenation of the gate sequences produced by the loop body for each counter value in order, or the empty sequence if the range is empty.
\item
  Condition \(a[\eta]=0\) or \(a[\eta]=1\): let \(k\) be the value of \(\eta\). If \(k\notin\{0,\dots,n-1\}\), the result is \(\bot\). Otherwise the result is the result of the body if the condition holds and the empty sequence if it does not.
\end{itemize}

A program defines a generator: \(P(n,a)\) is the circuit on \(N=n+e\) qubits given by the computed sequence, or \(\bot\).

\renewcommand{\snthmname}{Example 1}\begin{defx}

The reference \(R\) is the Bernstein--Vazirani oracle with one ancilla:

\end{defx}

\begin{verbatim}
R:  e = 1
    X(n); H(n);
    for i := 0 to n-1 step 1 do
        if a[i] = 1 then CX(i, n) fi
    od;
    H(n); X(n)
\end{verbatim}

The circuit \(R(n,a)\) is shown in Figure 3a. The candidate generator (the solution) \(G\) coincides with \(R\) but contains one more loop before the last line (Figure 3b):

\begin{verbatim}
    for i := 20 to n-1 step 1 do
        if a[i] = 1 then CX(i, n) fi
    od;
\end{verbatim}

Between the gates X, H and H, X the ancilla is in the state \(\lvert-\rangle\), and the gate CX\((i,n)\) multiplies a basis vector by \((-1)^{x_i}\) (phase kickback). Hence for \(x\in\{0,1\}^{n}\)
\[U_R(n,a)\bigl(\lvert x\rangle\otimes\lvert0\rangle\bigr)=(-1)^{a\cdot x}\,\lvert x\rangle\otimes\lvert0\rangle,\qquad a\cdot x=\sum_{i=0}^{n-1}a_ix_i .\]
For \(n\le20\) the second loop of the solution is empty, and \(G(n,a)=R(n,a)\). For \(n\ge21\) a qubit \(i\ge20\) with \(a_i=1\) receives a second gate CX\((i,n)\), and the factors \((-1)^{x_i}\) cancel. So \(G(n,a)\) is the oracle for the string \(a'\) with \(a'_i=a_i\) for \(i<20\) and \(a'_i=0\) for \(i\ge20\). The generators differ exactly at those \((n,a)\) where \(n\ge21\) and \(a_i=1\) for at least one \(i\ge20\). For example, for \(n=21\) and \(a_{20}=1\) the signs on an input with \(x_{20}=1\) differ.

The same oracle can be written without an ancilla:

\begin{verbatim}
G':  e = 0
     for i := 0 to n-1 step 1 do
         if a[i] = 1 then Z(i) fi
     od
\end{verbatim}

The circuit \(G'(n,a)\) multiplies \(\lvert x\rangle\) by \((-1)^{a\cdot x}\), that is, it acts on the working qubits as \(R(n,a)\) does.

\begin{figure}[tbp]
\centering
\includegraphics{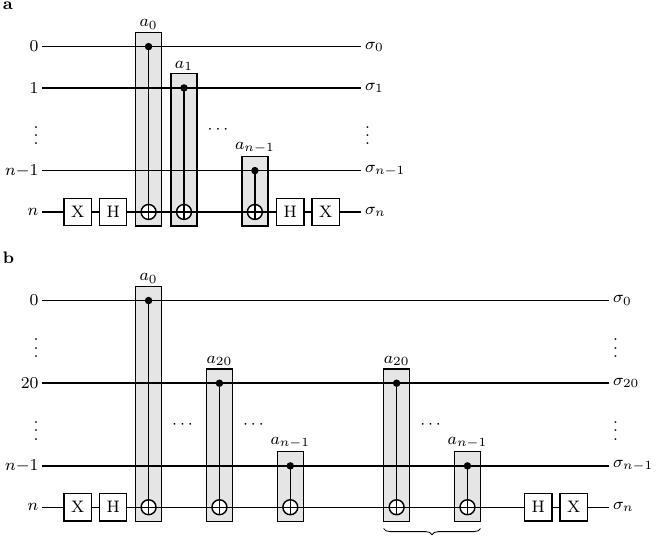}
\caption{The generators of Example 1. \textbf{a} The circuit \(R(n,a)\): the gray boxes are loop bodies, a body labeled \(a_i\) is applied if \(a_i=1\), and the letters \(\sigma_0,\dots,\sigma_n\) are read from top to bottom; \textbf{b} the circuit \(G(n,a)\); the brace marks the bodies of the second loop (\(i=20,\dots,n-1\))}\label{fig:2}
\end{figure}

\subsection{Relations}\label{sec:3.2}

Let \(R\) be the reference and \(G\) the candidate generator (the \emph{solution}), with \(e_R\) and \(e_G\) ancillas, let \((n,a)\in\mathrm{dom}(R)\cap\mathrm{dom}(G)\), \(U_R=U_R(n,a)\) and \(U_G=U_G(n,a)\). A generator \(P\) \emph{returns its ancillas} at \((n,a)\) if there is an operator \(V_P\) on the working qubits such that \(U_P\bigl(\lvert\psi\rangle\otimes\lvert0^{e_P}\rangle\bigr)=V_P\lvert\psi\rangle\otimes\lvert0^{e_P}\rangle\) for all \(\lvert\psi\rangle\).

\renewcommand{\snthmname}{Definition 2}\begin{defx}[relations]

The relations at a given \((n,a)\) are given by the table: the row states what is compared, the column whether exactly or up to phase. In the right column the condition means that there is \(\lambda\in\mathbb{C}\), \(\lvert\lambda\rvert=1\), with the stated equality, and \(\lambda\) may depend on \((n,a)\). The relations of the first two rows are defined for \(e_G=e_R\), those of the last row if \(R\) returns its ancillas.

\noindent
\begin{tabularx}{\textwidth}{@{}>{\raggedright\arraybackslash\hsize=0.84\hsize\hspace{0pt}}X>{\raggedright\arraybackslash\hsize=1.08\hsize\hspace{0pt}}X>{\raggedright\arraybackslash\hsize=1.08\hsize\hspace{0pt}}X@{}}
\toprule
Compared & Exact & Up to phase \\
\midrule
matrix $U$ & $G\equiv R$: $U_G=U_R$ & $G\simeq R$: $U_G=\lambda U_R$ \\
state $U\lvert0^{N}\rangle$ & $G\equiv_0R$: $U_G\lvert0^{N}\rangle=U_R\lvert0^{N}\rangle$ & $G\simeq_0R$: $U_G\lvert0^{N}\rangle=\lambda U_R\lvert0^{N}\rangle$ \\
operator $V$ on the working qubits & $G\equiv_{\mathrm{anc}}R$: $G$ returns its ancillas and $V_G=V_R$ & $G\simeq_{\mathrm{anc}}R$: $G$ returns its ancillas and $V_G=\lambda V_R$ \\
\botrule
\end{tabularx}

\end{defx}

A global phase is unobservable \cite[Sect.~2.2.7]{Nielsen2010_QCQI}, so equivalence checking tools for circuits check, besides exact equality, equality up to global phase \cite[Sect.~3]{Viamontes2007_EquivCheck}, \cite[Sect.~III-A]{Burgholzer2020_AdvancedEC}, \cite[Sect.~III]{Peham2022_ZX_EC}. The phase may depend on \((n,a)\). For example, the Grover diffusion operator \(D=2\lvert{+}^{n}\rangle\langle{+}^{n}\rvert-I\), written with a multi-controlled Z gate, equals
\[H^{\otimes n}X^{\otimes n}\bigl(I-2\lvert1^{n}\rangle\langle1^{n}\rvert\bigr)X^{\otimes n}H^{\otimes n}=-D,\]
and after \(r(n)\) iterations two versions of Grover's algorithm differ by the factor \((-1)^{r(n)}\).

The relations \(\equiv_0\) and \(\simeq_0\) concern state preparation, where only the column \(U\lvert0^{N}\rangle\) matters. The programs \(H(0);\ \mathbf{for}\ i:=1\ \mathbf{to}\ n-1\ \mathbf{step}\ 1\ \mathbf{do}\ \mathrm{CX}(0,i)\ \mathbf{od}\) and \(H(0);\ \mathbf{for}\ i:=1\ \mathbf{to}\ n-1\ \mathbf{step}\ 1\ \mathbf{do}\ \mathrm{CX}(i-1,i)\ \mathbf{od}\) prepare the same state \(\tfrac{1}{\sqrt2}\bigl(\lvert0^{n}\rangle+\lvert1^{n}\rangle\bigr)\), but for \(n\ge3\) their matrices differ (for \(n=3\) they map \(\lvert010\rangle\) to orthogonal vectors).

The relations \(\equiv_{\mathrm{anc}}\) and \(\simeq_{\mathrm{anc}}\) compare generators with different numbers of ancillas, which start in \(\lvert0\rangle\) and must return to it. In Example 1 the reference \(R\) returns its ancilla and \(V_R\lvert x\rangle=(-1)^{a\cdot x}\lvert x\rangle\), so \(G'\equiv_{\mathrm{anc}}R\) for all \((n,a)\), although \(e_{G'}=0\ne e_R=1\).

\subsection{The problem}\label{sec:3.3}

\renewcommand{\snthmname}{Problem 1}\begin{defx}

Given generators \(R\) (the reference) and \(G\) and a relation \(\sim\) of Definition 2, decide whether for all \((n,a)\in\mathrm{dom}(R)\) we have \((n,a)\in\mathrm{dom}(G)\) and \(G\sim R\) at \((n,a)\).

\end{defx}

This requirement is imposed only on \(\mathrm{dom}(R)\), as the reference generator defines the set of valid problem instances. For \(\equiv_{\mathrm{anc}}\) and \(\simeq_{\mathrm{anc}}\), Problem 1 is well-posed only when \(R\) cleanly returns its ancillas to \(\lvert0^{e_R}\rangle\) for all \((n,a)\in\mathrm{dom}(R)\). For a fixed \(n\) the condition of the problem is checked by computing the matrices \(U_R(n,a)\) and \(U_G(n,a)\) for all \(2^{n}\) strings \(a\); Problem 1 requires this for all \(n\).

In Example 1, \(G\not\simeq R\) and \(G\not\simeq_{\mathrm{anc}}R\) at \(n=21\), \(a_{20}=1\). At the same time \(G\equiv_0R\) for all \((n,a)\): on the input \(\lvert0^{N}\rangle\) both oracles give the sign \((-1)^{a\cdot0}=1\).

%% file: sections/04_ladders.tex
\section{Ladder programs}\label{sec:4}

\subsection{Definition}\label{sec:4.1}

\renewcommand{\snthmname}{Definition 3}\begin{defx}[ladder program]

A program \(P\) of the language \(\mathcal{L}\) with \(e\) ancillas is a \emph{ladder program} if \(P=P_1;\dots;P_L\) and each \emph{segment} \(P_\ell\) is of one of two kinds.

\end{defx}

\begin{enumerate}
\def\labelenumi{\arabic{enumi}.}
\tightlist
\item
  A \emph{fixed segment} is a command without loops in which every index expression has the form \(c\) or \(n+c\) with an integer \(c\).
\item
  A \emph{ladder} is a loop \(\mathbf{for}\ i:=c_1\ \mathbf{to}\ n+c_2\ \mathbf{step}\ s\ \mathbf{do}\ B\ \mathbf{od}\) or a loop \(\mathbf{for}\ i:=n+c_1\ \mathbf{to}\ c_2\ \mathbf{step}\ {-s}\ \mathbf{do}\ B\ \mathbf{od}\), where \(s\ge1\) and the \emph{body} \(B\) is a command without loops that satisfies the following.

  \begin{itemize}
  \tightlist
  \item
    (a) Every qubit index expression in \(B\) has the form \(i+o\), \(c\) or \(n+c\) with integers \(o\), \(c\). The qubits \(c\) and \(n+c\) are called \emph{fixed}.
  \item
    (b) Every condition in \(B\) is on a bit \(a[i+o]\) or, in a loop with step \(s\), on a bit \(a[c]\).
  \item
    (c) For every value of the counter the fixed positions lie outside the positions \(i+o\): positions \(c\) to the left, positions \(n+c\) to the right. Namely, let \(o_{\min}\) and \(o_{\max}\) be the least and the greatest \(o\) among the expressions \(i+o\) of qubits and bits in \(B\), and let \(c\) range over the constants of the fixed qubits and of the bits \(a[c]\). For a loop with step \(s\) we require \(c<c_1+o_{\min}\) for qubits and bits of the form \(c\) and \(c>c_2+o_{\max}\) for qubits of the form \(n+c\); for a loop with step \(-s\), the same inequalities hold with the roles of \(c_1\) and \(c_2\) interchanged.
  \end{itemize}
\end{enumerate}

In both kinds of segments, the qubit index expressions of one gate application are pairwise distinct as expressions. In addition, the program must be well-defined for all sufficiently large \(n\); specifically, every linear inequality generated by Definition 4 must hold for all \(n\) above some finite bound. The partition \(P=P_1;\dots;P_L\) into segments is part of the specification of a ladder program; the threshold (Definition 4) and the automaton of Theorem 1 are defined by this partition.

The positions \(i+o_{\min},\dots,i+o_{\max}\) form the \emph{window} of the body at counter value \(i\), and \(w=o_{\max}-o_{\min}+1\) is the \emph{window width}. Let \(S\) be the set of constants \(c\) of the fixed qubits \(c\) and bits \(a[c]\) of the body, \(E\) the set of positions \(n+c\) of the fixed qubits of the body, and \(h=\lvert S\rvert+\lvert E\rvert\). By condition (c), for every value of the counter the positions of \(S\) lie to the left of the window and those of \(E\) to the right.

A cut of the register shows why the family of a ladder program is represented by one automaton of finite dimension (Figure 4a). The automaton (Section 5) reads the register in qubit order, not in the order in which gates are applied. The \emph{intermediate value} of a qubit between two bodies is the index of the basis vector in a resolution of the identity inserted between them (Section 5.1). When qubits \(0,\dots,j\) have been read, the state keeps only the intermediate values of those window qubits and fixed positions on which the next bodies will act. There are at most \(w-1\) window qubits and \(h\) fixed positions among them, for every \(n\). Although the number of loop bodies acting on a fixed qubit \(c\) grows with \(n\), these bodies act sequentially, so the automaton state needs to track only a single intermediate value for qubit \(c\) at any cut. In the language of Section 2.2, the matrix of a ladder is an MPO whose bond dimension does not depend on \(n\). Theorem 1 builds it explicitly. For the mirror layer \(F_n=\prod_{i<n/2}\mathrm{CX}(i,n-1-i)\) (Section 4.2), \(n/2\) gates cross the middle cut, and the bond dimension grows with \(n\) (Figure 4b, Proposition 1).

\begin{figure}[tbp]
\centering
\includegraphics{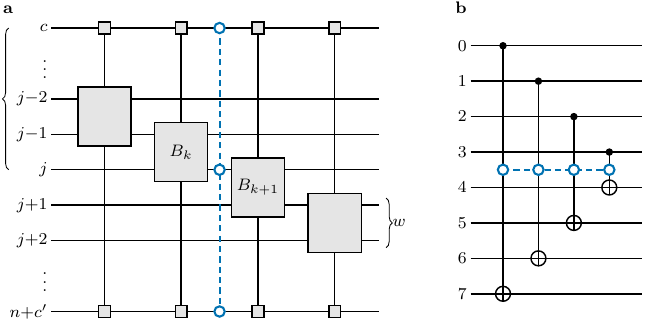}
\caption{The state of the automaton of a ladder and a cut of the mirror layer \(F_8\). \textbf{a} A ladder with window width \(w=2\), step \(s=1\) and fixed qubits \(c\) and \(n+c'\): the gray boxes and squares are bodies, \(B_k\) and \(B_{k+1}\) are adjacent bodies, the dashed line is the front between them, and the brace on the left marks the qubits \(0,\dots,j\) read so far; \textbf{b} the mirror layer \(F_8=\prod_{i=0}^{3}\mathrm{CX}(i,7-i)\), the dashed line is the cut between qubits 3 and 4. The circles are the values kept by the state of the automaton (a) and the gates that cross the cut (b)}\label{fig:3}
\end{figure}

An automaton defines a series on words of any length, while a program may be undefined for small \(n\): an index falls out of bounds, a ladder is empty, or positions \(c\) and \(n+c'\) coincide. The threshold separates such \(n\), and for them the method compares the programs directly (Section 5.5).

\renewcommand{\snthmname}{Definition 4}\begin{defx}[threshold]

For a ladder program \(P\) consider the conditions:

\begin{itemize}
\tightlist
\item
  (T1) qubit indices lie in \(\{0,\dots,N-1\}\), bit indices in \(\{0,\dots,n-1\}\);
\item
  (T2) each ladder executes its body at least once;
\item
  (T3) in each segment, positions (qubits and bits) of the form \(c\) are less than positions of the form \(n+c'\).
\end{itemize}

\end{defx}

Each condition is a conjunction of inequalities \(c+k_0n+k_1i\ge0\), where \(i\) is the counter of the ladder (\(k_1=0\) in a fixed segment). Substituting both ends of the range of the counter for \(i\) gives inequalities \(c'+k_0'n\ge0\). These imply (T1)--(T3), since an inequality linear in \(i\) holds on an interval if it holds at its ends. By Definition 3 each such inequality holds for all sufficiently large \(n\); let \(n'\) be the least number such that it holds for all \(n\ge n'\). The \emph{threshold} \(n_{\min}(P)\) is the maximum of 1 and all such bounds \(n'\).

For \(n\ge n_{\min}(P)\) and any \(a\) the program is defined, that is, \((n,a)\in\mathrm{dom}(P)\): the indices are admissible by (T1), and the qubits of one gate application are distinct by Definition 3, condition (c), and (T3).

\renewcommand{\snthmname}{Example 1}\begin{defx}[continued]

The reference \(R\) is a ladder program of three segments: the fixed segment \(\mathrm{X}(n);\mathrm{H}(n)\), the ladder \(\mathbf{for}\ i:=0\ \mathbf{to}\ n-1\ \mathbf{step}\ 1\) with window \(\{i\}\) (\(w=1\)), fixed qubit \(n\) (\(E=\{n\}\)) and condition \(a[i]=1\), and the fixed segment \(\mathrm{H}(n);\mathrm{X}(n)\). Condition (c) holds: for the fixed qubit \(n=n+0\) we have \(c=0>c_2+o_{\max}=-1+0\). The ladder is non-empty for \(n\ge1\), so \(n_{\min}(R)=1\). The solution \(G\) contains one more ladder of the same form with \(c_1=20\). It is non-empty only for \(n\ge21\), and \(n_{\min}(G)=21\).

\end{defx}

The layer of single-qubit gates \(\mathbf{for}\ i:=0\ \mathbf{to}\ n-1\ \mathbf{step}\ 1\ \mathbf{do}\ \mathrm{H}(i)\ \mathbf{od}\) (\(w=1\), \(h=0\)) and both GHZ preparation programs of Section 3.2 are also ladder programs: in the first \(S=\{0\}\) and \(w=1\), in the second \(w=2\) and \(h=0\).

\subsection{Families outside the class}\label{sec:4.2}

By Theorem 1 (Section 5), the family of any ladder program is represented by an automaton of finite dimension. Hence a family for which no such automaton exists is not given by any ladder program. An example is the \emph{mirror layer} \(F_n\): the product of the gates \(\mathrm{CX}(i,n-1-i)\) over \(i<n/2\), one gate for each pair of qubits symmetric about the middle of the register (Figure 4b).

\renewcommand{\snthmname}{Proposition 1}\begin{thmx}

Let \(F_n=\prod_{i=0}^{n/2-1}\mathrm{CX}(i,n-1-i)\) for even \(n\). For no \(n'\) is there a weighted automaton over \(\mathbb{C}\) with letters \((a_j,x_j,y_j)\) whose value on each word of length \(n\) equals \(\langle x\rvert F_n\lvert y\rangle\) for all even \(n\ge n'\).

\end{thmx}

\begin{proof}[Proof sketch]
For \(n=2k\), the block of the Hankel matrix over words of length \(k\) with \(a_j=0\) is the matricization of \(F_n\) along the cut between qubits \(k-1\) and \(k\). The cut is crossed by \(k\) CX gates on disjoint pairs of qubits, each of operator Schmidt rank 2, so the rank of the block is \(2^{k}\), and by (F4) the dimension of the automaton is unbounded.
\end{proof}

So no ladder program, with any number of ancillas, gives the family \(F_n\).\footnote{If a program has ancillas and the comparison is by \(\equiv_{\mathrm{anc}}\), one needs the series on the working qubits with the ancillas in \(\lvert0\rangle\); its automaton is built from the automaton of the program without increasing the dimension (Lemma S.7 in Online Resource 1), and by Proposition 1 there is no such automaton for \(F_n\).}

The quantum Fourier transform cannot be written in \(\mathcal{L}\): the rotation angles \(\pi/2^{j}\) depend on the loop counter, while in \(\mathcal{L}\) the angle is part of the gate and does not depend on the counter. For the cut of a register of \(2k\) qubits in half, the operator Schmidt rank of the quantum Fourier transform is \(2^{2k}\), that is, maximal \cite[Sect.~II]{Nielsen2003_QDynamicsResource}. This maximal rank arises from the bit-reversal permutation inherent in the transform \cite{Chen2023_QFTSmallEntanglement}. As in Proposition 1, it has no automaton of finite dimension.

%% file: sections/05_method.tex
\section{The method}\label{sec:5}

\subsection{The automaton of a program}\label{sec:5.1}

Let \(\Sigma=\{0,1\}^{3}\). To a word \(\sigma_0\cdots\sigma_{N-1}\) with letters \(\sigma_j=(a_j,x_j,y_j)\) and a given \(e<N\) we associate \(n=N-e\), the string \(a=(a_0,\dots,a_{n-1})\) and the vectors \(x,y\in\{0,1\}^{N}\); the bits \(a_j\) with \(j\ge n\) are not used.

\renewcommand{\snthmname}{Definition 5}\begin{defx}

An automaton \(\mathcal{A}\) over \(\mathbb{K}\) \emph{represents the matrix family} of a ladder program \(P\) with \(e\) ancillas if \(f_{\mathcal{A}}(\sigma_0\cdots\sigma_{N-1})=\langle x\rvert U_P(n,a)\lvert y\rangle\) for each word of length \(N=n+e\) with \(n\ge n_{\min}(P)\).

\end{defx}

The size \(n\) is given by the length of the word and the bit \(a_j\) by the letter of qubit \(j\), so one automaton describes all \(n\ge n_{\min}(P)\) and all \(a\). In the terms of Section 2.2 this is a uniform MPO whose letter contains, besides the output and input of qubit \(j\), the bit \(a_j\).

\renewcommand{\snthmname}{Example 1}\begin{defx}[continued]

The matrix family of the reference \(R\) is represented by the automaton \(\mathcal{A}_R\) with three states \(q_{\rm e},q_{\rm o},q_{\rm f}\) (Figure 5a). It reads the letters of the qubits in index order, not in the order in which gates are applied. The states \(q_{\rm e}\) and \(q_{\rm o}\) mean that the sum \(\sum_ja_jx_j\) over the letters read so far is even or odd, and \(q_{\rm f}\) that the last letter has been read. A letter \((a_j,x_j,y_j)\) with \(x_j\ne y_j\) has the zero matrix. Only \(q_{\rm f}\) has a final weight, and there are no transitions from \(q_{\rm f}\). So on a word of length \(n+1\) a path enters \(q_{\rm f}\) exactly on the last letter, the letter of the ancilla. Multiplying the matrices of the letters, we obtain
\[f_{\mathcal{A}_R}(\sigma_0\cdots\sigma_n)=[x=y]\,(-1)^{(a\cdot x)\,[y_n=0]},\]
where \([\cdot]\) is 1 if the condition holds and 0 otherwise. This is the action of \(U_R(n,a)\) from Section 3.1: on the working qubits with the ancilla in \(\lvert0\rangle\), and the identity with the ancilla in \(\lvert1\rangle\). Hence \(\mathcal{A}_R\) represents the matrix family of \(R\) for all \(n\ge n_{\min}(R)=1\), although the circuit \(R(n,a)\) grows with \(n\).
The automaton \(\mathcal{A}_R\) is written by hand. The automaton that the method builds for \(R\) (Theorem 1 and Lemma 1, Section 5.1) has \(16\cdot12\cdot16=3072\) states, and 4 after minimization (F3): its series differs from the series of \(\mathcal{A}_R\) only on words of length 1, which correspond to \(n=0\) outside \(\mathrm{dom}(R)\).

\end{defx}

\begin{figure}[tbp]
\centering
\includegraphics{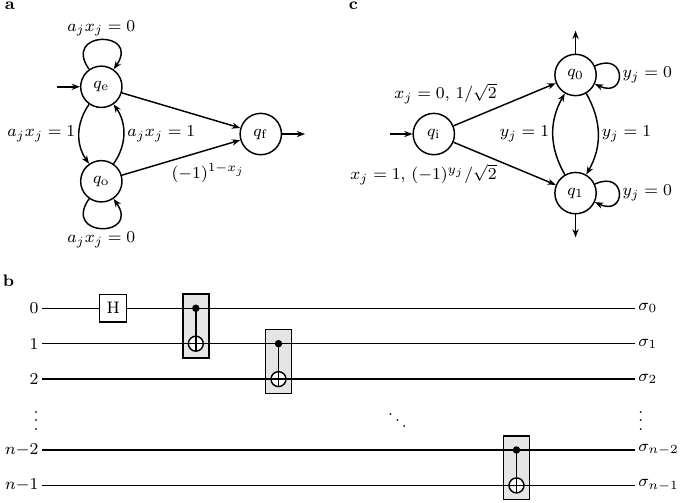}
\caption{The automata of Examples 1 and 2 and the circuit of Example 2. \textbf{a} The automaton \(\mathcal{A}_R\) of the reference of Example 1 (circuit in Figure 3a), letters \(\sigma_j=(a_j,x_j,x_j)\); \textbf{b} the circuit of the program of Example 2; \textbf{c} its automaton, letters \(\sigma_j=(a_j,x_j,y_j)\); a transition on \(\sigma_j\) leads only to \(q_{x_j}\). A label is a condition on the letter and a weight (no condition means any letter, no weight means weight 1); incoming arrows mark initial states, outgoing arrows final weight 1}\label{fig:ghz}
\end{figure}

\renewcommand{\snthmname}{Example 2}\begin{defx}

The second GHZ preparation program of Section 3.2, \(\mathrm{H}(0);\ \mathbf{for}\ i:=1\ \mathbf{to}\ n-1\ \mathbf{step}\ 1\ \mathbf{do}\ \mathrm{CX}(i-1,i)\ \mathbf{od}\), is a ladder program that consists of the fixed segment \(\mathrm{H}(0)\) and a ladder with window width \(w=2\) and \(h=0\) (Figure 5b). An entry of its matrix equals \(\langle x_0\rvert\mathrm{H}\lvert y_0\rangle\prod_{j\ge1}[x_j=x_{j-1}\oplus y_j]\). The family is represented by an automaton with three states (Figure 5c). The automaton that the method builds for this program has \(2\cdot6=12\) states, and 4 after minimization, for the same reason as in Example 1. After reading the letter of qubit \(j\), the automaton is in state \(q_{x_j}\): the output bit \(x_j\) controls the gate \(\mathrm{CX}(j,j+1)\) of the next iteration, so the state carries the value of a window qubit from body to body. In Example 1 the window has width 1, and the state carries only information about the fixed qubit \(n\) (the parity of \(a\cdot x\)). The initial state \(q_{\rm i}\) is needed because of the gate H on qubit 0. There is no smaller automaton: the block \(H[u,v]\) of the Hankel matrix with rows \(\lvert u\rvert\le3\) (including the empty word) and columns \(2\le\lvert v\rvert\le4\) is defined, since all words \(uv\) have length at least \(n_{\min}=2\). This block has rank 3, and by (F4) the rank of such a block is at most the dimension of any automaton that agrees with the series on these words.

\end{defx}

The automaton of a ladder program \(P=P_1;\dots;P_L\) is built segment by segment: the matrix of its circuit is the product \(V_L\cdots V_1\) of the matrices of the segments, and the product of matrices corresponds to the product of automata.

\renewcommand{\snthmname}{Lemma 1}\begin{thmx}[product]

Let the automata \((\alpha_1,A_1,\beta_1)\) and \((\alpha_2,A_2,\beta_2)\) represent the series \(\langle x\rvert U_1\lvert y\rangle\) and \(\langle x\rvert U_2\lvert y\rangle\). Then the automaton
\[\alpha=\alpha_2\otimes\alpha_1,\qquad A(a,x,y)=\sum_{z\in\{0,1\}}A_2(a,x,z)\otimes A_1(a,z,y),\qquad \beta=\beta_2\otimes\beta_1\]
represents the series \(\langle x\rvert U_2U_1\lvert y\rangle\); its dimension is the product of the dimensions.

\end{thmx}

\begin{proof}

This is the contraction of two MPOs over the common physical index \(z\) (Figure 2b): by the rule \((X_1\otimes Y_1)(X_2\otimes Y_2)=X_1X_2\otimes Y_1Y_2\), the value of the automaton on a word equals \(\sum_{z\in\{0,1\}^{N}}\langle x\rvert U_2\lvert z\rangle\langle z\rvert U_1\lvert y\rangle\).

\end{proof}

\begin{figure}[tbp]
\centering
\includegraphics{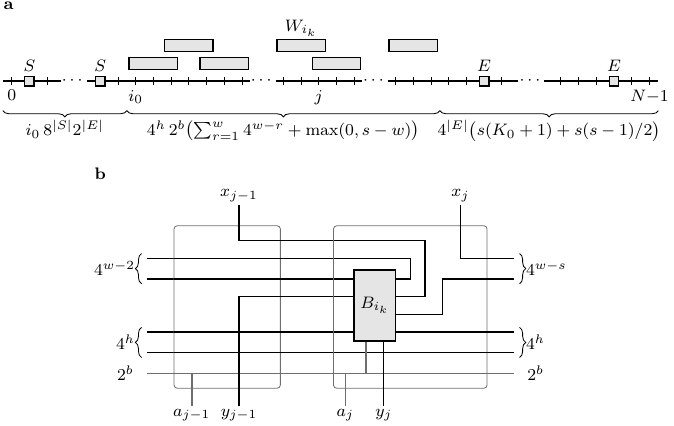}
\caption{The automaton of a ladder (Theorem 1) for \(w=3\), \(s=2\). \textbf{a} The gray boxes are the windows \(W_{i_1},\dots,W_{i_t}\), the braces indicate the three regions of the word and their respective state counts; \textbf{b} the tensors \(A(\sigma_{j-1})\) and \(A(\sigma_j)\) (frames). Bond lines from top to bottom: the outputs \(x\) and intermediate values \(\chi\) of the open window qubits, \(\chi_S,\chi_E\), the outputs \(x_S\) and the guessed inputs \(\tilde y_E\) of the positions of \(E\), the bits (gray); \(j\) is the last qubit of the window \(W_{i_k}\); the line from the output of \(B_{i_k}\) to the stored \(x\) denotes their equality}\label{fig:ladder}
\end{figure}

\renewcommand{\snthmname}{Theorem 1}\begin{thmx}[construction]

Let \(P=P_1;\dots;P_L\) be a ladder program with \(e\) ancillas. For each segment \(P_\ell\), we can construct an automaton \(\mathcal{A}_\ell\) over \(\mathbb{K}\) whose dimension does not depend on \(n\) and whose value on each word of length \(N=n+e\) with \(n\ge n_{\min}(P)\) equals \(\langle x\rvert V_\ell(n,a)\lvert y\rangle\), where \(V_\ell(n,a)\) is the matrix of the circuit of segment \(P_\ell\) on \(N\) qubits. The product of the automata \(\mathcal{A}_L,\dots,\mathcal{A}_1\) by Lemma 1 represents the matrix family of \(P\). The dimension of the automaton of a ladder is at most
\[4^{h}2^{b}\Bigl(\frac{4^{w}-1}{3}+\max(0,s-w)\Bigr)+D_0,\]
where \(w\), \(h\), \(s\) and \(S\) are as in Section 4.1, \(b\le w-1+\lvert S\rvert\) is the largest number of bits of \(a\) that the automaton keeps at the same time, and \(D_0\) is the number of states before the first and after the last window. The number \(D_0\) depends only on \(s\), \(\lvert S\rvert\), \(\lvert E\rvert\), the index \(i_0\) of the first qubit of the first window and the least number \(K_0\) of qubits to the right of the last window; explicit formulas for it and for the dimension of the automaton of a fixed segment are given in Online Resource 1, Sects. S1.2 and S1.3. In Figure~\ref{fig:ladder}a, \(D_0\) is the sum of the two outer state counts: before the first window no body has acted on the positions of \(E\), and only the guessed input is kept for them (factor \(2^{\lvert E\rvert}\)); afterwards the intermediate value is kept as well (\(4^{\lvert E\rvert}\)). For a loop with step \(-s\) the bound is the same after reflecting the qubit indices.

\end{thmx}

\begin{proof}[Proof sketch]

The state of the automaton of a fixed segment keeps the letters of the positions \(c\). The letters of the positions \(n+c\) come at the end of the word, so the state keeps their guessed values, triples \((a,x,y)\), \(8^{\lvert E\rvert}\) variants in all (the bit \(a\) exists only for working positions, \(c<0\)), and the weight is multiplied by the entry of the matrix of the segment at the bits read at the positions \(c\) and guessed at the positions \(n+c\). On the last letters a transition exists only if the letters read coincide with the stored ones, so only a single term survives in the sum over the bond index. In the language of MPOs, the bond index carries these values from qubit \(c\) to qubit \(n+c\) through the whole register. On the other qubits the matrix of a letter with \(x_j\ne y_j\) is zero.

Let the counter of a ladder with step \(s\) take the values \(i_1<\dots<i_t\), and let \(B_{i}\) be the matrix of the body at counter value \(i\). Then the matrix of the ladder equals \(B_{i_t}\cdots B_{i_1}\). Insert between \(B_{i_{k+1}}\) and \(B_{i_k}\cdots B_{i_1}\) a resolution of the identity \(\sum_{\chi}\lvert\chi\rangle\langle\chi\rvert\) over the window positions and the fixed positions; the sum over the intermediate values \(\chi\) becomes a sum over the states of the automaton. For each qubit on which a body will still act, the state keeps the output \(x_j\) read and the intermediate value \(\chi_j\) (in Figure 4a these are the circled values on the front between \(B_{i_k}\) and \(B_{i_{k+1}}\)), and also the bits \(a_j\) that the conditions will still read. The automaton reads the qubits from left to right. On the last qubit of the window of the body \(B_{i_k}\) it multiplies the weight by an entry of the matrix \(B_{i_k}\). Once no subsequent loop body acts on a given qubit, the transition enforces that its stored intermediate value equals the output bit read from the input word (Figure 6b). A ladder with step \(-s\) becomes a ladder with step \(s\) after renumbering the qubits \(j\mapsto N-1-j\), and its automaton becomes the transposed one.

We prove by induction on \(k\) that the sum of the weights of the paths that have applied \(k\) bodies equals the corresponding entry of the matrix \(B_{i_k}\cdots B_{i_1}\). A word of a given length is accepted by a unique path, which moves to the tail after the body with the last counter value. The value does not depend on the bits of the letters of the ancillas. Counting the states on three parts of the word (before the first window, the windows, after the last window) gives the dimension bound (Figure 6a). The full proof is in Online Resource 1, Sects. S1.1--S1.5 (Lemmas S.1--S.6).

\end{proof}

The first term of the bound of Theorem 1 is the number of states in the window part. A state there consists of three independent parts (Figure 6b): the pairs \((x_c,\chi_c)\) for the fixed positions \(c\in S\) and the pairs \((\tilde y_c,\chi_c)\), where \(\tilde y_c\) is the guessed input, for the positions \(n+c\in E\), at most \(4^{h}\) variants in all; the bits of \(a\) that the conditions will still read, at most \(2^{b}\) variants; and the state of the window. If \(r\le w\) letters remain to be read before the next body, \(w-r\) window qubits are open (a window qubit is open if its output has been read and the last body acting on it has not yet been applied), and the pair \((x_j,\chi_j)\) is kept for each of them. The sum over \(r\) gives \(\sum_{r=1}^{w}4^{w-r}=(4^{w}-1)/3\) window states, and the gap between windows for \(s>w\) gives another \(\max(0,s-w)\). When the window width grows while \(s\), \(\lvert S\rvert\), \(\lvert E\rvert\) and the lengths of the parts before the first and after the last window stay bounded, the leading term of the bound is \(4^{w+h}2^{b}/3\). For small \(w\) the term \(D_0\) may be larger. For the ladder of the reference of Example 1, \(w=h=s=1\), \(\lvert E\rvert=1\), \(i_0=0\), \(K_0=1\), so \(D_0=8\), the bound equals \(4+8=12\), and the automaton of this ladder built by the implementation has exactly 12 states. The dimension of the automaton of a program is at most the product of the dimensions of the automata of the segments (Lemma 1); minimization (F3) gives the least dimension.

\subsection{Relations as series}\label{sec:5.2}

We denote by \(f_P\) the series of the automaton built by Theorem 1. On words of length at least \(n_{\min}(P)+e\), it coincides with the series of any automaton that represents the matrix family of \(P\) (Definition 5).

Let \(R\) and \(G\) be ladder programs and \(n_0=\max\bigl(n_{\min}(R),n_{\min}(G)\bigr)\). For \(n\ge n_0\) both programs are defined for all \(a\) (Section 4.1), so Problem 1 requires \(G\sim R\) at all such \((n,a)\).

The relations up to phase require a series that does not change under \(U\mapsto\lambda U\), \(\lvert\lambda\rvert=1\). For a series \(f\) over letters \((a,x,y)\) we define the \emph{doubled series} \(f^{\otimes}\) over letters \((a,x,y,x',y')\): \(f^{\otimes}(w)=f(a,x,y)\,\overline{f(a,x',y')}\), where \((a,x,y)\) and \((a,x',y')\) are the words made of the corresponding components of the letters of \(w\). If \(f(w)=\langle x\rvert U\lvert y\rangle\), the values of \(f^{\otimes}\) are the entries of \(U\otimes\bar U\), and they do not change under \(U\mapsto\lambda U\). The automaton of the series \(f^{\otimes}\) is the double layer of the MPO (Figure 2c): the matrix of the letter \((a,x,y,x',y')\) is \(A(a,x,y)\otimes\bar A(a,x',y')\), as in (F1), where \(\bar A(\sigma)=\gamma_{-1}\bigl(A(\sigma)\bigr)\) and by (K1) \(\gamma_{-1}\) is complex conjugation. Its dimension is the square of the dimension of the automaton of \(f\). For \(\simeq\) no doubled series is needed: one layer \(U_GU_R^{\dagger}\) suffices (Lemma 2).

Besides the product (Lemma 1) and the difference (F1), the series of Table 1 use two operations. Let \(\mathcal{A}_P\) be the automaton of a program \(P\) by Theorem 1, and let an automaton \(\mathcal{A}=(\alpha,\{A(\sigma)\},\beta)\) represent the series \(\langle x\rvert U\lvert y\rangle\). The bar, as above, means applying \(\gamma_{-1}\). The \emph{adjoint} automaton \(\mathcal{A}^{\dagger}=\bigl(\bar\alpha,\{\bar A(a,y,x)\}_{(a,x,y)\in\Sigma},\bar\beta\bigr)\) represents the series \(\langle x\rvert U^{\dagger}\lvert y\rangle=\overline{\langle y\rvert U\lvert x\rangle}\), since \(\gamma_{-1}\) is a field automorphism and complex conjugation (K1). \emph{Fixing the input} replaces by zero the matrices of the letters with \(y\ne0\) (for the automaton of a doubled series, with \((y,y')\ne(0,0)\)), and the values give the column \(U\lvert0^{N}\rangle\). \emph{Fixing the ancilla states} replaces \(\beta\) by the vector \(\beta_z\) of Lemma S.7 (Online Resource 1). Neither taking the adjoint nor fixing these indices changes the dimension. Notation of Table 1: \(d_G\) and \(d_R\) are the dimensions of the automata \(\mathcal{A}_G\) and \(\mathcal{A}_R\); \(f_{GR^{\dagger}}\) is the series \(\langle x\rvert U_GU_R^{\dagger}\lvert y\rangle\), represented by the product (Lemma 1) of the automata \(\mathcal{A}_G\) and \(\mathcal{A}_R^{\dagger}\); \(f^{0}_{GR^{\dagger}}(w)=[x=y]\,f_{GR^{\dagger}}(w^{0})\), where \(w^{0}\) is the word \(w\) with zeros in place of all \(x_j\) and \(y_j\) (the same automaton in which the letter \((a,x,x)\) carries the matrix of the letter \((a,0,0)\) and a letter with \(x\ne y\) the zero matrix). For a program \(P\) with \(e_P\) ancillas and \(z\in\{0,1\}^{e_P}\), \(f_P^{(z)}\) denotes the series \(u\mapsto\langle x_u,z\rvert U_P\lvert y_u,0^{e_P}\rangle\) over words \(u\) of length \(n\) on the working qubits; it is represented by the automaton of \(P\) after fixing the ancillas (Lemma S.7).

\begin{table}[tbp]
\caption{Series whose zeroness on words of length at least \(N_0\) is equivalent to the relation; \(\lambda=\langle0^{N}\rvert U_GU_R^{\dagger}\lvert0^{N}\rangle\), \(\psi=U\lvert0^{N}\rangle\), and \(V\) is the operator on the working qubits of a generator that returns its ancillas (Section 3.2)}\label{tab:1}
\small

\noindent
\begin{tabularx}{\textwidth}{@{}>{\raggedright\arraybackslash\hsize=0.70\hsize\hspace{0pt}}X>{\raggedright\arraybackslash\hsize=1.45\hsize\hspace{0pt}}X>{\raggedright\arraybackslash\hsize=1.05\hsize\hspace{0pt}}Xl>{\raggedright\arraybackslash\hsize=0.80\hsize\hspace{0pt}}X@{}}
\toprule
Relation & Series $\Delta$ & Meaning & $N_0$ & Dimension \\
\midrule
$\equiv$ & $f_G-f_R$ & $U_G-U_R$ & $n_0+e$ & $d_G+d_R$ \\
$\simeq$ & $f_{GR^{\dagger}}-f^{0}_{GR^{\dagger}}$ & $U_GU_R^{\dagger}-\lambda I$ & $n_0+e$ & $2d_Gd_R$ \\
$\equiv_0$ & $f_G-f_R$ on letters with $y=0$ & $\psi_G-\psi_R$ & $n_0+e$ & $d_G+d_R$ \\
$\simeq_0$ & $f_G^{\otimes}-f_R^{\otimes}$ on letters with $y=y'=0$ & $\psi_G\otimes\bar\psi_G-\psi_R\otimes\bar\psi_R$ & $n_0+e$ & $d_G^{2}+d_R^{2}$ \\
$\equiv_{\mathrm{anc}}$ & $f_R^{(z)}$ for $z\ne0$; $f_G^{(0)}-f_R^{(0)}$ & return by $R$; $V_G-V_R$ & $n_0$ & $d_G+d_R$ \\
$\simeq_{\mathrm{anc}}$ & $f_R^{(z)}$ for $z\ne0$; $(f_G^{(0)})^{\otimes}-(f_R^{(0)})^{\otimes}$ & return by $R$; $V_G\otimes\bar V_G-V_R\otimes\bar V_R$ & $n_0$ & $d_G^{2}+d_R^{2}$ \\
\botrule
\end{tabularx}

\end{table}

The series \(f_R^{(z)}\), \(z\ne0\), test whether the reference properly uncomputes its ancillas; if any such series is non-zero, Problem 1 is ill-posed. For \(n\ge n_0\) the series \(f_G^{(z)}\) are not needed: if the reference returns its ancillas, the equality of the series of Table 1 implies this for the solution as well (Lemma 2). The dimension in Table 1 is a bound on the dimension of the difference automaton, and for relations with several series it bounds the largest of the dimensions.

\renewcommand{\snthmname}{Lemma 2}\begin{thmx}[relations as series]

Let automata represent the matrix families of \(R\) and \(G\), let \(\sim\) be a relation of Definition 2, let \(\Delta\) and \(N_0\) be as in Table 1, and let \(n\ge n_0\). For \(\equiv_{\mathrm{anc}}\) and \(\simeq_{\mathrm{anc}}\), the reference \(R\) returns its ancillas at \((n,a)\) if and only if the series \(f_R^{(z)}\), \(z\ne0\), are zero on the words with these \(n\) and \(a\). Under this condition (which holds vacuously for the other four relations), we have that \(G\sim R\) at \((n,a)\) if and only if the remaining series \(\Delta\) are zero on the words with these \(n\) and \(a\). In particular, \(G\sim R\) for all \(n\ge n_0\) and all \(a\) if and only if the series \(\Delta\) are zero on all words of length at least \(N_0\).

\end{thmx}

\begin{proof}

The words of length \(N\ge N_0\) are all quadruples \((n,a,x,y)\) with \(n\ge n_0\); the value of a series does not depend on the bits of the letters of the ancillas (Definition 5).

Cases \(\equiv\) and \(\equiv_0\): The statement follows immediately from the definitions.

Case \(\simeq\): For unitary \(U_G\) and \(U_R\), the equality \(U_G=\lambda U_R\) holds if and only if \(U_GU_R^{\dagger}=\lambda I\). Then \(\lambda=\langle0^{N}\rvert U_GU_R^{\dagger}\lvert0^{N}\rangle=f_{GR^{\dagger}}(w^{0})\), and \(\lvert\lambda\rvert=1\) follows from unitarity. Hence \(G\simeq R\) at \((n,a)\) if and only if \(f_{GR^{\dagger}}(w)=f^{0}_{GR^{\dagger}}(w)\) for all \(x,y\).

Case \(\equiv_{\mathrm{anc}}\): A generator \(P\) returns its ancillas at \((n,a)\) exactly when the vectors \(U_P\lvert y_u,0^{e_P}\rangle\) have no components with \(z\ne0\) on the ancillas, that is, \(f_P^{(z)}=0\) for \(z\ne0\). Then \(V_P\) is the matrix \(B_P\) with entries \(f_P^{(0)}\). Let \(R\) return its ancillas. Then \(V_R\) is unitary. If \(G\equiv_{\mathrm{anc}}R\), then \(B_G=V_G=V_R\). Conversely, let \(B_G=V_R\). Then \(\lVert B_G\lvert y_u\rangle\rVert=1\), while the column \(U_G\lvert y_u,0^{e_G}\rangle\) of a unitary matrix has norm 1. So its components with \(z\ne0\) are zero, \(G\) returns its ancillas, and \(V_G=B_G=V_R\). Thus \(G\equiv_{\mathrm{anc}}R\) if and only if \(f_G^{(0)}=f_R^{(0)}\).

Cases \(\simeq_0\) and \(\simeq_{\mathrm{anc}}\): We apply Lemma S.8: for matrices (or vectors) \(M_G\) and \(M_R\ne0\), the equality \(M_G=\lambda M_R\) with \(\lvert\lambda\rvert=1\) holds if and only if \(M_G[i,j]\,\overline{M_G[k,l]}=M_R[i,j]\,\overline{M_R[k,l]}\) for all \(i,j,k,l\). Here \(M_R\) is the vector \(U_R\lvert0^{N}\rangle\) of unit norm or the matrix \(V_R\), which is unitary since \(U_R\) is unitary and \(R\) returns its ancillas. Both sides are values of doubled series. For \(\simeq_{\mathrm{anc}}\), the equality of the doubled series gives \(B_G=\lambda V_R\) with \(\lvert\lambda\rvert=1\); then \(B_G\) is unitary, and, as for \(\equiv_{\mathrm{anc}}\), \(G\) returns its ancillas and \(V_G=B_G\).

\end{proof}

\subsection{Zeroness check and witness length}\label{sec:5.3}

Let a series \(\Delta\) be given by an automaton of dimension \(d\). It is checked for zeroness on words of length at least \(N_0\) in the same way as in (F2). First, bases of the spans \(\operatorname{span}\{\alpha^{\top}A(u):\lvert u\rvert=k\}\) are computed for \(k=0,1,\dots,N_0\). Each has at most \(d\) vectors. Then a basis of the span \(\operatorname{span}\{\alpha^{\top}A(u):\lvert u\rvert\ge N_0\}\) is extended by the vectors \(vA(\sigma)\) while it grows. The series is zero on words of length at least \(N_0\) if and only if \(v\beta=0\) for each vector \(v\) of the basis. Otherwise the word of a vector with \(v\beta\ne0\) is a witness. The number of field operations is \(O\bigl((N_0+1)\lvert\Sigma\rvert d^{3}\bigr)\).

\renewcommand{\snthmname}{Lemma 3}\begin{thmx}[witness length]

If the series of an automaton of dimension \(d\) is non-zero on some word of length at least \(N_0\), then it is non-zero on some word of length between \(N_0\) and \(N_0+d-1\).

\end{thmx}

\begin{proof}

Put \(W_k=\operatorname{span}\{\alpha^{\top}A(u):N_0\le\lvert u\rvert\le N_0+k\}\). Then \(W_0\subseteq W_1\subseteq\cdots\) and \(W_{k+1}=W_0+W_kA(\Sigma)\), where \(W_kA(\Sigma)=\operatorname{span}\{vA(\sigma):v\in W_k,\ \sigma\in\Sigma\}\). If \(W_{k+1}=W_k\), then \(W_{k+2}=W_0+W_{k+1}A(\Sigma)=W_{k+1}\), and the chain stabilizes. If \(W_0=0\), then all \(W_k=0\) and the series is zero on all words of length at least \(N_0\), contrary to the assumption. So \(W_0\ne0\). Since \(\dim W_k\le d\), the chain stabilizes no later than at \(k=d-1\). For a word \(u\) of length at least \(N_0\) with a non-zero value, the vector \(\alpha^{\top}A(u)\) lies in \(W_{d-1}\), and \(\alpha^{\top}A(u)\beta\ne0\). Hence \(\alpha^{\top}A(u')\beta\ne0\) for one of the words \(u'\) that span \(W_{d-1}\), and their lengths are at most \(N_0+d-1\).

\end{proof}

\renewcommand{\snthmname}{Corollary 1}\begin{thmx}

Let \(d\) be at least the dimension of some automaton of the series \(\Delta\) (for example, \(d\) is the bound of Table 1 or the dimension after minimization (F3)), and let \(n_{\rm cut}=N_0+d-e\) (for \(\equiv_{\mathrm{anc}}\) and \(\simeq_{\mathrm{anc}}\), \(N_0+d\)). If \(G\not\sim R\) at some \((n,a)\) with \(n\ge n_0\), then \(G\not\sim R\) at some \((n,a)\) with \(n_0\le n\le n_{\rm cut}\).

\end{thmx}

\begin{proof}

By Lemma 3 the series \(\Delta\) has a witness of length at most \(N_0+d-1\), that is, with \(n<n_{\rm cut}\).

\end{proof}

If the cutoff \(n_{\rm cut}\) is known, instance checking for all \(n\le n_{\rm cut}\) and all \(a\) gives the same answer as Problem 1. The cutoff \(n_{\rm cut}\) is only an upper bound, and Section 7 shows that it can be large.

\subsection{Computation modulo $p$}\label{sec:5.4}

Over the cyclotomic field \(\mathbb{K}\) the zeroness check can be done exactly (Section 2.3), but the numerators and denominators of the coefficients grow. The kernel therefore computes modulo primes \(p\) in 64-bit integers (Section 6.2). Computation modulo \(p\) (Section 2.3) can give a false zero: a non-zero value of the series \(\Delta\) becomes zero. Lemma 4 below uses the field norm of a series value to bound the number of primes that can yield a false zero, while Lemma 5 bounds this norm using the fact that gate matrices remain unitary under every Galois automorphism \(\gamma_k\).

Choose \(m\) as the least common multiple of 8 and the numbers \(4\,\mathrm{den}(q)\) over all angles \(q\pi\) of the gates of the programs \(R\) and \(G\). Then for each angle \(\theta\) the number \(e^{\mathrm{i}\theta/2}\) is a power of \(\zeta_m\), \(\mathrm{i}=\zeta_m^{m/4}\), and \(\sqrt2=\zeta_8+\zeta_8^{-1}\). The entries of the gate matrices of \(\mathcal{G}\) are \(0\), \(\pm1\), \(\pm\mathrm{i}\), \(\pm1/\sqrt2\), powers of \(e^{\mathrm{i}\theta/2}\), \(\cos(\theta/2)=(e^{\mathrm{i}\theta/2}+e^{-\mathrm{i}\theta/2})/2\) and \(\sin(\theta/2)=(e^{\mathrm{i}\theta/2}-e^{-\mathrm{i}\theta/2})/(2\mathrm{i})\). All of them lie in \(\mathbb{Z}[\zeta_m,\tfrac12]\), and the denominator is a power of \(\sqrt2\). Put \(\kappa_g=1\) for H and CH, \(\kappa_g=2\) for RX and RY, and \(\kappa_g=0\) for the other gates of \(\mathcal{G}\). Then the entries of the matrix \(\sqrt2^{\,\kappa_g}M_g\) lie in \(\mathcal{O}\). For a program \(P\), let \(\kappa_P(n)\) be the sum of \(\kappa_g\) over the gates of the fixed segments and over \(n+e_P\) copies of the body of each ladder, ignoring the conditions on bits. A body is executed at most \(n+e_P\) times, so \(\sqrt2^{\,\kappa_P(n)}\langle x\rvert U_P(n,a)\lvert y\rangle\in\mathcal{O}\) for all \(a\). The number \(\kappa_P(n)\) is non-decreasing in \(n\) and grows with \(n\) if a gate with \(\kappa_g>0\) appears in the body of a ladder (for example, a layer of H in a loop). Let \(\kappa(C)\) be the sum of \(\kappa_g\) over the gate applications of a circuit \(C\).

The automata of Sections 5.1--5.2 are built from the entries of gate matrices by addition, multiplication and application of \(\gamma_{-1}\). Hence, applying \(\pi_p\) (K3) to the entries of the automata, we obtain automata over \(\mathbb{F}_p\) that represent the series \(\pi_p\circ f\). Applying \(\gamma_{-1}\) corresponds to the substitution \(\zeta_m\mapsto\omega_p^{-1}\). Minimization (F3) and the zeroness check (Section 5.3) are performed over \(\mathbb{F}_p\) and preserve the series \(\pi_p\circ f\). The transition graph of a segment (Section 2.2) can be built directly over \(\mathbb{F}_p\): it is the subgraph of the graph over \(\mathbb{K}\) without the transitions whose weight is zero modulo \(p\), and it represents the same series \(\pi_p\circ f\). Only the number of reachable states depends on \(p\) (Lemma S.9 in Online Resource 1). If \(\pi_p(\Delta(w))\ne0\), then \(\Delta(w)\ne0\).

\renewcommand{\snthmname}{Lemma 4}\begin{thmx}[norms]

Let \(t\ge3\), \(\xi\in\mathbb{Z}[\zeta_m,\tfrac12]\), \(\xi\ne0\), \(\sqrt2^{\,\kappa}\xi\in\mathcal{O}\), and \(\lvert\gamma_k(\xi)\rvert\le c\) for all \(k\in(\mathbb{Z}/m\mathbb{Z})^{\times}\). Then there are at most \(\varphi(m)\log_2\bigl(c\,2^{\kappa/2}\bigr)/\log_2t\) primes \(p\ge t\), \(p\equiv1\pmod m\), with \(\pi_p(\xi)=0\).

\end{thmx}

\begin{proof}

Put \(\xi'=\sqrt2^{\,\kappa}\xi\in\mathcal{O}\). Since \(\pi_p(\sqrt2)^{2}=\pi_p(2)\ne0\), the equality \(\pi_p(\xi)=0\) implies \(\pi_p(\xi')=0\). Then \(\pi_p\bigl(\mathrm{Nm}(\xi')\bigr)=\prod_k\pi_p\bigl(\gamma_k(\xi')\bigr)=0\): the factor with \(k=1\) is zero. By (K2), \(\mathrm{Nm}(\xi')\) is a non-zero integer, and \(\pi_p\) on \(\mathbb{Z}\) is reduction modulo \(p\); hence \(p\) divides \(\mathrm{Nm}(\xi')\). Since \(\gamma_k(\sqrt2)=\pm\sqrt2\), we have \(\lvert\mathrm{Nm}(\xi')\rvert=\prod_k\lvert\gamma_k(\xi')\rvert\le\bigl(c\,2^{\kappa/2}\bigr)^{\varphi(m)}\). A non-zero integer of absolute value at most \(X\) has at most \(\log_2X/\log_2t\) distinct prime divisors not less than \(t\).

\end{proof}

\renewcommand{\snthmname}{Lemma 5}\begin{thmx}[automorphisms]

For each \(k\in(\mathbb{Z}/m\mathbb{Z})^{\times}\) and each gate \(g\in\mathcal{G}\), the matrix \(\gamma_k(M_g)\) obtained by applying \(\gamma_k\) to the entries is unitary. Hence for the matrix \(U\) of any circuit over \(\mathcal{G}\) the matrix \(\gamma_k(U)\) is unitary and \(\lvert\gamma_k(\langle x\rvert U\lvert y\rangle)\rvert\le1\).

\end{thmx}

The proof is a check over the list of gates of \(\mathcal{G}\) (Online Resource 1, Sect. S1.8).

Lemmas 4 and 5 give a bound for the values of the series of Table 1. For words of length \(N\) (corresponding to register size \(n\)), put \(\kappa(N)=\kappa_R(n)+\kappa_G(n)\) (multiplied by 2 for doubled series). It is non-decreasing in \(N\). The value \(\Delta(w)\) on a word of length \(N\) is a difference of two values, each of which by Lemma 5 and (K1) has absolute value at most 1 after any \(\gamma_k\). So \(\lvert\gamma_k(\Delta(w))\rvert\le2\) and \(\sqrt2^{\,\kappa(N)}\Delta(w)\in\mathcal{O}\). Let \(p_0\ge3\) be the lower bound on the primes modulo which the automata are built, a parameter of the method. By Lemma 4 with \(t=p_0\) and \(c=2\), a non-zero \(\Delta(w)\) vanishes modulo at most \(k_\Delta(N)=\bigl\lfloor\varphi(m)\bigl(1+\kappa(N)/2\bigr)/\log_2p_0\bigr\rfloor\) primes \(p\ge p_0\). The same holds for words of length less than \(N\).

\subsection{The algorithm and its correctness}\label{sec:5.5}

The answer ``equal'' requires the series \(\Delta\) to vanish modulo every prime of a \emph{plan}. The plan depends on two numbers. The first is a bound \(d\) on the dimension of the automaton of the series \(\Delta\) over \(\mathbb{K}\): by Lemma 3 a witness, if there is one, is shorter than \(N^{*}=N_0+d\), and at most \(k_\Delta(N^{*})\) primes give a false zero on it (Section 5.4). The second is the number \(k_{\rm gate}\). The algorithm obtains the bound \(d\) from the dimensions of the automata of the segments computed modulo \(p\). A prime is called \emph{bad} for a segment \(P_\ell\) if modulo it a non-zero entry of the matrix of the body (for a fixed segment, of the matrix of the segment) vanishes for some bits, and \emph{good} otherwise. By Lemma S.9 (Online Resource 1), there are at most \(k_\ell=\nu_\ell\lfloor\varphi(m)\kappa_{B_\ell}/(2\log_2p_0)\rfloor\) bad primes \(p\ge p_0\), where \(\nu_\ell\) is the number of non-zero entries of the matrices of the body and \(\kappa_{B_\ell}\) is the sum of \(\kappa_g\) over the gates of the body. Modulo a good prime, the transition graph of a segment has the same reachable states as over \(\mathbb{K}\). Modulo a bad one, it may have fewer, and then the dimension of the automaton of the segment, and with it the bound \(d\), may be underestimated. Put \(k_{\rm gate}=\max_\ell k_\ell\) over the segments of both programs.

The \emph{exact plan} is the first \(T=\max(k_{\rm gate},k_\Delta(N^{*}))+1\) primes \(p\equiv1\pmod m\), \(p\ge p_0\), in increasing order. By Dirichlet's theorem there are infinitely many such primes. Among the first \(k_{\rm gate}+1\) of them, for each segment there is a prime that does not annihilate the entries of the matrix of the body, so the numbers of states and the bound \(d\) obtained from them are correct (if \(d\) is built from the exact ranks of the segments, their correctness follows from Corollary S.11). Among all \(T\) primes, one gives no false zero on the witness. Hence the answer of the exact plan does not depend on chance (Theorem S.12). The exact plan is used if \(T\le T_{\max}\).

The \emph{probabilistic plan} is used if \(T>T_{\max}\). This happens when \(\kappa_R(n)\) or \(\kappa_G(n)\) grows with \(n\), since then \(k_\Delta(N^{*})\) grows as well. The primes are chosen at random, independently and uniformly from the set \(\mathcal{P}_m\) of primes \(p\equiv1\pmod m\) in \([p_0,2p_0)\). The plan requires \(\mathcal{P}_m\ne\varnothing\).
Random primes may underestimate the numbers of states or the least dimensions of the segments, or give a false zero on the witness. A separate batch is chosen for each of the three errors. Since the primes are independent, all \(T_k\) primes of batch \(k\) fail with probability at most the \(T_k\)-th power of the single-prime failure probability.

\begin{itemize}
\tightlist
\item
  Batch 0 of \(T_0\) primes gives the numbers \(\hat M_\ell\) of reachable states of the transition graphs of the segments. A number is underestimated only if all primes of the batch are bad for the segment.
\item
  Batch 1 of \(T_1\) primes gives the least dimensions \(\hat r_\ell\) of the automata of the segments (ranks of Hankel matrices modulo \(p\), (F4)). A rank is underestimated only if every prime of the batch annihilates a non-zero minor of the Hankel matrix over \(\mathbb{K}\).
\item
  Batch 2 of \(T_2\) primes checks \(\Delta\) for zeroness with the bound \(d\) computed from \(\prod_\ell\hat r_\ell\). One prime gives a false zero on the witness with probability at most \(q_0=\bigl(k_\Delta(N^{*})+1\bigr)/\lvert\mathcal{P}_m\rvert\) (the added 1 is a margin for the rounding in \(k_\Delta\)).
\end{itemize}
Let \(\Delta\ne0\) on words of length at least \(N_0\), and let the lengths \(T_1\), \(T_2\) and the number \(N^{*}\) be computed from the true numbers of states \(M_\ell\) and the least dimensions \(r_\ell\) of the segments. Then the probability of the answer ``equal'' is at most
\[\varepsilon=\varepsilon^{(0)}+\varepsilon^{(1)}+\varepsilon^{(2)},\qquad \varepsilon^{(0)}=\sum_\ell\Bigl(\frac{k_\ell}{\lvert\mathcal{P}_m\rvert}\Bigr)^{T_0},\qquad \varepsilon^{(2)}=q_0^{\,T_2},\]
where \(\varepsilon^{(1)}=\sum_\ell q_\ell^{\,T_1}\), and \(q_\ell\) is the bound of Lemma S.10 on the probability that one prime underestimates the rank of segment \(\ell\) (Theorem S.13). From the \(\hat M_\ell\) and \(d\) it has found, the algorithm computes the bound \(\hat\varepsilon=\hat\varepsilon^{(0)}+\hat\varepsilon^{(1)}+\hat\varepsilon^{(2)}\) by the same formulas and chooses the least batch lengths, not exceeding \(t_{\max}\), for which \(\hat\varepsilon^{(0)},\hat\varepsilon^{(1)}\le\varepsilon_0/4\) and \(\hat\varepsilon^{(2)}\le\varepsilon_0/2\). If the lengths computed from the true numbers do not reach \(t_{\max}\), then \(\varepsilon\le\varepsilon_0\) (Corollary S.14). Otherwise the algorithm reports \(\hat\varepsilon\), a bound computed from the dimensions found rather than a proved bound. For \(\hat\varepsilon\ge1\) it makes no claim and returns ``no answer''. The values of \(p_0\), \(T_{\max}\), \(\varepsilon_0\) and \(t_{\max}\) in the implementation are given in Section 6.2.

For \(n<n_0\) the automaton may not represent the program, so these sizes are checked by computing the matrices modulo \(p\) directly.

\renewcommand{\snthmname}{Proposition 2}\begin{thmx}[small $n$]

Let \(p_1\ge3\) be the lower bound on the primes for the sizes \(n<n_0\), let \(\kappa_{<}\) be the largest value of the sum \(\kappa(R(n,a))+\kappa(G(n,a))\) over \((n,a)\) with \(n<n_0\), and let \(T'=\bigl\lfloor\varphi(m)(1+\kappa_{<}/2)/\log_2p_1\bigr\rfloor+1\). Suppose that for each \((n,a)\in\mathrm{dom}(R)\) with \(n<n_0\) we have \((n,a)\in\mathrm{dom}(G)\), and that modulo each of the first \(T'\) primes \(p\equiv1\pmod m\), \(p\ge p_1\), the matrices \(U_R(n,a)\) and \(U_G(n,a)\) satisfy the condition:

\begin{itemize}
\tightlist
\item
  for \(\equiv\) and \(\equiv_0\), the entries of the compared matrices (vectors) are equal;
\item
  for \(\simeq\) and \(\simeq_0\), all \(2\times2\) minors of the two-column matrix formed by the entries of the compared matrices (vectors) are zero;
\item
  for \(\equiv_{\mathrm{anc}}\) and \(\simeq_{\mathrm{anc}}\), the entries \(\langle x,z\rvert U\lvert y,0^{e}\rangle\) with \(z\ne0\) are zero for \(U=U_R\) (the reference returns its ancillas) and for \(U=U_G\), and the blocks \(\langle x,0^{e}\rvert U\lvert y,0^{e}\rangle\) satisfy the condition for \(\equiv\) (respectively \(\simeq\)).
\end{itemize}
Then \(G\sim R\) at all \((n,a)\in\mathrm{dom}(R)\) with \(n<n_0\).

\end{thmx}

The proof applies Lemmas 4 (with \(t=p_1\)) and 5 to a non-zero difference of entries, a minor, or an entry of \(U_R\) or \(U_G\) with \(z\ne0\) (Online Resource 1, Sect. S1.8).

\begin{algorithm}[tbp]
\caption{Checking $G\sim R$ for ladder programs}\label{alg:1}
\small

\begin{enumerate}
\def\labelenumi{\arabic{enumi}.}
\tightlist
\item
  If \(e_G\ne e_R\) and \(\sim\) is one of the relations \(\equiv,\simeq,\equiv_0,\simeq_0\), return ``not equal'' and any \((n,a)\in\mathrm{dom}(R)\). Compute \(n_0=\max\bigl(n_{\min}(R),n_{\min}(G)\bigr)\) and \(m\) (Section 5.4).
\item
  Modulo primes \(p\equiv1\pmod m\), \(p\ge p_0\), build the automata of the segments of \(R\) and \(G\) (Theorem 1), compute the bound \(d\), and choose a plan. If the exact plan is not applicable and \(\mathcal{P}_m=\varnothing\), return ``no answer''.
\item
  For each prime \(p\) of the plan, multiply the automata of the segments (Lemma 1), minimizing the automaton after each product (F3), and build the automata of the series \(\Delta\) (Table 1).

  \begin{itemize}
  \tightlist
  \item
    (a) Modulo each prime of the plan, check the series \(f_R^{(z)}\), \(z\ne0\): if for one of them a word of length at least \(N_0\) with a non-zero value is found (Section 5.3), return ``reference does not return its ancillas''.
  \item
    (b) Then modulo each prime of the plan check the remaining series: if a word \(w\) with \(\pi_p(\Delta(w))\ne0\) is found, return ``not equal'' and the pair \((n,a)\) of the word \(w\).
  \end{itemize}
\item
  For \(n=1,\dots,n_0-1\) and all \(a\in\{0,1\}^{n}\) with \((n,a)\in\mathrm{dom}(R)\), check the conditions of Proposition 2. If \(R\) does not return its ancillas at \((n,a)\) (for \(\equiv_{\mathrm{anc}}\), \(\simeq_{\mathrm{anc}}\)), return ``reference does not return its ancillas''. If \((n,a)\notin\mathrm{dom}(G)\) or the check fails, return ``not equal'' and \((n,a)\).
\item
  If \(\hat\varepsilon<1\), return ``equal'' and the bound \(\hat\varepsilon\) (\(\hat\varepsilon=0\) for the exact plan); otherwise return ``no answer''.
\end{enumerate}

\end{algorithm}

\renewcommand{\snthmname}{Theorem 2}\begin{thmx}[correctness]

Let \(R\) and \(G\) be ladder programs and \(\sim\) a relation of Definition 2. The probability below is over the random choice of the primes of the probabilistic plan.

\begin{enumerate}
\def\labelenumi{\arabic{enumi}.}
\tightlist
\item
  If Algorithm 1 returns ``not equal'' and \((n,a)\), then \((n,a)\in\mathrm{dom}(R)\), and either \((n,a)\notin\mathrm{dom}(G)\) or \(G\not\sim R\) at \((n,a)\). For \(\equiv_{\mathrm{anc}}\) and \(\simeq_{\mathrm{anc}}\) this holds for the exact plan, and for the probabilistic plan under the condition that \(R\) returns its ancillas at all \((n,a)\in\mathrm{dom}(R)\).
\item
  If Algorithm 1 returns ``reference does not return its ancillas'', then \(R\) does not return them at some \((n,a)\in\mathrm{dom}(R)\).
\item
  If Algorithm 1 returns ``equal'' by the exact plan, then the answer to Problem 1 is positive.
\item
  If the answer to Problem 1 is negative or (for \(\equiv_{\mathrm{anc}}\), \(\simeq_{\mathrm{anc}}\)) \(R\) does not return its ancillas at some \((n,a)\in\mathrm{dom}(R)\), then the probability that Algorithm 1 returns ``equal'' by the probabilistic plan is at most the sum \(\varepsilon\) of Section 5.5 computed from the true numbers of states \(M_\ell\) and the least dimensions \(r_\ell\) of the segments. If the batch lengths computed from these numbers do not reach \(t_{\max}\), then \(\varepsilon\le\varepsilon_0\) (Corollary S.14).
\end{enumerate}

\end{thmx}

The proof is in Appendix A.

\emph{Remark.} The condition of item 1 is needed only for the probabilistic plan, where a non-zero series \(f_R^{(z)}\) may vanish modulo all primes of the plan, and the algorithm answers ``not equal'' instead of ``reference does not return its ancillas''; for the exact plan such a series is non-zero modulo at least one of the \(T\) primes (Appendix A, item 1).

\emph{Cost.} Let \(d\) be the dimension bound of the series \(\Delta\) and \(T_{\rm all}\) the number of primes of the plan. Steps 2--3 take \(O\bigl(T_{\rm all}(N_0+1)\lvert\Sigma\rvert d^{3}\bigr)\) operations in \(\mathbb{F}_p\) for the zeroness check, plus building and minimizing the automata of the segments; \(d\) is at most the product of the dimensions of the automata of the segments and grows exponentially with \(w+h\) (Theorem 1). Step 4 enumerates all \(2^{n}\) strings \(a\) for each \(n<n_0\) and computes \(2^{n+e}\times2^{n+e}\) matrices, so its cost grows as \(2^{n_0}\,4^{n_0+e}\) up to a polynomial factor. Condition (T2) implies that the ladder \(\mathbf{for}\ i:=K\ \mathbf{to}\ n+c_2\) gives \(n_0\ge K-c_2\); for large \(K\), step 4 is infeasible, and the method cannot establish the equality of such a pair.

\renewcommand{\snthmname}{Example 1}\begin{defx}[concluded]

For the pair \((R,G)\) and the relation \(\equiv_{\mathrm{anc}}\), Algorithm 1 finds a word with \(n=21\) in step 3 and answers ``not equal''. For the pair \((R,G')\) the exact plan gives the answer ``equal''; this statement holds for all \(n\ge1\) and all \(a\).

\end{defx}

%% file: sections/06_implementation.tex
\section{Implementation}\label{sec:6}

The method is implemented in Python. The input is two functions \texttt{solve(n,\ a)} (or \texttt{solve(n)} without a bit parameter) that build a Qiskit \cite{JavadiAbhari2024_Qiskit} circuit of class \texttt{QuantumCircuit} on \(n+e\) qubits; the ancillas are the last \(e\) qubits.
The translator (Section 6.1) builds a ladder program and a threshold from each function, and the kernel (Section 6.2) runs Algorithm 1. The stages run in this order: the translator, step 1 of Algorithm 1, the quick search (Section 6.2), steps 2--5; Table 2 lists which stage gives which answer.

\subsection{Translator}\label{sec:6.1}

The translator executes the code of the function symbolically: \(n\) stays a symbol, and the loop counter becomes the symbol \(i\). The value of an expression is an affine form \(c+k_0n+k_1i\) with integer coefficients, an interval of qubit indices (the value \texttt{list(range(b,\ b\ +\ l))} with affine start \(\beta\) and length \(\mu\)), or a \texttt{range} with affine bounds and a constant step.

The translator derives judgments \(\Gamma\vdash E\Rightarrow v\mid\Phi\): under the variable environment \(\Gamma\), the expression \(E\) has the abstract value \(v\) provided the \emph{proof obligations} \(\Phi\) hold; these are inequalities \(c+k_0n+k_1i\ge0\) under which the value of the expression in Python equals \(v\). The four main inference rules are given below; the full list is in Table S1 and Sects. S2.3--S2.5 of Online Resource 1.

\medskip
\noindent\fbox{\begin{minipage}{\dimexpr\textwidth-2\fboxsep-2\fboxrule\relax}
\small
\[
\frac{\Gamma\vdash E_1\Rightarrow\eta_1\mid\Phi_1\qquad\Gamma\vdash E_2\Rightarrow\eta_2\mid\Phi_2}{\Gamma\vdash E_1+E_2\Rightarrow\eta_1+\eta_2\mid\Phi_1\cup\Phi_2}\ \text{(sum)}
\]
\[
\frac{\Gamma\vdash Q\Rightarrow[\beta,\beta+\mu)\mid\Phi_1\qquad\Gamma\vdash k\Rightarrow\eta\mid\Phi_2}{\Gamma\vdash Q[k]\Rightarrow\beta+\eta\mid\Phi_1\cup\Phi_2\cup\{\eta\ge0,\ \mu-1-\eta\ge0\}}\ \text{(index)}
\]
\[
\frac{\Gamma\vdash k\Rightarrow\eta\mid\Phi_1\qquad\Gamma\vdash \mathit{body}\Rightarrow B\mid\Phi_2}{\Gamma\vdash \texttt{if a[k]:}\ \mathit{body}\Rightarrow \mathbf{if}\ a[\eta]=1\ \mathbf{then}\ B\ \mathbf{fi}\mid\Phi_1\cup\Phi_2\cup\{\eta\ge0,\ n-1-\eta\ge0\}}\ \text{(bit)}
\]
\[
\frac{\Gamma\vdash E_1\Rightarrow c_1\mid\Phi_1\qquad\Gamma\vdash E_2\Rightarrow n+c_2\mid\Phi_2\qquad s>0\qquad\Gamma[i]\vdash \mathit{body}\Rightarrow B\mid\Phi}{\begin{array}{c}\Gamma\vdash \texttt{for i in range(}E_1\texttt{, }E_2\texttt{, s):}\ \mathit{body}\\ \Rightarrow \mathbf{for}\ i:=c_1\ \mathbf{to}\ n+c_2-1\ \mathbf{step}\ s\ \mathbf{do}\ B\ \mathbf{od}\mid\Phi_1\cup\Phi_2\cup\Phi|_{\partial}\end{array}}\ \text{(loop)}
\]
Here \(\eta,\eta_1,\eta_2\) are affine forms, \(\Gamma[i]\) is \(\Gamma\) with the counter symbol \(i\), and \(\Phi|_{\partial}\) are the conditions \(\Phi\) at \(i=c_1\) and \(i=n+c_2-1\); since the conditions are linear in \(i\), they imply the conditions at all values of the counter.
\end{minipage}}
\medskip

The condition of the rule ``index'' excludes a negative index, for which Python takes an element from the other end of the list, and an index past the end of the list, for which Python raises an exception.

A \texttt{for} loop over a \texttt{range} with bounds of the form \(c\) and \(n+c\) becomes a ladder, a call \texttt{qc.g(...)} becomes a gate application, and a bit test (\texttt{a{[}k{]}}, \texttt{(a\ \textgreater{}\textgreater{}\ k)\ \&\ 1}) becomes a condition on a bit. The \emph{translator threshold} \(n_{\rm tr}(P)\) is the largest of the thresholds of the proof obligations and of the inequalities of Definition 4, so \(n_{\rm tr}(P)\ge n_{\min}(P)\). On other code the translator returns a refusal with a reason code (Table S2 in Online Resource 1). Refusals are of two kinds. An \emph{out-of-class} refusal means that the code uses a construct absent from ladder programs (for example, branching on \(n\) or an angle depending on \(n\) or \(i\)). An \emph{out-of-fragment} refusal means that the translator does not parse the syntax of the code (for example, a default parameter), although the same circuit may be given by a ladder program. A refusal code does not always determine the kind (mark ``both'' in Table S2).

The translator computes the expression of a rotation angle symbolically, in the form \(q\pi+r\) with rational \(q\) and \(r\). If \(r=0\), the gate of the program \(P\) gets the angle \(q\pi\). If \(r\ne0\) or the expression has another form, the translator refuses with the code \texttt{nonconst\_angle} (the rule is in Online Resource 1, Sect. S2.2).

\renewcommand{\snthmname}{Theorem 3}\begin{thmx}[translator]

Assume (A1), (A2) on the semantics of Python and Qiskit and (A4) on the parameter \(a\) (the translator checks (A3) itself):

\begin{itemize}
\tightlist
\item
  (A1) Python executes the supported fragment by its standard semantics, and module constants do not change at run time;
\item
  (A2) a call \texttt{qc.g(q\_1,\ ...,\ q\_k)} with pairwise distinct qubits of the register appends to the circuit one application of the gate \(g\) to the qubits \(q_1,\dots,q_k\) and raises no exception;
\item
  (A3) the symbolic value of the angle expression of each gate has the form \(q\pi\), that is, \(r=0\) (for \(r\ne0\) the translator refuses, Sect. S2.2);
\item
  (A4) the parameter \(a\) is a list of zeros and ones or an integer whose bits are read by shifts.
\end{itemize}
If the translator returns a program \(P\) satisfying Definition 3 and a threshold \(n_{\rm tr}(P)\) for a function, then for all \(n\ge n_{\rm tr}(P)\) and all \(a\in\{0,1\}^{n}\) the call \texttt{solve(n,\ a)} terminates without an exception and builds the circuit \(P(n,a)\), up to replacing the angle \(\theta\) of each gate, computed by Python in floating-point arithmetic, with the symbolic value \(q\pi\) of its expression.

\end{thmx}

\begin{proof}[Proof idea]

Induction on the structure of the expression (Lemma S.15), consistency of names (Lemma S.16), and the same execution path of the body of a ladder at all values of the counter (Lemma S.17). The full proof is in Online Resource 1, Sect. S2.

\end{proof}

Theorem 3 proves the rules of the translator, not their implementation. We cross-checked the implementation against the Qiskit circuits: for each function of the set of Section 7 accepted by the translator (Section 7.2), the program \(P\) at \(n_{\rm tr}(P)\le n\le n_{\rm tr}(P)+4\) and four random strings \(a\) per \(n\) matched the Qiskit circuit gate application by gate application.

For \(n<n_0\) the implementation performs step 4 of Algorithm 1 by running the functions themselves: \((n,a)\in\mathrm{dom}\) if the function terminates without an exception. The angles of the built circuit are converted to rational multiples of \(\pi\) with denominator at most \(2^{12}\). If an angle differs from the nearest such number by more than \(10^{-12}\), the implementation refuses. This requires one more assumption:

\begin{itemize}
\tightlist
\item
  (A5) for \(n<n_0\), the angle expression of each gate of the circuit built by the function has the symbolic value \(q\pi\) with \(\mathrm{den}(q)\le2^{12}\) and \(4\,\mathrm{den}(q)\mid m\) (\(m\) from step 1 of Algorithm 1), and its floating-point value differs from \(q\pi\) by less than \(10^{-12}\).
\end{itemize}
Two distinct fractions \(q,q'\) with denominators at most \(2^{12}\) differ by more than \(2^{-24}\), so the angles \(q\pi\) and \(q'\pi\) differ by more than \(\pi\cdot2^{-24}\), which is much larger than \(2\cdot10^{-12}\). Under (A5), rounding therefore recovers the symbolic value of the angle.

\renewcommand{\snthmname}{Corollary 2}\begin{thmx}

Suppose the translator accepts the functions of the reference and the solution and returns programs satisfying Definition 3, (A1)--(A5) hold, Algorithm 1 is run with the threshold \(n_0=\max\bigl(n_{\rm tr}(R),n_{\rm tr}(G)\bigr)\), and step 4 is performed by running the functions. Then the statements of Theorem 2 hold for the generators given by these functions with exact angles: the angle of each gate is replaced with the symbolic value \(q\pi\) of its expression.

\end{thmx}

\begin{proof}

Algorithm 1 and Theorem 2 remain valid for any \(n_0\ge\max\bigl(n_{\min}(R),n_{\min}(G)\bigr)\): Lemma 2 and Table 1 require only \(n\ge n_{\min}\). For \(n\ge n_0\) the functions build the same circuits as the programs, up to this replacement of angles (Theorem 3), and the programs are defined for all \(a\) (Section 4.1). Hence steps 2--3 and Lemma 2 apply to the generators with exact angles. For \(n<n_0\), step 4 checks the functions themselves. By (A5), rounding gives the symbolic values of the angles, and Proposition 2 applies to the circuits with these angles.

\end{proof}

\subsection{Kernel}\label{sec:6.2}

The kernel computes in \(\mathbb{F}_p\) with 64-bit integers and sums products of residues without intermediate reduction modulo \(p\). Before each matrix multiplication with inner dimension \(D\), it checks the condition \(D(p-1)^{2}<2^{63}\) and stops with an error if it fails. On the set of Section 7 this never happened.\footnote{The primes of the probabilistic plan are less than \(2^{25}\), so for them the condition holds for \(D\le2^{13}\). The primes of the exact plan and of the exact ranks are taken in increasing order starting from \(2^{24}\) and may exceed \(2^{25}\). For them the condition is checked in the same way.} The lower bound on the primes for the automata (Section 5.4) is \(p_0=2^{24}\): for \(p<2^{25}\) and \(\mathcal{P}_m\subset[p_0,2p_0)\) this is the largest admissible value, and it gives the largest \(\log_2p_0\) in the bounds of Sections 5.4--5.5. So \(\mathcal{P}_m\) is the set of primes \(p\equiv1\pmod m\) in \([2^{24},2^{25})\). Step 4 of Algorithm 1 reduces each product modulo \(p\) before summation and checks that \(p<2^{31}\). Then the product is less than \(2^{62}\), so in step 4 \(p_1=2^{30}\) (Proposition 2).

The parameters of the plans (Section 5.5) are \(T_{\max}=8\), \(\varepsilon_0=2^{-40}\), \(t_{\max}=200\). The least dimension of the automaton of a segment or a program is computed exactly by Corollary S.11 if this needs at most 64 primes; otherwise the bound by the number of states of the transition graph is used. The primes of the probabilistic plan are chosen by a pseudorandom number generator. We assume that its choice is uniform and independent, as Theorem 2 requires. The implementation returns ``no answer'' if \(\hat\varepsilon\ge1\), or if the exact plan is not applicable and \(\mathcal{P}_m=\varnothing\). It computes the batch lengths of the probabilistic plan before choosing the primes, and before each batch it computes \(\hat\varepsilon\) from the numbers already found. If the bound is at least 1, the remaining batches are skipped and the answer is ``no answer''.

The cutoff \(n_{\rm cut}\) reported by the implementation is the cutoff of Table 1 with bounds in place of \(d_G\) and \(d_R\): in the exact plan, with the least dimensions by Corollary S.11 or, if these need more than 64 primes, with the product of the ranks of the segments; in the probabilistic plan, with the product of the largest dimensions of the automata of the segments after minimization over the primes of the plan (the probability that such a dimension is less than the rank is included in \(\varepsilon\)).

The kernel assembles the series \(f_{GR^{\dagger}}\) of the relation \(\simeq\) segment by segment, starting from the middle: first the product of the first segment of \(G\) and the first segment of \(R^{\dagger}\), then the next segment of \(G\) on the left and the next segment of \(R^{\dagger}\) on the right, with minimization after each step. If the first segments of the reference and the solution coincide, their product equals \(I\), and after minimization it does not increase the dimension before the other segments are added. The same order is used for decision diagrams in \cite[Sect.~IV-A]{Burgholzer2020_AdvancedEC}.

Before building the automata, the implementation searches for a counterexample by running the functions (\emph{quick search}): for all \(n\) with \(n+\max(e_R,e_G)\le13\) and at most eight strings \(a\) per \(n\), it compares, by the relation of the task, the images of two random vectors (for \(\equiv_0\) and \(\simeq_0\), the image of \(\lvert0^{N}\rangle\)) under both circuits modulo a random prime \(p\equiv1\pmod m\). For \(\equiv_{\mathrm{anc}}\) and \(\simeq_{\mathrm{anc}}\) it also checks whether the reference and the solution return their ancillas. Pairs \((n,a)\) at which the reference raises an exception are skipped, and an exception of the solution gives a counterexample. A violation of the relation modulo \(p\) (the images differ or are not proportional) implies a violation over \(\mathbb{K}\), since \(\pi_p\) is a homomorphism. If there is no violation, the search makes no claim. The answer is determined by the first pair found: for \(\equiv_{\mathrm{anc}}\) and \(\simeq_{\mathrm{anc}}\) it is ``reference does not return its ancillas'' if the reference does not return them at this pair, and ``not equal'' otherwise. The return of the ancillas by the reference is checked here modulo one prime, so for \(\equiv_{\mathrm{anc}}\) and \(\simeq_{\mathrm{anc}}\) the answer ``not equal'' is correct under the same condition as in item 1 of Theorem 2. If the translator accepts the reference and refuses the solution only because a condition reads a bit \(a_k\) with \(k\ge n\) (Table S2), the quick search is still run on this pair. A difference found is a counterexample, since it is obtained by running the functions.

If \(e_G\ne e_R\) and the relation is one of \(\equiv,\simeq,\equiv_0,\simeq_0\), the implementation answers ``not equal'' at \(n=n_{\min}(R)\) (step 1 of Algorithm 1). In step 3(b), after the series \(f_R^{(z)}\), the implementation also checks the series \(f_G^{(z)}\), \(z\ne0\), for zeroness. By Lemma 2 they are not needed for the answer, but their automata have dimension \(d_G\) rather than \(d_G^{2}+d_R^{2}\) as for the doubled series, so a solution that does not return its ancillas is filtered out at lower cost. If the reference returns its ancillas, a non-zero value implies \(G\not\sim R\), so Theorem 2 remains valid. For \(\equiv_{\mathrm{anc}}\) and \(\simeq_{\mathrm{anc}}\) in the probabilistic plan, before answering ``not equal'' and \((n,a)\), the implementation checks exactly that the reference returns its ancillas at these \(n\) and \(a\): the series \(f_R^{(z)}\), \(z\ne0\), on the words with the bits \(a\) are checked for zeroness modulo the first \(\lfloor\varphi(m)(1+\kappa_R(n)/2)/\log_2p_0\rfloor+1\) primes \(p\equiv1\pmod m\), \(p\ge p_0\) (Lemmas 4 and 5). Hence the answer ``not equal'' of step 3(b) is correct even without the condition of item 1 of Theorem 2. Together with a counterexample, the implementation reports the stage at which it was found: the comparison of the numbers of ancillas, the quick search, step 3 or step 4.

Table 2 lists the answers of the implementation. Theorem 2 and Corollary 2 give the meaning of the first three.

\begin{table}[tbp]
\caption{Answers of the implementation.}\label{tab:2}
\small

\noindent
\begin{tabularx}{\textwidth}{@{}>{\raggedright\arraybackslash\hsize=0.50\hsize\hspace{0pt}}X>{\raggedright\arraybackslash\hsize=1.10\hsize\hspace{0pt}}X>{\raggedright\arraybackslash\hsize=1.40\hsize\hspace{0pt}}X@{}}
\toprule
Answer & When & What is claimed \\
\midrule
not equal, $(n,a)$ & $e_G\ne e_R$ (step 1), the quick search (including after a refusal of the translator on the solution, Section 6.2) or steps 3--4 found a non-zero value modulo $p$, or the solution raised an exception & $(n,a)\in\mathrm{dom}(R)$, and either $(n,a)\notin\mathrm{dom}(G)$ or the generators are not in the relation at $(n,a)$ (for $\equiv_{\mathrm{anc}}$, $\simeq_{\mathrm{anc}}$, if the counterexample was found by the quick search, under the condition that the reference returns its ancillas at all $(n,a)\in\mathrm{dom}(R)$; Theorem 2, item 1) \\
equal, $\hat\varepsilon$ & all zeroness checks passed, $\hat\varepsilon<1$ & for $\hat\varepsilon=0$ (exact plan) the answer to Problem 1 is positive; for the probabilistic plan $\hat\varepsilon$ is a bound on the error probability (Section 5.5; Theorem 2, item 4) \\
reference does not return its ancillas & the series $f_R^{(z)}$, $z\ne0$ (step 3(a)), or the return check in the quick search, at a counterexample of the probabilistic plan, or for $n<n_0$ (step 4) gave a non-zero value (only $\equiv_{\mathrm{anc}}$, $\simeq_{\mathrm{anc}}$) & Problem 1 is ill-posed (precondition violated) \\
no answer & the bound $\hat\varepsilon$ of the probabilistic plan is at least 1 (the remaining batches are skipped), or the exact plan is not applicable and $\mathcal{P}_m=\varnothing$ (step 2) & nothing \\
refusal & the translator did not accept a function (Section 6.1) & nothing \\
timeout & the time limit expired at any stage & nothing \\
\botrule
\end{tabularx}

\end{table}

%% file: sections/07_evaluation.tex
\section{Experiments}\label{sec:7}

We answer four research questions. RQ1: are the answers of the method correct? RQ2: which errors does the method find that instance checks miss? RQ3: how long does the method take to answer for all \(n\), compared with enumeration of sizes \(n\) by the QCEC tool and with the AutoQ-Para tool? RQ4: what code does the translator accept?

\subsection{Experimental setup}\label{sec:7.1}

\emph{Data.} The benchmark contains 169 tasks and 2200 reference--solution pairs. The tasks fall into categories: sixteen categories differ in the placement of the repeated block (neighboring qubits with a shift, even and odd pairs, from one qubit to all others, a ring, a condition on the bits \(a\), a gate with a growing number of controls, and others), and ten differ in the kind of algorithm (circuits similar to the Fourier transform \cite{Coppersmith1994_ApproxQFT} and to Grover's algorithm \cite{Grover1996_Search}, phase estimation \cite{Kitaev1995_PhaseEstimation}, an adder \cite{Cuccaro2004_RippleAdder}, an oracle on the bits \(a\), preparation of entangled and other states, parameterized circuits of rotation layers, a Trotter step for the Hamiltonian of a qubit chain, and others). For the 26 tasks whose reference the translator does not accept (among them the mirror layer \(F_n\) and the Fourier transform with qubit reversal, which have no automaton of finite dimension, Section 4.2), the benchmark contains a reference--reference pair (26 pairs). These pairs test the refusal. The bits \(a\) are used by 26 tasks (429 pairs).

Each task has a relation: \(\simeq\) for 2059 pairs, \(\equiv_{\mathrm{anc}}\) for 48, \(\simeq_0\) for 93. Three other relations are checked separately on the same pairs. For this, the relation of the task is replaced with the relation that differs from it only in the treatment of the global phase, the \emph{second relation} of the task: \(\simeq\) is replaced with \(\equiv\), \(\simeq_0\) with \(\equiv_0\), \(\equiv_{\mathrm{anc}}\) with \(\simeq_{\mathrm{anc}}\) (Section 7.2).

The correct solutions are the 26 reference--reference pairs and 420 equivalent rewritings of the reference: another order of commuting gates, a helper function, a loop traversed backward, a shifted counter, and so on. The remaining 1754 solutions are \emph{mutants}: each is obtained by one \emph{mutation}, that is, one edit of the code of a correct solution (Table~\ref{tab:mut}).

\begin{table}[tbp]
\caption{Mutation kinds.}\label{tab:mut}
\small

\footnotesize\noindent
\begin{tabularx}{\textwidth}{@{}l>{\raggedright\arraybackslash}Xr@{}}
\toprule
Kind & Examples & Pairs \\
\midrule
gate & extra, missing or replaced gate; swapped control and target; wrong angle; gate that changes only the global phase & 318 \\
loop & bound \(\pm1\), including with step 2; wrong step; reversed traversal order; missing or extra loop, including an extra loop with the counter from \(K\in\{3,11,20\}\) & 957 \\
index & index counted from the end of the register & 120 \\
condition on bits \(a\) & neighboring bit, negation, reversed order, special case \(a=1\cdots1\) & 84 \\
branch on \(n\) & comparison with a constant, on a residue of \(n\), explicit circuits for small \(n\), missing edge case & 164 \\
angle & wrong angle formula; extra rotation ladder of the approximate Fourier transform & 15 \\
ancilla & not returned to \(\lvert0\rangle\) or prepared in a wrong state & 6 \\
phase & extra relative phase at the start of the circuit, lost sign of an amplitude & 90 \\
\botrule
\end{tabularx}

\end{table}

The gate mutations are those generated by the mutation tools for quantum programs Muskit \cite[Sect.~III]{Mendiluze2021_Muskit} and QMutPy \cite[Sects.~2.1--2.2]{Fortunato2022_QMutPy}. The mutations of loops, bit conditions and branches on \(n\) are ours. The missing edge case corresponds to the most frequent bug pattern in the code of quantum platforms \cite[Sect.~4.4]{Paltenghi2022_BugsPlatforms}. A mutation that changes only the global phase gives a solution equal to the reference under \(\simeq\) (37 pairs). So the benchmark has 483 equal pairs and 1717 faulty ones.

\emph{Ground truth.} The ground truth for each pair is given by construction and checked by computation: by full matrices at \(n\le7\) (for tasks with bits, at \(n\le6\), with all \(a\) at \(n\le4\) and eight strings \(a\) at larger \(n\)); at \(n>7\), by the state vector on \(\lvert0^{N}\rangle\) and on a random product of single-qubit states at those \(n\le24\) at which some loop from a constant bound to a bound of the form \(n+c\) executes its first, second or third iteration. For a faulty solution, the smallest tested size at which it differs from the reference is found. For correct solutions, the statement ``equal for all \(n\)'' follows from the construction. The ground truth does not depend on the answer of the method.

\emph{Instance checks.} We compare the method with three fixed-size checks. The first two extend to several sizes the comparison of the output state \cite[Sect.~3.2]{Mikuriya2025_QCoderBenchmark} and the comparison of the action of the circuit at a fixed \(n\) \cite{Guo2025_QuanBench} (there by process fidelity, here on random inputs). The third is a sampled check of variable-size circuits \cite{Pontolillo2025_QuCheck}. The bounds \(n\le10\) and \(n\le20\) are our choice. Each check compares the circuits by the relation of the task and stops at the first difference found. If the reference raises an exception, the case is skipped, and if the solution does, the error counts as found. The \emph{state check} compares the output on \(\lvert0^{N}\rangle\) at each \(n\le10\); the \emph{random-input check} compares the action on two random products of single-qubit states at each \(n\le10\) \cite[Sect.~3.2, Thm.~1]{Burgholzer2020_RandomStimuli}. Both take all strings \(a\) at \(n\le3\) and five strings (\(0^{n}\), \(1^{n}\) and three random ones) at larger \(n\). The \emph{sampled check} takes 20 cases with \(n\) chosen uniformly from \(\{1,\dots,20\}\), \(a\) uniformly from \(\{0,1\}^{n}\), and one random product of single-qubit states as input. For \(\simeq_0\), the input of all three checks is \(\lvert0^{N}\rangle\).

\emph{Enumeration.} For RQ3 we use the QCEC tool \cite{Burgholzer2021_QCEC,Wille2024_MQTHandbook}, version 3.11, which implements the methods of \cite{Burgholzer2020_AdvancedEC,Peham2022_ZX_EC}. For each equal pair, it checks all \(n\) from 1 to the cutoff \(n_{\rm cut}\) reported by the method, with a time limit of 60~s on the whole enumeration and, for tasks with bits, at most 4096 strings \(a\) at each \(n\). QCEC was run with the alternating and simulation methods and the default tolerance. The global phase of the circuit was moved into the gates, and the answer ``equivalent up to global phase'' was accepted only for the relations up to phase. For \(\equiv_0\) and \(\simeq_0\), all qubits were marked as ancillas in the state \(\lvert0\rangle\). For \(\equiv_{\mathrm{anc}}\), the qubits beyond the number of qubits of the smaller circuit were marked as ancillas, and with equal numbers of qubits the full matrices were compared, which is stronger than \(\equiv_{\mathrm{anc}}\). Our QCEC evaluation harness skips the sizes at which a generator raises an exception. The time limit of the method is 300~s per pair. The method and QCEC run single-threaded on a machine with an Intel Xeon 2.1~GHz processor (4 cores, 15~GB of memory).
The times in Sections 7.2--7.5 were measured with one process per machine. Each time is from a single run, except in Table~\ref{tab:6}, where it is the median of three. Software versions and reproduction commands are in the README of the data archive. The language model Claude Opus 5.5 (Anthropic) assisted the author in developing the code and running the experiments (see ``Declarations'').

\subsection{RQ1: correctness}\label{sec:7.2}

\begin{table}[tbp]
\caption{Answers of the method and ground truth: number of pairs for the task relation and for the second relation; the ground truth is for the same relation. There were no answers ``equal'' with \(2^{-40}<\hat\varepsilon<1\) and no answers ``reference does not return its ancillas'' (Table~\ref{tab:2}).}\label{tab:3}
\small

\noindent
\begin{tabularx}{\textwidth}{@{}l>{\raggedright\arraybackslash}Xrr@{}}
\toprule
\begin{tabular}[b]{@{}l@{}}Ground\\truth\end{tabular} & Answer of the method & \begin{tabular}[b]{@{}r@{}}Task\\relation\end{tabular} & \begin{tabular}[b]{@{}r@{}}Second\\relation\end{tabular} \\
\midrule
not equal & ``not equal'' with counterexample & 1432 & 1469 \\
not equal & refusal: mutation takes the code out of the class & 284 & 284 \\
not equal & timeout (300~s) & 1 & 1 \\
equal & ``equal'', exact plan & 220 & 278 \\
equal & ``equal'', \(\hat\varepsilon\le2^{-40}\) & 224 & 138 \\
equal & ``no answer'', \(\hat\varepsilon\ge1\) & 12 & 3 \\
equal & refusal on the reference & 26 & 26 \\
equal & refusal: solution loop not of ladder form & 1 & 1 \\
 & \textbf{wrong answers} & \textbf{0} & \textbf{0} \\
\botrule
\end{tabularx}

\end{table}

The method produced no incorrect verdicts, neither for the task relations nor for the second relations (Table~\ref{tab:3}). By Theorem 2 and Section 6.2, each answer ``not equal'' with a counterexample is correct (for \(\equiv_{\mathrm{anc}}\) and \(\simeq_{\mathrm{anc}}\), if the counterexample came from the quick search, under the condition of item 1 of Theorem 2, which was checked for the references of the benchmark at small \(n\) together with the ground truth), so the agreement of these answers with the ground truth tests the implementation. The method gave a conclusive answer (``not equal'' with a counterexample, or ``equal'') on 1876 of 2200 pairs (1885 for the second relations). On all 37 pairs with a global-phase mutation, equal under \(\simeq\), the method answered ``not equal'' under \(\equiv\).

On 12 equal pairs the computed bound \(\hat\varepsilon\) is at least 1, and the method returns ``no answer''. In these pairs, gates with \(\kappa_g>0\) lie in the body of a ladder, and the dimension bound obtained as a product over segments is large (Section 5.5). All 284 refusals on mutants are caused by out-of-class constructs. Of the 26 refusals on references, 21 are out of class. For the other five, as for the refusal on a correct solution, the refusal code does not determine the kind (Table S2). The translator accepted 2023 of the 2369 functions of the benchmark (169 references and 2200 solutions), and each accepted function matched its program at the checked sizes (Section 6.1). Step 4 never refused because of an angle (assumption (A5), Section 6.1).

The number of states of the automaton of a ladder built by the implementation does not exceed the bound of Theorem 1 on all 1054 distinct ladders of the benchmark (a check of the implementation; Online Resource 1, Sect. S4).

\subsection{RQ2: which errors the method finds}\label{sec:7.3}

The method found 1432 of the 1717 errors of the benchmark, including 139 of the 172 that appear only at \(n>20\) and that no instance check found. For a faulty solution, let \(n_{\rm d}\) denote the smallest tested size (Section 7.1) at which it differs from the reference. Table~\ref{tab:4} splits the 1717 faulty solutions into five groups. Groups 4 and 5 are the solutions with \(11\le n_{\rm d}\le20\) and with \(n_{\rm d}>20\). Among the solutions with \(n_{\rm d}\le10\), we single out the mutations that add a relative phase at the start of the circuit (group 2: the difference shows only on a superposition of basis inputs) and the mutations of a condition on the bits \(a\) (group 3: the difference depends on the string \(a\)). The remaining solutions form group 1.

The fractions in the table are determined by the composition of the benchmark, so they show which errors each check finds, not how frequent the errors are. Group 4 contains 140 solutions with an extra loop from \(K=11\), 32 with a branch on \(n\) at the constant 11, and 12 others. Group 5 contains 140 solutions with an extra loop from \(K=20\) and 32 with a branch on \(n\) at the constant 20. A solution with an extra loop differs from the reference starting from \(n=K+1\) or \(n=K+2\), just beyond the largest sizes enumerated by the instance checks (\(n\le10\) and \(n\le20\)). The zero fractions of the instance checks in these groups are therefore set by the construction of the benchmark. The quick search checks the sizes with \(n+e\le13\) and therefore finds 141 errors of group 4 missed by the checks with \(n\le10\).

\begin{table}[tbp]
\caption{Fraction of errors found. ``State'', ``Rand.'', ``Samp.'' are the state check, the random-input check and the sampled check (Section 7.1); ``Found'': the method found the error; ``Automata'': the number of pairs where the counterexample came from the automata; ``Refusal, timeout'': a refusal of the translator or a timeout.}\label{tab:4}
\small

\footnotesize\setlength{\tabcolsep}{4pt}\noindent
\begin{tabularx}{\textwidth}{@{}>{\raggedright\arraybackslash}Xrrrrrrr@{}}
\toprule
Group & Pairs & State & Rand. & Samp. & Found & Automata & \begin{tabular}[b]{@{}r@{}}Refusal,\\timeout\end{tabular} \\
\midrule
1. at small $n$ & 1188 & 45\% & 100\% & 98\% & 86\% & 0 & 14\% \\
2. only on superpositions & 89 & 0\% & 100\% & 100\% & 83\% & 0 & 17\% \\
3. bit-condition mutation & 84 & 43\% & 100\% & 94\% & 60\% & 0 & 40\% \\
4. only at $10<n\le20$ & 184 & 0\% & 0\% & 100\% & 82\% & 9 & 18\% \\
5. only at $n>20$ & 172 & 0\% & 0\% & 0\% & 81\% & 139 & 19\% \\
\botrule
\end{tabularx}

\end{table}

If the translator did not refuse the solution, or refused it only because it reads a nonexistent bit, the method found the error at any \(n_{\rm d}\): 1432 of 1433 such errors, with a timeout on one. There are 9 solutions with a nonexistent bit: they read the bit \(a_n\) in a loop up to \(n\). The translator refuses them, and the counterexample comes from the quick search (Section 6.2). The other 284 refusals concern code that the mutation took out of the class, for example by adding a branch on \(n\). In group 3 these are the mutations of reversed bit order (an index counted from the end) and of the special case \(a=1\cdots1\). In group 3 the random-input check found all errors, because at \(n\le10\) each of them shows on at least one of the five checked strings \(a\), although this check does not enumerate all strings \(a\) at \(n>3\).

Without the quick search, the automata find 1418 of the 1432 counterexamples (on five pairs the time limit expires, and the 9 solutions with a nonexistent bit remain refusals). With it, 1284 counterexamples come from the quick search, that is, from instance checking inside the implementation (9 of them on solutions refused by the translator), and 148 from the automata (step 3). Step 4 and the comparison of the numbers of ancillas gave no counterexamples. The automata found counterexamples only in groups 4 and 5 (Table~\ref{tab:4}), so on the benchmark the automata provide an advantage over instance checking specifically on errors that appear at large \(n\). The quick search reduces the median time on the pairs where it found a counterexample from 0.021 to 0.008~s and increases it on the equal pairs from 0.19 to 0.29~s.

\subsection{RQ3: time}\label{sec:7.4}

On the 444 equal pairs with the answer ``equal'', the method takes 0.36~s on average (median 0.29~s, maximum 1.63~s), including translation. The quick search takes 28\% of this time. On the same pairs, the state check takes 0.015~s on average, the random-input check 0.029~s, and the sampled check 2.1~s.

\begin{figure}[tbp]
\centering
\includegraphics{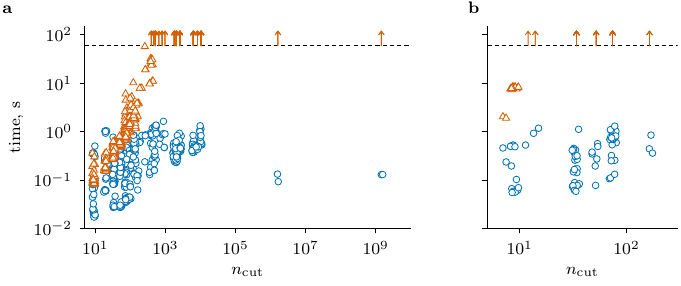}
\caption{Time of the method and of QCEC enumeration up to the cutoff \(n_{\rm cut}\) on the 444 equal pairs with the answer ``equal''. \textbf{a} Tasks without bits \(a\) (370 pairs); \textbf{b} with bits \(a\) (74 pairs); the ranges of the \(n_{\rm cut}\) axis differ between the panels, the time axis is shared. Blue circles: the method; orange triangles: enumeration finished within 60~s; upward arrows from the dashed 60~s line: enumeration not finished within this time (117 and 59 pairs). Both axes are logarithmic; circles and triangles with the same \(n_{\rm cut}\) are spread horizontally by at most 10\%, arrows are not spread}\label{fig:5}
\end{figure}

The same answer as the method is given by enumerating all \(n\) from 1 to \(n_{\rm cut}\) (Corollary 1) with all strings \(a\). Without the method the cutoff \(n_{\rm cut}\) is unknown, so the comparison shows how long enumeration with the same guarantee would take if the cutoff were given. Table~\ref{tab:5} compares the method with enumeration by QCEC on the 444 equal pairs with the answer ``equal'', and Figure~\ref{fig:5} shows the same pairs individually. An unfinished enumeration counts as 60~s, the limit. On tasks without the bit parameter \(a\), QCEC finished the enumeration on all pairs with \(n_{\rm cut}<395\) and on none with \(n_{\rm cut}>454\), while the time of the method over the whole range up to \(n_{\rm cut}=1.5\cdot10^{9}\) does not exceed 1.63~s (Figure~\ref{fig:5}a). For small \(n_{\rm cut}\), enumeration can be faster: on tasks without the bit parameter \(a\), QCEC was faster than the method on 20 of the 253 pairs with a finished enumeration, and all these pairs have \(n_{\rm cut}\le34\). On tasks with bits, enumeration also grows with the number of strings \(a\) and exceeds 60~s already at \(n_{\rm cut}>9\) (Figure~\ref{fig:5}b).

\begin{table}[tbp]
\caption{Equal pairs with the answer ``equal'': number of finished QCEC enumerations and median times. In the cells ``\(x\) / \(y\)'': enumeration up to \(n_{\rm cut}\) and up to the smaller cutoff of Table 1 computed from the dimensions after minimization; the time of an unfinished enumeration is 60~s; QCEC enumeration is limited to sizes \(n\le10^{5}\).}\label{tab:5}
\small

\footnotesize\noindent
\begin{tabularx}{\textwidth}{@{}>{\raggedright\arraybackslash\hsize=0.95\hsize\hspace{0pt}}X>{\raggedright\arraybackslash\hsize=1.25\hsize\hspace{0pt}}Xr>{\raggedright\arraybackslash\hsize=0.95\hsize\hspace{0pt}}X>{\raggedright\arraybackslash\hsize=0.80\hsize\hspace{0pt}}X>{\raggedright\arraybackslash\hsize=1.05\hsize\hspace{0pt}}X@{}}
\toprule
Tasks & $n_{\rm cut}$ & Pairs & QCEC finished & Method, median, s & QCEC, median, s \\
\midrule
without bits $a$ & $9\ldots34$ & 100 & 100 / 100 & 0.09 & 0.3 / 0.3 \\
without bits $a$ & $52\ldots971$ & 183 & 153 / 167 & 0.27 & 1.6 / 1.3 \\
without bits $a$ & $1803\ldots1.5\cdot10^{9}$ & 87 & 0 / 77 & 0.47 & 60 / 1.5 \\
with bits $a$ & $7\ldots164$ & 74 & 15 / 15 & 0.33 & 60 / 60 \\
\botrule
\end{tabularx}

\end{table}

On the 15 pairs with bits where QCEC finished the enumeration, \(n_{\rm cut}\le9\), so the enumeration covered all strings \(a\). QCEC enumeration is limited to sizes \(n\le10^{5}\). The cutoff \(n_{\rm cut}\) is larger for four pairs (up to \(1.5\cdot10^{9}\)). All four lie in the last row for size-only tasks, where QCEC finished no enumeration up to \(n_{\rm cut}\), and their smaller cutoff is at most 803, so the limit does not affect the numbers of Table~\ref{tab:5}.

The cutoff \(n_{\rm cut}\) of the implementation is computed from dimension bounds (Section 6.2) and may exceed the cutoff of Table 1 computed from the dimensions of the automata of the families after minimization. This smaller cutoff is less than \(n_{\rm cut}\) for 108 of the 444 pairs, including all 87 pairs without bits with \(n_{\rm cut}>1000\). With it, QCEC finishes the enumeration on 359 pairs instead of 268 (Table~\ref{tab:5}). The dimensions after minimization are computed modulo one prime, so they are at most the dimensions over the field, and they are unknown without building the automata. The time of enumeration up to the smaller cutoff is therefore a lower bound on the time of an enumeration that gives the guarantee of Corollary 1.

On unequal pairs, enumeration needs to reach only the least differing size \(n_{\rm d}\). Table~\ref{tab:neq} compares the method with QCEC enumeration that is given \(n_{\rm d}\) in advance and checks only the sizes \(n\le n_{\rm d}\). Even then, on the pairs where both tools found the difference, the median time of QCEC is almost 6 times that of the method, and QCEC is faster on only 93 of 1339 pairs. On 1217 of these pairs, however, the counterexample of the method came from the quick search, not from the automata. On out-of-class mutants, where the method refuses, QCEC up to the known \(n_{\rm d}\) finds the difference on 265 of 284 pairs. Our QCEC evaluation harness skips the sizes at which the solution raises an exception (Section 7.1). So on 31 of the 67 pairs where the solution raises an exception at \(n_{\rm d}\), QCEC enumeration answers ``equivalent'', and on 36 it checks no instance at all. On these 67 pairs the method answered ``not equal'' on 50, and the translator refused 17.

\begin{table}[tbp]
\caption{Unequal pairs: answers of the method and of QCEC enumeration up to the known \(n_{\rm d}\), number of pairs (1717 in total). Asterisk: the solution raises an exception at \(n_{\rm d}\), and the QCEC evaluation harness skips this size. On the 1339 pairs where both found the difference, the median and 90th percentile of time: method 0.009 and 0.082~s, QCEC 0.052 and 0.225~s.}\label{tab:neq}
\small

\noindent
\begin{tabularx}{\textwidth}{@{}Xrrrr@{}}
\toprule
Answer of the method & ``not equivalent'' & ``equivalent''* & not checked* & timeout 60~s \\
\midrule
``not equal'' & 1339 & 14 & 36 & 43 \\
refusal & 265 & 17 & 0 & 2 \\
timeout 300~s & 1 & 0 & 0 & 0 \\
\botrule
\end{tabularx}

\end{table}

Step 4 of Algorithm 1 (the small-size check \(n<n_0\)) is exponential in \(n_0\) (Section 5.5) but takes little time on the benchmark: on the 444 equal pairs with the answer ``equal'', \(n_0\le4\) for 432 pairs and \(n_0\in\{7,11\}\) for the rest, and step 4 takes 0.001~s in the median and at most 0.033~s. A threshold \(n_0\ge11\) occurs in the benchmark only for solutions with an extra loop from \(K=11\) or \(K=20\), on which the method answers ``not equal'' before step 4, and for two tasks with a loop from the constant 9 in the reference.

To find the \(n_0\) at which step 4 becomes infeasible, we built 13 correct solutions with \(n_0\) from 7 to 22: to the references of three tasks (a chain of CX gates, a chain of RZZ\((\pi/4)\) gates after a layer of H, and the Bernstein--Vazirani oracle, BV below) we added a loop from a constant \(K\) whose body is a gate followed by its inverse, so that the solution equals the reference and \(n_0=K+2\) (Table~\ref{tab:step4}). Without the parameter \(a\), step 4 is feasible up to \(n_0=13\), and on the oracle, where the strings \(a\) are also enumerated, up to \(n_0=10\). For larger \(n_0\) the run did not finish within 600~s and 8~GB of memory. QCEC checks the same sizes \(n<n_0\) without bits in fractions of a second, and on the oracle at \(n_0=13\) in 75~s, but in floating-point arithmetic with a tolerance, so such an answer does not prove equality at \(n<n_0\).

\begin{table}[tbp]
\caption{Constructed correct solutions: time of step 4 of the method and of QCEC enumeration of the same sizes \(n<n_0\), s. A dash: the run did not finish within 600~s and 8~GB of memory; the oracle solutions were built for \(n_0\le13\).}\label{tab:step4}
\small

\footnotesize\noindent
\begin{tabular}{@{}llrrrrr@{}}
\toprule
Reference & & \(n_0=7\) & 10 & 13 & 16 & 22 \\
\midrule
CX chain & method & 0.002 & 0.023 & 2.6 & --- & --- \\
 & QCEC & 0.27 & 0.12 & 0.16 & 0.17 & 0.32 \\
RZZ chain & method & 0.012 & 0.37 & 80 & --- & --- \\
 & QCEC & 0.10 & 0.29 & 0.21 & 0.19 & 0.25 \\
BV oracle & method & 0.33 & 230 & --- & & \\
 & QCEC & 1.2 & 9.1 & 75 & & \\
\botrule
\end{tabular}

\end{table}

The probabilistic plan is chosen because it is faster than the exact one. If the restriction \(T\le T_{\max}\) is lifted, the exact plan (Theorem S.12) gives the answer ``equal'' with \(\varepsilon=0\) on 222 of the 224 equal pairs on which the probabilistic plan gave \(\hat\varepsilon\le2^{-40}\), and finds no difference on any of them. Two pairs, which need \(34\,525\) and \(42\,127\) primes, did not finish within the 30-minute time limit. The exact plan needs a median of 126.5 primes and 2.0~s per pair, the probabilistic plan 3 to 13 primes and 0.26~s (Figure S1 in Online Resource 1). For 90\% of the pairs the exact plan takes at most 152~s, and the longest time is 22~min. For the 12 pairs returning ``no answer'', the exact plan was not built: for eight the construction exceeded the 30-minute limit, and for four the plan would need more than \(10^{9}\) primes. The times in this paragraph and in Figure S1 cover only the computations with automata, without translation and step 4.

The method of \cite{Abdulla2025_ParamVerification} is implemented in the tool AutoQ-Para; compiled from source by us, it gives the same answers on the five examples of Table 1 of \cite{Abdulla2025_ParamVerification} in time of the same order.
The tool has no input language, and the model of each example is written as a C++ function that defines automata and transducers. The tool builds the composition of transducers and the transducer of the family with a given repetition pattern \cite[Sect.~6]{Abdulla2025_ParamVerification}, but the author of the model gives the pattern, the blocks, the precondition and the postcondition. For GHZ state preparation our model took about 250 lines, and the tool does not check that the model corresponds to the code. Our method builds the model from the code of the reference and the solution automatically, but only for ladder programs.

We compared the method and AutoQ-Para on the families of \cite[Sects.~4.1--4.5]{Abdulla2025_ParamVerification} and on GHZ state preparation (Table~\ref{tab:6}). In the BV, adder and syndrome extraction examples, AutoQ-Para checks a circuit against a precondition and a postcondition given by hand as automata, while the method compares two generators. So for BV we wrote the postcondition as a generator of a circuit that prepares \(\lvert a\rangle\lvert-\rangle\), and for the adder and the syndrome extraction we compared two versions of one circuit. The times in these rows therefore refer to different tasks. The generators are written from the description of the families in \cite{Abdulla2025_ParamVerification}. The ground truth of all 17 pairs was checked by full matrices at \(n\) from 3 to 9. Both tools answered correctly on all common families, including circuits with an inserted error (the method on 8 pairs, AutoQ-Para in 4 runs on GHZ and the CX chain). Where both tools compare two circuits, AutoQ-Para is faster on GHZ and the CX chains (0.01--0.02~s versus 0.09--0.14~s for the method, including translation). On the Trotter step the difference is less than a factor of 1.5. Grover's algorithm with a gate with a growing number of controls is out of the class of the method. The AutoQ-Para times in Table~\ref{tab:6} do not include writing the models.

\begin{table}[tbp]
\caption{The method and AutoQ-Para on common families: median time of three runs, s. All answers are correct; the answers ``equal'' of the method are exact, except BV and the Trotter step (\(\hat\varepsilon\le2^{-41}\)).}\label{tab:6}
\small

\footnotesize\setlength{\tabcolsep}{4pt}\noindent
\begin{tabularx}{\textwidth}{@{}>{\raggedright\arraybackslash}Xllrr@{}}
\toprule
Family and comparison & Relation & \begin{tabular}[b]{@{}l@{}}Reference\\in AutoQ-Para\end{tabular} & \begin{tabular}[b]{@{}r@{}}AutoQ-Para,\\s\end{tabular} & \begin{tabular}[b]{@{}r@{}}Method,\\s\end{tabular} \\
\midrule
GHZ: CX chain versus CX from the first qubit to all others & $\equiv_0$ & circuit & 0.02 & 0.09 \\
CX chain versus chain of $(I\otimes H)\,\mathrm{CZ}\,(I\otimes H)$ & $\equiv$ & circuit & 0.01 & 0.10 \\
chain of reversed CX versus chain of CX surrounded by $H$ gates & $\equiv$ & circuit & 0.01 & 0.14 \\
BV, $a=1010\cdots$, versus postcondition & $\equiv_0$ & specification & $<0.01$ & 0.33 \\
BV, any $a$, versus postcondition & $\equiv_0$ & --- & --- & 0.47 \\
Cuccaro ripple-carry adder \cite{Cuccaro2004_RippleAdder}: two versions of the UMA (UnMajority and Add) block & $\equiv$ & specification & 10.0 & 0.71 \\
repetition-code syndrome extraction: two versions & $\equiv$ & specification & 0.30 & 1.65 \\
Trotter step for the Heisenberg chain: two versions & $\equiv$ & circuit & 0.63 & 0.93 \\
one Grover iteration & --- & circuit & 0.08 & out of class \\
\botrule
\end{tabularx}

\end{table}

\subsection{RQ4: applicability}\label{sec:7.5}

The translator accepts a narrow fragment of Python: outside the benchmark it accepted 40 of 216 functions with a parameter \(n\), and none from MQT Bench \cite{Quetschlich2023_MQTBench} or QuanBench (Table~\ref{tab:rq4}; refusal kinds in Section 6.1). Inspection of the code shows that 34 of the 36 generators of MQT Bench give families outside the class (free parameters, random structure, mid-circuit measurements, the Fourier transform, angles that depend on the counter or are not multiples of \(\pi\), several registers, a growing number of controls, and others). The other two, GHZ and BV, are given by ladder programs, and after their syntax is rewritten the translator accepts them. Our 62 generators were written independently of the benchmark for other tasks (well-known quantum algorithms and tasks of quantum programming contests, 2--3 versions of the solution per task).

We tested third-party code on the Qiskit QuantumKatas set of 350 tasks with references \cite{CruzBenito2026_QiskitQuantumKatas}. By inspecting the code prior to running the translator, we selected 99 tasks in which the function extends a given circuit and takes a list of qubits of arbitrary length. Of these, 52 give a family in \(n\) by their signature (a register of \(n\) qubits, single qubits and at most one bit string \(a\)). We wrapped each of the 52 references in a function \texttt{solve(n,\ a)} that creates the circuit and calls the reference. The translator accepts 12 of the 52 references, and for all 12 the programs of the translator matched the Qiskit circuits at five sizes starting from the threshold.

Solutions of the same 52 tasks were written by the language model Claude Opus 5.5 (Anthropic), one solution per task. The model received only the task statements, without access to the method or the references. A comparison by full matrices up to global phase at \(n\le5\) (for tasks with bits, at \(n=5\) on three strings \(a\)) showed that 35 solutions equal the reference, 9 differ, and for 8 the comparison was not performed because the reference or the solution raises an exception at small \(n\). The translator accepted 15 solutions and both functions on 12 pairs. On all 12 the solution equals the reference at \(n\le5\), and the method answered ``equal'' by the exact plan, in at most 1.2~s per pair. Of the 9 differing solutions, the translator refused 8, and for the ninth it did not accept the reference. So there are no wrong answers, but the method could not check these solutions either.

In QuanBench \cite{Guo2025_QuanBench}, only 14 of the 966 functions (references and stored model answers) have a size parameter, and all 14 call the library Fourier transform, which is out of the class. The fraction of the class among model-written code therefore cannot be measured on this set.

\begin{table}[tbp]
\caption{Generators outside the benchmark: how many functions the translator accepts and why it refuses. Parameter \(n\): by the signature of the function. Refusal kind (Section 6.1): by the refusal code and Table S2; ``undetermined'': the code may mean either kind; seven refusals of the model's solutions on a \texttt{return} statement not at the end of the function (there is no such code in Table S2) are counted as out of fragment; for MQT Bench and QuanBench the kind was determined by reading the code.}\label{tab:rq4}
\small

\footnotesize\noindent
\begin{tabularx}{\textwidth}{@{}>{\raggedright\arraybackslash}Xrrrrrr@{}}
\toprule
Source & Functions & \begin{tabular}[b]{@{}r@{}}With para-\\meter \(n\)\end{tabular} & Accepted & \begin{tabular}[b]{@{}r@{}}Out of\\class\end{tabular} & \begin{tabular}[b]{@{}r@{}}Out of\\fragment\end{tabular} & \begin{tabular}[b]{@{}r@{}}Kind un-\\determined\end{tabular} \\
\midrule
MQT Bench & 36 & 36 & 0 & 34 & 2 & 0 \\
our generators for other tasks & 62 & 62 & 13 & 20 & 10 & 19 \\
QuantumKatas, references & 99 & 52 & 12 & 26 & 10 & 4 \\
QuantumKatas, language-model solutions & 52 & 52 & 15 & 17 & 18 & 2 \\
QuanBench & 966 & 14 & 0 & 14 & 0 & 0 \\
\botrule
\end{tabularx}

\end{table}

\subsection{Threats to validity}\label{sec:7.6}

The class of ladder programs, the translator, the benchmark and the error model were developed by us. On code not written for the benchmark, the translator accepts a small fraction of the generators (Table~\ref{tab:rq4}). In addition, the QuantumKatas tasks were selected by signature, one rule of the translator (Online Resource 1, Sect. S2.1) was formulated from these tasks, and the solutions were written by one language model, one per task. The results of RQ1--RQ3 therefore concern code that falls into the class, and the fraction of such code among student solutions remains to be measured (Section 9). The generators for the comparison with AutoQ-Para were written by us from the description of the families, and the limit of step 4 was measured on three constructed families.

The errors of the benchmark are single mutations of the kinds of Table~\ref{tab:mut}, and the conclusions of RQ1 and RQ2 concern them. The instance checks are our implementations of the approaches of Section 7.1, not the systems themselves. Time depends on the run conditions: the same code on two machines of the same type with different numbers of processes gave median times of the method differing by a factor of 1.38, so we do not interpret time differences of less than a factor of 1.5. The ground truth ``equal'' is given by construction and checked at small \(n\) (Section 7.1), not proved for each pair.

%% file: sections/08_related.tex
\section{Related work}\label{sec:8}

\textbf{Equivalence checking at a fixed size.} The QCEC tool combines checking on decision diagrams \cite{Niemann2016_QMDD}, in which \(U\cdot U'^{\dagger}\) is built starting from the middle, with random simulations. The answer is ``equivalent'', ``not equivalent'' or ``timeout'' \cite[Sects.~IV--VII]{Burgholzer2020_AdvancedEC}. QCEC also implements checking on ZX-diagrams \cite{Peham2022_ZX_EC}. The relation there is equality up to global phase, that is, our \(\simeq\). Ancillas are handled by fixing their initial state \cite[Sect.~III]{Peham2022_ZX_EC}.

Exact answers are also given by checking on tensor decision diagrams for circuits with real-valued angles \cite{Hong2024_ParamEC_TDD}, checking via an intermediate MPO \cite{Sander2024_MPO_EC}, checking via path sums \cite{Amy2018_PathSums} and weighted model counting \cite{Huang2026_PathSumWMC}, and checking of constant-depth circuits in time linear in \(n\) at fixed depth, for the output on \(\lvert0\rangle^{\otimes n}\) and for all inputs up to phase \cite{Yu2025_ShallowEC}. Probabilistic identity testing by Schwartz--Zippel \cite[Lemma~1, Cor.~1]{Schwartz1980_PolyIdentities}, \cite[Thm.~I]{Zippel1979_SparsePoly} gives a ``not equal'' answer that is always correct \cite[Thm.~4.1]{Xu2023_Queso}. All these methods check circuits of one size. For a family they give instance checking, which Section 7 compares with our method.

The answer ``not equal'' is correct both here and in \cite{Xu2023_Queso} because a homomorphism maps zero to zero. Xu et al.\ \cite{Xu2023_Queso} bound the error of the answer ``equal'' by the Schwartz--Zippel lemma, and we bound it through the norm of an element of a cyclotomic field (Lemma 4).

\textbf{Parameters as rotation angles.} For circuits with continuous parameters, cutoff theorems are known: equivalence reduces to finitely many circuits without parameters, with decidability for cyclotomic gate sets, a probabilistic variant, and a version up to phase \cite{Ross2025_Cutoff}. Our parameter is discrete, the size \(n\) and the bits \(a\), but the result has the same structure: a finite cutoff, elements of a cyclotomic field, a probabilistic variant. Unlike \cite{Ross2025_Cutoff}, we compute modulo primes rather than in the cyclotomic field, and we bound the number of primes at which the check may err.

\textbf{The parameter as size.} The work closest to ours is \cite{Abdulla2025_ParamVerification}: parameterized circuits are represented by synchronized weighted tree automata and transducers. Equivalence of two parameterized programs reduces to applying the transducers of both programs to a tree automaton representing all basis states and checking the resulting tree automata for functional equivalence \cite[Sects.~1, 5.3]{Abdulla2025_ParamVerification}. The models are built from a circuit with a repetition pattern. For the Bernstein--Vazirani oracle this pattern does not fit, and the transducer is built by hand for the hidden strings of the form \(1010\cdots10\) and \(1010\cdots101\) \cite[Sects.~4.1, 6]{Abdulla2025_ParamVerification}. The QFT transducer is built for a fixed number of qubits \(n\), has \(O(n^{2})\) states and does not include the reversal of the qubit order: modeling it would lead to an exponential growth of size \cite[Sect.~6.1]{Abdulla2025_ParamVerification}. We build the model automatically from a ladder program (and from Python code, Section 6), and the bits \(a\) enter the letter of the automaton, so one check covers all hidden strings. The two tools are compared on common families in Section 7.4 (Table~\ref{tab:6}).

Other approaches to circuit families require human effort or solve a different problem. QSynth checks a synthesized recursive program for all \(n\) against a specification rather than a reference generator \cite{Deng2023_QSynth}. The proof system for recursively defined circuits \cite{Ying2024_RecursiveCircuits} is based on Hoare triples. Proofs in it are done by hand, and the authors leave their automation to future work. Other deductive systems also require manual proofs: in SQIR, the proofs of Grover's algorithm and of phase estimation took one and two person-weeks \cite{Paltenghi2024_SurveyQSTesting}. The loop summarization of \cite{ChenTF2024_LoopSummarization} is built for a concrete input state with a constant number of iterations \(K\), and the authors leave a closed form in \(K\) to future work. In AutoQ 2.0, the loop invariant is given by the user \cite{Chen2024_AutoQ2}.

\textbf{Weighted automata and tensor networks.} Weighted automata and uniform MPS with boundary vectors define the same functions (Section 2.2) \cite[Sect.~II.C]{Crosswhite2008_FiniteAutomataMPS}. In \cite[Example~23]{Styliaris2025_MPUBeyondQCA}, the tensor defining a subclass of MPS that correspond to phase matrix product unitaries (MPU) is interpreted as a deterministic weighted automaton. There, unitarity and translation invariance for all \(N\) are described algebraically, and the equality of two families is not checked. Sequential generation with an auxiliary system of dimension \(D\) gives an MPS of dimension \(D\) \cite{Schon2005_SequentialGeneration}. The automaton of a ladder with fixed qubits of the form \(c\) works the same way, and for fixed qubits of the form \(n+c\) it keeps guessed values and checks them at the end of the word (Theorem 1). In MPS theory, the equality of families for all \(N\) is the premise of the fundamental theorem \cite[Thm.~IV.4, Cor.~IV.5]{Cirac2021_MPS_PEPS_Review}, and an algorithm to check it from two tensors is not discussed there. Finite cutoffs are known for the uniqueness of the canonical form, where one sufficiently large size suffices \cite[Thm.~7]{PerezGarcia2007_MPSRepresentations}, and in the power-sum technique, where the agreement of the sums \(\sum_l\mu_l^{N}\) and \(\sum_l\nu_l^{N}\) for all \(N\) up to the number of terms implies agreement for all \(N\) \cite[Lemma~12]{DeLasCuevas2017_IrreducibleMPS}. We build the automaton from a program with a bit parameter automatically and reduce six equivalence relations to a zeroness check.

\textbf{Undecidable generalizations.} Related questions about families of tensor networks of all sizes are undecidable. For a translation-invariant MPO it is undecidable whether it is positive for all \(n\) \cite[Thm.~5]{Kliesch2014_MPO_NPhard}. For PEPS in 2D, both the non-zeroness problem (whether there is a boundary condition under which the state is non-zero for all sizes) and the problem of whether two tensors give the same PEPS (in the sense of equal reduced density matrices in the limit of infinitely distant boundaries) are undecidable \cite{Scarpa2020_PEPSLimitations}. Our problem differs from the first in the kind of question and from the second in dimension: all six equivalence relations reduce to the zeroness of the series of a one-dimensional automaton (Table~1), which is decidable (F2). The zeroness check takes a number of operations polynomial in the dimension of the automaton (Section 5.5), but the dimension itself grows exponentially with the window width (Theorem 1). The QFT without qubit reversal admits approximate compression \cite{Chen2023_QFTSmallEntanglement}, but an exact method cannot use this (Section 4.2).

\textbf{Autograding and benchmarks.} Autograding of Q\# assignments that implement oracles runs the solution on each basis state and, after checking that the solution uses no measurements, concludes correctness on superpositions by linearity \cite[Sect.~II]{Mykhailova2022_Assignments}. A grading system with hidden references compares the solution with the reference on one variant of the assignment \cite{Kaltchenko2026_ChatGPTQiskitHW}. Benchmarks of quantum code generation compare the solution with the reference on problems with a given number of qubits: by the output state \cite[Sect.~3.2]{Mikuriya2025_QCoderBenchmark} or by process fidelity \cite{Guo2025_QuanBench}. The two GHZ preparation programs of Section 3.2 are given in \cite[Fig.~5]{Guo2025_QuanBench} as an example of functionally correct solutions whose process fidelity with respect to the reference may be less than one. For \(n\ge3\) it is less than one, while under the relation \(\equiv_0\) the programs are equal. Property-based testing supports circuits of variable size but checks them on a sample \cite{Pontolillo2025_QuCheck}.

%% file: sections/09_conclusion.tex
\section{Limitations and conclusion}\label{sec:9}

\textbf{Class.} The method applies only to ladder programs: the body of a loop with one counter acts on a window of constant width, the gates have constant arity, and their angles have the form \(q\pi\) and do not depend on \(n\) or the counter. Many well-known families remain outside the class: of the 36 generators of MQT Bench, two are in the class (Section 7.5), and the others use the Fourier transform and other counter-dependent angles, gates with a growing number of controls, several registers growing with \(n\), mid-circuit measurements, or free parameters. For some families there is no automaton of finite dimension reading the qubits in index order at all, among them the mirror layer \(F_n\) and the Fourier transform with qubit reversal (Section 4.2). Apart from the boundary of the class, there is a limitation of the implementation: the translator parses a narrow fragment of Python (Section 6.1) and accepts a small fraction of the generators in external code (Table~\ref{tab:rq4}).

\textbf{Parameters and relations.} The classical parameter is a bit string \(a\), one bit per working qubit. Arithmetic circuits with a numerical parameter (a multiplier, a modulus) are not in the class; whether they are represented by an automaton with another order of reading the qubits is an open question. Equality of measurement outcome distributions and relations with a tolerance are outside the problem statement.

\textbf{Cost and guarantee.} The dimension of the automaton grows exponentially with the window width and the number of fixed positions (Theorem 1) and, in the worst case, with the number of segments.
Then the probabilistic plan may give ``no answer'', and the exact plan needs many primes (Sections 7.2, 7.4). The guarantee of the probabilistic plan assumes a uniform and independent choice of primes (Section 6.2). The small-size check \(n<n_0\) is exponential in \(n_0\) (Section 5.5): on the constructed solutions it did not finish for \(n_0\ge16\), and with bits \(a\) already at \(n_0=13\) (Section 7.4). The method does not establish the equality of such pairs.

\textbf{Implementation and data.} The rules of the translator are proved under assumptions on the semantics of Python and Qiskit (Theorem 3), and the program matches the circuit of the code up to replacing the angles with the exact values \(q\pi\). The implementation of the translator and the kernel was tested by cross-checking (Section 6.1) and by experiments (Section 7.2), but not proved. The benchmark and the error model were compiled by the author, and the errors of the benchmark are single mutations (Section 7.6).

\textbf{Conclusion.} We stated the problem of checking a quantum circuit generator against a reference at all \((n,a)\) at which the reference is defined, and solved it for ladder programs. The matrix family of a ladder program is a uniform MPO, or weighted automaton, in which the bit parameter enters the physical index. Each of the six relations reduces to a zeroness check of its series. The algorithm answers ``not equal'' with a counterexample, and ``equal'' exactly or with a computed bound \(\hat\varepsilon\) on the error probability. In the experiments, the method gave no wrong answers, confirmed equality on 444 of 483 equal pairs (QCEC enumeration up to the same cutoff did not finish within 60~s on 176 of them), and found 139 of the 172 errors inserted into the benchmark that appear only at \(n>20\) and are therefore beyond the reach of instance checking with the bound \(n\le20\) (Section 7.3).

Future work includes extending the syntax of the translator (default parameters, a \texttt{return} statement not at the end of the function, Table~\ref{tab:rq4}) and measuring the fraction of the class in external code, including student solutions; lowering the threshold \(n_0\) from which the automata are compared, so as to shorten the enumeration of step 4; moving from ladders to tree automata for recursive generators; and finding a class of numerical parameters that can be written in the bits of the letters of the automaton.

%% file: sections/A_proofs.tex
\section{Proofs of Theorem 2 and of facts (K2), (K3)}\label{sec:A}

Statements S.1--S.17 are stated and proved in Online Resource 1, which also gives the notation (Sect. S1.1).

\textbf{Theorem 2.}
\begin{enumerate}
\def\labelenumi{\arabic{enumi}.}
\tightlist
\item
  \emph{``Not equal''.} The answer is obtained modulo some \(p\): \(\pi_p(\Delta(w))\ne0\) or, for \(n<n_0\), an entry of the difference or a minor that is non-zero modulo \(p\). Then the corresponding value is non-zero over \(\mathbb{K}\). For \(n\ge n_0\), the pair \((n,a)\) of the word lies in \(\mathrm{dom}(R)\cap\mathrm{dom}(G)\) (Section 4.1), and by Theorem 1, Lemma 2 and Lemmas S.6--S.8 the relation is violated at \((n,a)\); for \(\equiv_{\mathrm{anc}}\) and \(\simeq_{\mathrm{anc}}\), the series \(f_R^{(z)}\) have by then been checked modulo all primes of the plan (step 3(a) of Algorithm 1). In the exact plan this implies that \(R\) returns its ancillas for all \(n\ge n_0\): the series \(f_R^{(z)}\) is given by an automaton of dimension at most \(d_R\le d\), its values after any \(\gamma_k\) have absolute value at most 1, and \(\kappa_R\le\kappa\); hence a series that is non-zero on words of length at least \(N_0\) has a witness shorter than \(N^{*}\) (Lemma 3), which vanishes modulo at most \(k_\Delta(N^{*})<T\) primes (Lemma 4), and step 3(a) would have found a non-zero value. For \(n<n_0\), by the construction of step 4 of Algorithm 1: the return of the ancillas by \(R\) at the same \((n,a)\) is checked exactly (Proposition 2) before the answer ``not equal''. The answer of step 1 (\(e_G\ne e_R\)) is correct by Definition 2.
\item
  \emph{``Reference does not return its ancillas''.} The same for the series \(f_R^{(z)}\), \(z\ne0\), or for the return check at \(n<n_0\).
\item
  \emph{``Equal'', exact plan.} For \(n\ge n_0\): Theorem 1 (the automata represent the programs), Lemma 2 and Lemmas S.6--S.8 (the relation holds if and only if \(\Delta=0\) on words of length at least \(N_0\)), Theorem S.12. For \(n<n_0\): Proposition 2.
\item
  \emph{``Equal'', probabilistic plan.} Suppose the answer to Problem 1 is negative or \(R\) does not return its ancillas at some \((n,a)\in\mathrm{dom}(R)\). For \(n<n_0\), step 4 finds the violation exactly (Proposition 2). Otherwise, by Theorem 1 and Lemma 2, on words of length at least \(N_0\) the series \(\Delta\) or a series \(f_R^{(z)}\), \(z\ne0\), is non-zero; the latter is given by an automaton of dimension at most \(d_R\le d\) with values whose absolute value after any \(\gamma_k\) is at most 1, and \(\kappa_R\le\kappa\). Hence Theorem S.13 applies to both series, and the probability of the answer ``equal'' is at most \(\varepsilon\); if the batch lengths computed from the true numbers do not reach \(t_{\max}\), then \(\varepsilon\le\varepsilon_0\) (Corollary S.14).
\end{enumerate}

\textbf{Fact (K2).} Let \(\xi\in\mathcal{O}\), that is, \(\xi=g(\zeta_m)\) for some \(g\in\mathbb{Z}[X]\). By (K1), \(\gamma_k(\xi)=g(\zeta_m^{k})\), and the numbers \(\zeta_m^{k}\), \(k\in(\mathbb{Z}/m\mathbb{Z})^{\times}\), are all the roots of \(\Phi_m\). Hence \(\mathrm{Nm}(\xi)\) is the product of the values of \(g\) at the roots of the monic polynomial \(\Phi_m\in\mathbb{Z}[X]\), that is, the resultant of \(\Phi_m\) and \(g\), and lies in \(\mathbb{Z}\). Each \(\gamma_k\) is injective, so for \(\xi\ne0\) all factors are non-zero.

\textbf{Fact (K3).} The group \(\mathbb{F}_p^{\times}\) is cyclic of order \(p-1\), and \(m\mid p-1\); hence there is an element \(\omega_p\) of order \(m\). The identity \(X^{m}-1=\prod_{t\mid m}\Phi_t\) in \(\mathbb{Z}[X]\) also holds modulo \(p\), and \(\omega_p^{m}=1\), so \(\Phi_t(\omega_p)=0\) for some \(t\mid m\). If \(t<m\), then \(\Phi_t\mid X^{t}-1\) gives \(\omega_p^{t}=1\), contrary to the order of \(\omega_p\); hence \(\Phi_m(\omega_p)=0\). By (K1), the monic polynomial \(\Phi_m\) is the minimal polynomial of \(\zeta_m\). If \(g\in\mathbb{Z}[X]\) and \(g(\zeta_m)=0\), then \(g=h\Phi_m+g_0\), where the quotient \(h\) and the remainder \(g_0\) of division by a monic polynomial lie in \(\mathbb{Z}[X]\) and \(\deg g_0<\varphi(m)\), so \(g_0(\zeta_m)=0\) implies \(g_0=0\). Hence \(\mathbb{Z}[\zeta_m]\cong\mathbb{Z}[X]/(\Phi_m)\), and the substitution \(X\mapsto\omega_p\) defines a homomorphism on \(\mathbb{Z}[\zeta_m]\). The prime \(p\) is odd, so two is invertible in \(\mathbb{F}_p\), and the homomorphism extends to \(\mathbb{Z}[\zeta_m,\tfrac12]\). \(\square\)

%% file: supplement/S1_lemmas.tex
\section{Proof of Theorem 1 and lemmas for Theorem 2}\label{sec:S1}

\subsection{Notation}\label{sec:S1.1}

We consider a register of \(N=n+e\) qubits, where the last \(e\) are ancillas, and words \(w=\sigma_0\sigma_1\cdots\sigma_{N-1}\) with letters \(\sigma_j=(a_j,x_j,y_j)\). An automaton \((\alpha,\{A(\sigma)\},\beta)\) over a field \(F\) defines the series \(\alpha^{\top}A(\sigma_0)\cdots A(\sigma_{N-1})\beta\). The field is \(\mathbb{K}=\mathbb{Q}(\zeta_m)\) with \(8\mid m\), so that \(\sqrt2=\zeta_8+\zeta_8^{-1}\in\mathcal{O}\); \(\pi_p:\mathbb{Z}[\zeta_m,\tfrac12]\to\mathbb{F}_p\) is the homomorphism \(\zeta_m\mapsto\omega_p\) of (K3); \(p_0\) is the lower bound on the primes modulo which the automata are built (Section 5.4 of the article; its value in the implementation is in Section 6.2), and \(\mathcal{P}_m\) is the set of primes \(p\equiv1\pmod m\) in \([p_0,2p_0)\).

\textbf{Transition graph} (Section 2.2). A finite set of states \(Q\), initial weights \(\iota:Q\to F\), transitions \(\delta(q,\sigma)\subseteq Q\times F\) and final weights \(\tau:Q\to F\). The weight of a path is the product of the weights, and the value on a word is the sum of the weights of the paths that read the word. A traversal of the reachable states turns a transition graph into an automaton of the same series: \(\alpha=\iota\), \(A(\sigma)_{qq'}\) is the sum of the weights of the transitions \(q\to q'\) on \(\sigma\), \(\beta=\tau\). All weights are entries of the matrices of the bodies of ladders and of fixed segments, or ones, so the image of a transition graph over \(\mathbb{K}\) modulo \(p\) defines the series \(\pi_p\circ f\).

\emph{Segments after normalization.} The positions of a segment are the indices of its qubits and of the bits read by its conditions. For a fixed segment this is the set \(J=S\cup E\), where \(S\) are the positions \(c\ge0\) at the start of the register and \(E\) the positions \(N+c\), \(c<0\), at the end (the position \(n+c\) of Definition 3 is \(N+(c-e)\)); \(c_{\min}\) is the least \(c\) among the positions of \(E\), and \(c_{\min}=0\) if \(E\) is empty. A ladder with step \(s\ge1\), after shifting the offsets (\(o_{\min}=0\)), consists of the bodies \(B_i\) for \(i=i_0,i_0+s,\dots\), \(i<N+c_0\), where \(B_i\) acts on the window \(W_i=\{i,\dots,i+w-1\}\) and on the fixed positions \(S\subseteq[0,i_0)\) and \(E\) (to the right of the last window), and its conditions read the bits of the positions of the window and of \(S\). The number \(K_0=1-c_0-w\) is the least length of the tail after the last window, and \(\lvert c\rvert\le K_0\) for all positions \(N+c\in E\). In the notation of Definition 3: \(i_0=c_1+o_{\min}\), \(N+c_0-1=n+c_2+o_{\min}\), \(K_0=e-1-c_2-o_{\max}\). A ladder with step \(-s\) reduces to this one (Lemma S.4).

\subsection{Fixed segment}\label{sec:S1.2}

\renewcommand{\snthmname}{Lemma S.1}\begin{thmx}

Let a fixed segment be the product \(V\) of gates on the positions \(J=S\cup E\), and \(N\ge\max S+1+\lvert c_{\min}\rvert\) (with \(\max S+1=0\) for empty \(S\)). Then its transition graph defines the series \(\langle x\rvert V\otimes I\lvert y\rangle\) on words of length \(N\); for smaller \(N\) it accepts no word.

\end{thmx}

\begin{proof}

The transition graph reads the letters of the positions of \(S\) and stores them in the state. After reading the last of them (or at once, if \(S\) is empty), it guesses the letters \(g=(a,x,y)_E\) of the end positions, multiplies the weight by \(V[x_J,y_J]\), where \(V\) is computed for the bits \(a_J\) (read on \(S\) and guessed on \(E\)), and from then on keeps only \(g\). At positions outside \(J\) it requires \(x_j=y_j\). Before each letter, a path may nondeterministically move to the tail: then the last \(\lvert c_{\min}\rvert\) letters are checked against \(g\) at the positions of \(E\). At a position of \(S\) or \(E\) where the segment has only a bit and no qubit, the graph also requires \(x_j=y_j\). The positions of \(S\) and the last \(\lvert c_{\min}\rvert\) positions are disjoint exactly when \(N\ge\max S+1+\lvert c_{\min}\rvert\); otherwise the path does not reach the check of the end. For such \(N\) the accepted path is unique: the guess must agree with what is read. Its weight is \(V[x_J,y_J]\prod_{j\notin J}[x_j=y_j]=\langle x\rvert V\otimes I\lvert y\rangle\).

\end{proof}

The states of the transition graph are: a position \(j\le\max S\) with the letters of \(S\) read so far, at most \((\max S+1)\,8^{\lvert S\rvert}\); the guess \(g\) in the gap and in the tail with position \(1,\dots,\lvert c_{\min}\rvert\), at most \((\lvert c_{\min}\rvert+1)\,8^{\lvert E\rvert}\). So the dimension of the automaton of a fixed segment is at most \((\max S+1)\,8^{\lvert S\rvert}+(\lvert c_{\min}\rvert+1)\,8^{\lvert E\rvert}\) (Theorem 1).

\subsection{Ladder and the proof of Theorem 1}\label{sec:S1.3}

\textbf{Transition graph of a ladder.} A position of \(S\) at which the body has only a bit is treated as a qubit on which the body acts as the identity. The state stores: the part of the word (start, windows, tail), a position counter, the \emph{open} qubits of the window, each with its stored output \(x_j\), bit \(a_j\) (if a condition will still read it) and intermediate value \(\chi_j\); the same triples for the positions of \(S\); and pairs (guess \(g_c\), intermediate value \(\chi_c\)) for the positions \(N+c\in E\).

\begin{itemize}
\tightlist
\item
  \emph{Start.} Guess the inputs \(g\in\{0,1\}^{E}\) of the end positions and set their intermediate values to \(g\). At a position \(j<i_0\) of \(S\), store \((x_j,a_j)\) and set \(\chi_j:=y_j\); the other positions require \(x_j=y_j\).
\item
  \emph{Windows.} The counter \(r\) is the number of letters until the next body; initially \(r=w\), after a body \(r=s\). If \(r>w\), the letter lies in a gap between windows and requires \(x_j=y_j\). Otherwise the letter is opened: \(x_j\) is stored, \(\chi_j:=y_j\), and the bit \(a_j\) is stored only if a condition of one of the next bodies will read it (otherwise 0 is stored in its place).
\item
  \emph{Body.} When the \(w\)-th letter of the window \(W_{i_k}\) is opened (position \(j_k=i_k+w-1\)), the transition graph has, for each basis output \(\upsilon\) of the body, a transition with weight \(B_{i_k}(a)[\upsilon,\mathrm{in}]\), where \(\mathrm{in}\) are the intermediate values of the window and of the fixed positions, and \(a\) are the stored bits of the window and of \(S\). The first \(t_{\rm f}=\min(s,w)\) qubits of the window are finished: the path continues only if their outputs in \(\upsilon\) agree with the stored \(x\). The other \(w-t_{\rm f}\) qubits stay open with intermediate values from \(\upsilon\); the values of the fixed positions are also taken from \(\upsilon\).
\item
  \emph{End.} After any body, a path may end the window part: the open qubits and the positions of \(S\) are checked against the stored \(x\), then \(\rho\in\{0,\dots,s-1\}\) is chosen and exactly \(K_0+\rho\) letters of the tail are read. A position \(N+c\in E\) requires \(y=g_c\) and \(x=\chi_c\), and the other positions of the tail require \(x=y\).
\end{itemize}

The state stops storing the bit of a window letter (replaces it with zero) when the letter has moved to a position to the left of all positions whose bits are read by the conditions. After a body, the open qubits of the window move only to the left, so a forgotten bit is no longer needed.

Let \(\Pi_k=B_{i_k}\cdots B_{i_1}\), \(\mathrm{Fin}_k=[0,i_k+t_{\rm f})\setminus S\) (finished positions), \(\mathrm{Op}_k=[i_k+t_{\rm f},\,j_k]\) (open positions), and let \(Z_k\) be the positions of \([0,j_k]\) on which no body acts (the positions before \(i_0\) outside \(S\) and the gaps between windows).

\renewcommand{\snthmname}{Lemma S.2}\begin{thmx}[invariant]

The sum of the weights of the paths that have read \(\sigma_0\cdots\sigma_{j_k}\), applied \(k\) bodies and are in a state with intermediate values \(\chi_{\mathrm{Op}_k},\chi_S,\chi_E\) and guess \(g\) equals
\[\Bigl(\prod_{j\in Z_k}[x_j=y_j]\Bigr)\cdot\bigl\langle x_{\mathrm{Fin}_k\setminus Z_k},\,\chi_{\mathrm{Op}_k},\,\chi_S,\,\chi_E\bigr\rvert\,\Pi_k\,\bigl\lvert y_{[0,j_k]\setminus Z_k},\,g\bigr\rangle,\]
where \(\Pi_k\) is regarded as an operator on the positions \(([0,j_k]\setminus Z_k)\cup E\).

\end{thmx}

The sets of positions in the bra and in the ket coincide: \(Z_k\subseteq\mathrm{Fin}_k\) and \((\mathrm{Fin}_k\setminus Z_k)\cup\mathrm{Op}_k\cup S=[0,j_k]\setminus Z_k\).

\begin{proof}

Induction on \(k\).

\emph{Base \(k=1\).} Before the body \(B_{i_1}\), the transitions only store letters, and the path with intermediate values \(\chi=y\) is unique; the positions \([0,i_0)\setminus S\) give the factor \(\prod_{Z_1}[x_j=y_j]\). The transition of the body gives the weight \(B_{i_1}[\upsilon,(y_{W_{i_1}},y_S,g)]\), and the check of the first \(t_{\rm f}\) outputs gives the contraction with \(\langle x_{[i_1,i_1+t_{\rm f})}\rvert\). Since \(\mathrm{Fin}_1\setminus Z_1=[i_1,i_1+t_{\rm f})\), this is the statement for \(k=1\).

\emph{Step \(k\to k+1\).}

\begin{enumerate}
\def\labelenumi{\arabic{enumi}.}
\tightlist
\item
  The letters \(j_k<j\le j_{k+1}\) either lie in a gap (for \(s>w\) these are the positions \([i_k+w,\,i_{k+1})\); no body acts on them, and they give factors \([x_j=y_j]\)) or enter the window \(W_{i_{k+1}}\) for the first time; their input \(y_j\) becomes the intermediate value.
\item
  \(\mathrm{Op}_k\subseteq W_{i_{k+1}}\): for \(s\le w\) we have \(\mathrm{Op}_k=[i_k+s,\,i_k+w-1]=[i_{k+1},\,j_k]\), and for \(s>w\) the set \(\mathrm{Op}_k\) is empty. Hence the window \(W_{i_{k+1}}\) consists of \(\mathrm{Op}_k\) and new positions.
\item
  Insert the resolution of the identity \(\sum_{\chi}\lvert\chi\rangle\langle\chi\rvert\) on the positions \(W_{i_{k+1}}\cup S\cup E\) between \(B_{i_{k+1}}\) and \(\Pi_k\). The sum over the intermediate values \(\chi\) is the sum over the states of the transition graph with intermediate values \(\chi\), and the weight of the transition is \(B_{i_{k+1}}[\upsilon,\chi]\). The positions on which \(B_{i_{k+1}}\) does not act enter \(\Pi_{k+1}\) as the identity.
\item
  The positions \([i_{k+1},i_{k+1}+t_{\rm f})\) are not touched after \(B_{i_{k+1}}\): the next body starts at \(i_{k+1}+s\ge i_{k+1}+t_{\rm f}\). Their output is contracted with \(\langle x\rvert\), which is what the transitions of the graph do, and they move to \(\mathrm{Fin}_{k+1}\).
\end{enumerate}

Bits of conditions. The body \(B_{i_k}\) reads bits only on its window and on \(S\) (Definition 3, (b); \(S\) contains the positions of the bits \(a[c]\)) and may depend on them arbitrarily (nested conditions give a conjunction, a pair of conditions \(a[\eta]=0\) and \(a[\eta]=1\) gives an exclusive or): the state stores the bits themselves, not the value of the condition.

\end{proof}

\renewcommand{\snthmname}{Lemma S.3}\begin{thmx}[end and uniqueness]

Let the ladder be non-empty for the given \(N\), that is, \(i_0<N+c_0\), and let \(t\) be the number of its bodies. Then the transition graph defines the series \(\langle x\rvert\Pi_t\otimes I\lvert y\rangle\) on words of length \(N\).

\end{thmx}

\begin{proof}

A path that ended the window part after body \(k\) and chose \(\rho\) accepts a word of length exactly \(j_k+1+K_0+\rho=i_k+1-c_0+\rho\). For a word of length \(N\) this gives \(\rho=N+c_0-1-i_k\), and the condition \(0\le\rho\le s-1\) is equivalent to \(i_k\le N+c_0-1<i_k+s\), that is, \(k=t\). Take Lemma S.2 with \(k=t\). The check of the open qubits and of the positions of \(S\) against the stored \(x\) gives the bra \(\langle x\rvert\) on these positions. In the tail, the condition \(y=g_c\) removes the sum over the guesses, the condition \(x=\chi_c\) gives the bra \(\langle x_E\rvert\), and the other positions of the tail give \([x_j=y_j]\). We obtain \(\langle x\rvert\Pi_t\otimes I\lvert y\rangle\).

\end{proof}

\emph{Proof of Theorem 1.} Correctness follows from Lemma S.1 for a fixed segment, Lemma S.3 for a ladder with step \(s>0\), and Lemma S.4 for step \(-s<0\). The conditions of the lemmas (\(N\ge\max S+1+\lvert c_{\min}\rvert\), the ladder is non-empty, the positions of \(S\) lie to the left of the first window, the positions of \(E\) to the right of the last, \(K_0\ge0\)) hold for \(n\ge n_{\min}(P)\) by Definition 3, (c), and by the inequalities at the ends of the range of the counter obtained from (T1)--(T3) of Definition 4 (Section 4.1); in particular, \(K_0\ge0\) is inequality (T1) for the position \(i+o_{\max}\) at the end \(i=n+c_2\). The automaton of the program is the product of the automata of the segments (Lemma 1), and Lemma S.6 gives the independence of the bits of the letters of the ancillas.

The dimension bound is a count of states by parts.

\begin{itemize}
\tightlist
\item
  \emph{Start:} the position counter \(j<i_0\), the positions of \(S\) read so far (at most 8 values of the triple per position) and the guess for the positions of \(E\) (2 per position): at most \(i_0\,8^{\lvert S\rvert}2^{\lvert E\rvert}\).
\item
  \emph{Windows:} with the counter \(r\le w\), exactly \(w-r\) qubits are open; with \(r>w\) (a gap, possible only for \(s>w\)), none. The pair (output, intermediate value) of an open qubit and of a fixed position has 4 values, and the stored bits \(2^{b}\), \(b\le w-1+\lvert S\rvert\). In total at most \(4^{h}2^{b}\bigl(\sum_{r=1}^{w}4^{\,w-r}+\max(0,s-w)\bigr)=4^{h}2^{b}\bigl(\tfrac{4^{w}-1}{3}+\max(0,s-w)\bigr)\).
\item
  \emph{Tail:} for each \(\rho\in\{0,\dots,s-1\}\), the position \(0,\dots,K_0+\rho\) and the pairs of the positions of \(E\): at most \(4^{\lvert E\rvert}\sum_{\rho=0}^{s-1}(K_0+\rho+1)=4^{\lvert E\rvert}\bigl(s(K_0+1)+\tfrac{s(s-1)}{2}\bigr)\).
\end{itemize}

The sum of the three terms gives the bound of Theorem 1:
\[4^{h}2^{b}\Bigl(\frac{4^{w}-1}{3}+\max(0,s-w)\Bigr)+D_0,\qquad D_0=i_0\,8^{\lvert S\rvert}2^{\lvert E\rvert}+4^{\lvert E\rvert}\Bigl(s(K_0+1)+\frac{s(s-1)}{2}\Bigr),\]
where \(i_0=c_1+o_{\min}\) is the first qubit of the first window and \(K_0=e-1-c_2-o_{\max}\) is the least number of qubits to the right of the last window (Definition 3). \(\square\)

\subsection{Ladder with step $-s$}\label{sec:S1.4}

\renewcommand{\snthmname}{Lemma S.4}\begin{thmx}

If the automaton \((\alpha,\{A(\sigma)\},\beta)\) defines the series \(f\), then the automaton \((\beta,\{A(\sigma)^{\top}\},\alpha)\) defines the series \(\sigma_0\cdots\sigma_{N-1}\mapsto f(\sigma_{N-1}\cdots\sigma_0)\). The ladder \(\mathbf{for}\ i:=n+c_1\ \mathbf{to}\ c_2\ \mathbf{step}\ {-s}\) becomes a ladder with step \(s\) after the change of positions \(j\mapsto N-1-j\); its automaton is the transposed automaton of the reflected ladder. The conditions of such a ladder read only bits of the window (Definition 3, (b)), so after reflection there are no conditions on the positions of \(E\).

\end{thmx}

\begin{proof}

Transposition of the product: \(\bigl(\alpha^{\top}A(\sigma_0)\cdots A(\sigma_{N-1})\beta\bigr)^{\top}=\beta^{\top}A(\sigma_{N-1})^{\top}\cdots A(\sigma_0)^{\top}\alpha\). Under reflection, the window qubit \(i+o\) goes to \(i'-o\), where \(i'=N-1-i\) is the counter value of the reflected ladder, the position \(c\) goes to the position \(N-1-c\), and the position \(N+c\) to the position \(-1-c\); the order of the bodies is preserved.

\end{proof}

\subsection{Words of any length}\label{sec:S1.5}

The bound on the automorphisms in Lemma S.10 is needed on all words of a block of the Hankel matrix, including short ones, where a ladder is still empty or a fixed segment does not yet fit.

\renewcommand{\snthmname}{Lemma S.5}\begin{thmx}

For any \(N\) and any bits \(a\), the matrix \(\bigl(f(w)\bigr)_{x,y}\) defined by the transition graph of a segment on words of length \(N\) is zero or unitary. The same holds for the automaton of the program and after applying any automorphism \(\gamma_k\).

\end{thmx}

\begin{proof}

In an accepted path each position has exactly one role: a position of \(S\), a position of a window or a gap, or a position of the tail (the positions of \(E\) among them). The roles come in reading order and therefore do not overlap. The proof of Lemmas S.2--S.3 uses only this and the fact that a body acts on its own positions. Hence for any \(N\) for which an accepted path exists, the value equals \(\langle x\rvert\Pi\otimes I\lvert y\rangle\), where \(\Pi\) is the product of the bodies placed on these pairwise distinct positions, that is, the matrix of a circuit. If there are no accepted paths (the ladder is empty, the fixed segment does not fit), the matrix is zero. The automaton of the program defines the product of the matrices of the segments (Lemma 1), and a product of unitary and zero matrices is unitary or zero. The automorphism \(\gamma_k\) maps the matrix of each gate to a unitary matrix (Lemma 5), and the structure of the transition graph does not depend on \(\gamma_k\).

\end{proof}

\subsection{Relations}\label{sec:S1.6}

\renewcommand{\snthmname}{Lemma S.6}\begin{thmx}[bits of the ancillas]

On words with \(n\ge n_{\min}(P)\), the value of the series of a program \(P\) does not depend on the bits \(a_j\) on the letters of the ancillas \(j\ge n\).

\end{thmx}

\begin{proof}

For \(n\ge n_{\min}(P)\), the conditions read only bits \(a_k\) with \(0\le k\le n-1\) (condition (T1) of Definition 4), and the conditions in the bodies of ladders only the bits of the window and of the positions of \(S\) (Definition 3, (b)). The weight of a path in the transition graph of a segment depends on the bit of a letter only at positions whose bit is read by a condition. At the other positions, the transitions do not read the bit or (at the positions of \(E\)) check it against their own guess; for any value of the bit, the accepted path has the same weight (Lemmas S.1--S.3).

\end{proof}

\renewcommand{\snthmname}{Lemma S.7}\begin{thmx}[ancillas]

Let the automaton \((\alpha,\{A(\sigma)\},\beta)\) define the series \(\langle x\rvert U\lvert y\rangle\) on words of length \(n+e\), and \(z\in\{0,1\}^{e}\). Then the automaton \((\alpha,\{A(\sigma)\},\beta_z)\) with
\[\beta_z=A(0,z_0,0)\cdots A(0,z_{e-1},0)\,\beta\]
defines, on words \(u\) of length \(n\) over the working qubits, the series \(\langle x_u,z\rvert U\lvert y_u,0^{e}\rangle\).

\end{thmx}

\begin{proof}

\(\alpha^{\top}A(u)\beta_z=\alpha^{\top}A\bigl(u\,(0,z_0,0)\cdots(0,z_{e-1},0)\bigr)\beta\); the bit of an ancilla is taken to be 0 (Lemma S.6). The dimension of the automaton does not change.

\end{proof}

\renewcommand{\snthmname}{Lemma S.8}\begin{thmx}[up to phase]

Let \(M_G,M_R\) be matrices of the same size (or vectors), \(M_R\ne0\). Then \(M_G=\lambda M_R\) with \(\lvert\lambda\rvert=1\) if and only if
\[M_G[i,j]\,\overline{M_G[k,l]}=M_R[i,j]\,\overline{M_R[k,l]}\quad\text{for all }i,j,k,l.\]

\end{thmx}

\begin{proof}

The forward direction is obvious. Conversely, choose \((k,l)\) with \(M_R[k,l]\ne0\). For \((i,j)=(k,l)\) we get \(\lvert M_G[k,l]\rvert=\lvert M_R[k,l]\rvert\ne0\). Then for all \((i,j)\)
\[M_G[i,j]=M_R[i,j]\cdot\frac{\overline{M_R[k,l]}}{\overline{M_G[k,l]}}=\lambda\,M_R[i,j],\qquad \lvert\lambda\rvert=1.\ \square\]

The lemma is applied to the vectors \(U\lvert0^{N}\rangle\) (relation \(\simeq_0\)) and to the matrices \(V\) (relation \(\simeq_{\mathrm{anc}}\)). Both sides define doubled series (Section 5.2); the automaton obtained by applying \(\gamma_{-1}\) is built modulo \(p\) by the substitution \(\zeta_m\mapsto\omega_p^{-1}\).

\subsection{\texorpdfstring{Computation modulo \(p\) and plans}{Computation modulo p and plans}}\label{sec:S1.7}

A prime is \emph{bad} for a segment if modulo it at least one non-zero entry of the matrix of its body (for a fixed segment, of the matrix \(V\)) vanishes for some bits.

\textbf{Lemma S.9 (size of the transition graph).} Let the matrices of the body of a segment \(P_\ell\) (over all values of the bits) have \(\nu_\ell\) non-zero entries, and let \(\kappa_{B_\ell}\) be the sum of \(\kappa_g\) over the gates of the body. Then there are at most \(k_\ell=\nu_\ell\lfloor\varphi(m)\kappa_{B_\ell}/(2\log_2p_0)\rfloor\) bad primes \(p\ge p_0\). If \(p\) is not bad, the set of reachable states of the transition graph modulo \(p\) coincides with the set over \(\mathbb{K}\); if it is bad, it is contained in it.

\emph{Proof.} An entry of the body multiplied by \(\sqrt2^{\,\kappa_{B_\ell}}\) lies in \(\mathcal{O}\), and its images under \(\gamma_k\) have absolute value at most 1 (Lemma 5). By Lemma 4 with \(t=p_0\), at most \(\lfloor\varphi(m)\kappa_{B_\ell}/(2\log_2p_0)\rfloor\) primes make it vanish. Summing over the entries gives the bound. A transition of the graph exists if its weight is non-zero. A zero over \(\mathbb{K}\) stays zero modulo \(p\), and a non-zero weight becomes zero only for a bad \(p\). Hence the transition graph modulo \(p\) is a subgraph of the graph over \(\mathbb{K}\) and coincides with it for a good \(p\).

\end{proof}

Let \(k_{\rm gate}=\max_\ell k_\ell\) over the segments of both programs. The maximum of the number of reachable states over any \(k_{\rm gate}+1\) distinct primes equals the number \(M_\ell\) of states over \(\mathbb{K}\).

\renewcommand{\snthmname}{Lemma S.10}\begin{thmx}[rank modulo $p$]

Let a series be given by a transition graph with at most \(M\) states over \(\mathbb{K}\), and let its value on any word of length \(N'\), after any \(\gamma_k\), be an entry of a zero or unitary matrix, with \(\sqrt2^{\,\kappa(N')}\) times the value lying in \(\mathcal{O}\) (\(\kappa\) non-decreasing). Let \(r\) be the least dimension of an automaton of the series over \(\mathbb{K}\), and \(\hat r\) the largest of the least dimensions modulo \(T_1\) primes chosen independently and uniformly from \(\mathcal{P}_m\). Then
\[\begin{gathered}\Pr[\hat r<r]\le q(r-1)^{T_1}\le q(M-1)^{T_1},\qquad q(s)=\frac{\Lambda(s)/\log_2p_0+1}{\lvert\mathcal{P}_m\rvert},\\ \Lambda(s)=\varphi(m)\Bigl(\frac{s+1}{2}\log_2(s+1)+(s+1)\frac{\kappa(2s)}{2}\Bigr).\end{gathered}\]

\end{thmx}

\begin{proof}

The number \(r\) is not random. For an automaton of least dimension over \(\mathbb{K}\), the vectors \(\alpha^{\top}A(u)\) of words of length at most \(r-1\) span \(\mathbb{K}^{r}\), and so do the vectors \(A(v)\beta\). Indeed, both spaces have dimension \(r\) by minimality \cite[proof of Thm.~3.7]{Kiefer2020_WANotes}, and each has a basis of vectors of words of length at most \(r-1\) \cite[Prop.~2.1 and the analogous statement for the vectors \(A(v)\beta\)]{Kiefer2020_WANotes}. Hence the block of the Hankel matrix \(H[u,v]=f(uv)\) on words of length at most \(r-1\) has rank \(r\) and contains a non-zero minor \(\mu\) of order \(r\). Its entries are values of the series on words of length at most \(2(r-1)\); after any \(\gamma_k\) they have absolute value at most 1, and \(\sqrt2^{\,\kappa(2(r-1))}\) times each of them lies in \(\mathcal{O}\). By Hadamard's inequality \(\lvert\gamma_k(\mu)\rvert\le r^{r/2}\), so Lemma 4 applies to \(\mu\) with \(t=p_0\), \(c=r^{r/2}\) and \(\kappa=r\kappa(2(r-1))\), and at most \(\Lambda(r-1)/\log_2p_0\) primes make \(\mu\) vanish. The least dimension modulo \(p\) equals the rank of the Hankel matrix modulo \(p\) (F4), and if \(\hat r<r\), then each chosen \(p\) made \(\mu\) vanish. The primes are independent, and the probability for one is at most \(q(r-1)\); \(q\) is increasing, and \(r\le M\).

\end{proof}

The condition on the values holds for the segments and for the program by Lemma S.5.
For the series of a single segment and for words of any length \(N'\), including short ones, \(\kappa\) in Lemma S.10, Corollary S.11 and Theorem S.13 is understood as follows: \(\kappa_\ell(N')=N'\kappa_{B_\ell}\) for a ladder whose body has the sum of \(\kappa_g\) equal to \(\kappa_{B_\ell}\), and \(\kappa_\ell(N')=\kappa(P_\ell)\) for a fixed segment; for a program, the sum over the segments. On a word of length \(N'\), an accepted path in the transition graph of a ladder applies at most \(N'\) bodies, so \(\sqrt2^{\,\kappa_\ell(N')}\) times the value of the series lies in \(\mathcal{O}\).

\renewcommand{\snthmname}{Corollary S.11}\begin{thmx}[exact rank]

Under the conditions of Lemma S.10, let \(k=\lfloor\Lambda(M-1)/\log_2p_0\rfloor\). Then \(r\) equals the largest of the least dimensions modulo the first \(k+1\) primes \(p\equiv1\pmod m\), \(p\ge p_0\), in increasing order.

\end{thmx}

\begin{proof}

Reduction does not increase the rank: minors that are zero over \(\mathbb{K}\) are also zero modulo \(p\). The minor \(\mu\) from the proof of Lemma S.10 vanishes modulo at most \(\Lambda(r-1)/\log_2p_0\le k\) primes, so one of the \(k+1\) primes does not make it vanish, and modulo that prime the rank is at least \(r\).

\end{proof}

For a segment, \(M=M_\ell\) (Lemma S.9); for a program, \(M=\prod_\ell r_\ell\), the product of the exact ranks of the segments (Lemma 1). If \(k\) is large, the bound \(r\le M\) can be used instead of \(r\); the function \(q\) and the number \(k\) with \(M\) instead of \(M-1\) give bounds with a margin.

\renewcommand{\snthmname}{Theorem S.12}\begin{thmx}[exact plan]

Let \(M_\ell\) be the maximum of the number of reachable states over the first \(k_{\rm gate}+1\) primes, \(d_G,d_R\) the exact ranks of the programs by Corollary S.11 or their upper bounds \(\prod_\ell r_\ell\), \(d\) the dimension bound of the series \(\Delta\) of Table 1, \(N^{*}=N_0+d\), \(\kappa(N^{*})\) as in Section 5.4 (the sum of \(\kappa_R\) and \(\kappa_G\), doubled for pair series; non-decreasing in the length), and \(k_\Delta=\lfloor\varphi(m)(1+\kappa(N^{*})/2)/\log_2p_0\rfloor\). If \(\Delta\) is zero modulo each of the first \(T=\max(k_{\rm gate},k_\Delta)+1\) primes \(p\equiv1\pmod m\), \(p\ge p_0\), then \(\Delta=0\) over \(\mathbb{K}\) on all words of length at least \(N_0\).

\end{thmx}

\begin{proof}

By Lemma S.9, among \(k_{\rm gate}+1\) primes there is a good one for each segment, so the \(M_\ell\) are exact, and \(d_G,d_R\) are valid bounds over \(\mathbb{K}\) (Corollary S.11, or \(r_\ell\le M_\ell\) and Lemma 1). If \(\Delta\ne0\), then by Lemma 3 it has a witness \(w^{*}\) of length from \(N_0\) to \(N_0+d-1<N^{*}\). The value \(\Delta(w^{*})\) is a non-zero element of \(\mathbb{K}\); since the length of \(w^{*}\) is less than \(N^{*}\) and \(\kappa\) is non-decreasing, \(\sqrt2^{\,\kappa(N^{*})}\Delta(w^{*})\in\mathcal{O}\), and its images under \(\gamma_k\) have absolute value at most 2 (Section 5.4), so it vanishes modulo at most \(k_\Delta\) primes (Lemma 4 with \(t=p_0\)). Among \(T>k_\Delta\) distinct primes there is a \(p\) with \(\pi_p(\Delta(w^{*}))\ne0\), and the zeroness check modulo this \(p\), which is complete over \(\mathbb{F}_p\) (Section 5.3), would have found a non-zero word.

\end{proof}

If the bodies of the ladders contain no gates with \(\kappa_g>0\), then \(\kappa\) does not grow with the length; if in addition \(\varphi(m)\bigl(1+\kappa(N^{*})/2\bigr)<\log_2p_0\) and \(k_{\rm gate}=0\), then \(T=1\).

\renewcommand{\snthmname}{Theorem S.13}\begin{thmx}[probabilistic plan]

Let \(\mathcal{P}_m\ne\varnothing\). The primes are chosen independently and uniformly from \(\mathcal{P}_m\) in three batches. The length of each batch is fixed before its primes are chosen.

\begin{itemize}
\tightlist
\item
  Batch 0 of \(T_0\) primes gives \(\hat M_\ell\), the maximum of the number of reachable states of the transition graph of a segment; \(T_0\) is chosen from \(k_\ell\) and is therefore not random.
\item
  Batch 1 of \(T_1\) primes gives \(\hat r_\ell\), the maximum of the least dimensions of the automata of the segments; \(T_1\) is chosen from \(\hat M_\ell\). The batch stops early only if all \(\hat r_\ell=\hat M_\ell\).
\item
  Batch 2 of \(T_2\) primes checks \(\Delta\) for zeroness; \(d\) is computed from \(\prod_\ell\hat r_\ell\), and \(T_2\) from \(d\).
\end{itemize}
Let
\[\begin{gathered}\varepsilon^{(0)}=\sum_\ell\Bigl(\frac{k_\ell}{\lvert\mathcal{P}_m\rvert}\Bigr)^{T_0},\qquad \varepsilon^{(1)}=\sum_\ell\bigl(q_\ell(M_\ell-1)\bigr)^{T_1(M)},\\ \varepsilon^{(2)}=q_0^{\,T_2(d_{\rm true})},\qquad q_0=\frac{\varphi(m)(1+\kappa(N^{*})/2)/\log_2p_0+1}{\lvert\mathcal{P}_m\rvert},\end{gathered}\]
where \(q_\ell\) is the function \(q\) of Lemma S.10 for the segment \(P_\ell\), \(M_\ell\) is the number of states over \(\mathbb{K}\), \(T_1(M)\) and \(T_2(d_{\rm true})\) are the batch lengths computed from the exact \(M_\ell\) and from the bound \(d_{\rm true}\) obtained from the exact least dimensions \(r_\ell\), and \(N^{*}\) is computed from \(d_{\rm true}\). Let \(\mathcal{E}_0\) be the event ``for some segment all primes of batch 0 are bad'', \(\mathcal{E}_1\) the event ``\(\hat r_\ell<r_\ell\) for some \(\ell\)'', and \(\mathcal{E}_2\) the event ``all checks of batch 2 gave zero''. If \(\Delta\ne0\) on words of length at least \(N_0\), then
\[\Pr[\mathcal{E}_0]\le\varepsilon^{(0)},\qquad \Pr[\mathcal{E}_1\setminus\mathcal{E}_0]\le\varepsilon^{(1)},\qquad \Pr[\mathcal{E}_2\setminus(\mathcal{E}_0\cup\mathcal{E}_1)]\le\varepsilon^{(2)},\]
and the probability of the answer ``equal'' is at most \(\varepsilon=\varepsilon^{(0)}+\varepsilon^{(1)}+\varepsilon^{(2)}\).

\end{thmx}

\begin{proof}

\(\Pr[\mathcal{E}_0]\le\varepsilon^{(0)}\): for each segment there are at most \(k_\ell\) bad primes (Lemma S.9), and the primes are independent; apply the union bound over the segments. Outside \(\mathcal{E}_0\), all \(\hat M_\ell=M_\ell\) (Lemma S.9), so the length of batch 1 equals the non-random number \(T_1(M)\), and \(\Pr[\mathcal{E}_1\setminus \mathcal{E}_0]\le\varepsilon^{(1)}\) by Lemma S.10 and the union bound over the segments. Early stopping gives no error: outside \(\mathcal{E}_0\), \(\hat r_\ell=\hat M_\ell\) implies \(\hat r_\ell=r_\ell\), since \(\hat r_\ell\le r_\ell\le M_\ell\). Outside \(\mathcal{E}_0\cup \mathcal{E}_1\), all \(\hat r_\ell=r_\ell\), the bound \(d\) and the length of batch 2 equal \(d_{\rm true}\) and \(T_2(d_{\rm true})\), the witness \(w^{*}\) of Lemma 3 is a fixed word independent of batch 2, and each prime of the batch makes \(\Delta(w^{*})\) vanish with probability at most \(q_0\) (Lemma 4; the length of \(w^{*}\) is less than \(N^{*}\), and \(\kappa\) is non-decreasing). The batches are independent, so \(\Pr[\mathcal{E}_2\setminus(\mathcal{E}_0\cup \mathcal{E}_1)]\le\varepsilon^{(2)}\). The answer ``equal'' requires \(\mathcal{E}_2\), and \(\mathcal{E}_2\subseteq \mathcal{E}_0\cup(\mathcal{E}_1\setminus \mathcal{E}_0)\cup(\mathcal{E}_2\setminus(\mathcal{E}_0\cup \mathcal{E}_1))\).

\end{proof}

The algorithm computes the bound \(\hat\varepsilon=\hat\varepsilon^{(0)}+\hat\varepsilon^{(1)}+\hat\varepsilon^{(2)}\) by the same formulas with \(\hat M_\ell\) and \(d\) instead of \(M_\ell\) and \(d_{\rm true}\); if \(\hat r_\ell=\hat M_\ell\) for all \(\ell\), it sets \(\hat\varepsilon^{(1)}=0\). Outside \(\mathcal{E}_0\cup \mathcal{E}_1\), the computed \(\hat\varepsilon^{(0)},\hat\varepsilon^{(2)}\) equal \(\varepsilon^{(0)},\varepsilon^{(2)}\).

\renewcommand{\snthmname}{Corollary S.14}\begin{thmx}

Let the lengths \(T_0\), \(T_1(M)\) and \(T_2(d_{\rm true})\) be chosen so that \(\varepsilon^{(0)},\varepsilon^{(1)}\le\varepsilon_0/4\) and \(\varepsilon^{(2)}\le\varepsilon_0/2\). If \(\Delta\ne0\) on words of length at least \(N_0\), then the probability of the answer ``equal'' is at most \(\varepsilon_0\).

\end{thmx}

\begin{proof}

By Theorem S.13, the probability is at most \(\varepsilon^{(0)}+\varepsilon^{(1)}+\varepsilon^{(2)}\le\varepsilon_0\).

\end{proof}

The choice of batch lengths at most \(t_{\max}\) and the outcome ``no answer'' for \(\hat\varepsilon\ge1\) are in Section 5.5 of the article; the values of \(\varepsilon_0\), \(t_{\max}\) and the way the primes are chosen are in Section 6.2.

\subsection{Proofs of Lemma 5 and Proposition 2}\label{sec:S1.8}

\emph{Lemma 5} (Section 5.4 of the article): for each automorphism \(\gamma_k\) and gate \(g\in\mathcal{G}\), the matrix \(\gamma_k(M_g)\) is unitary, so \(\lvert\gamma_k(\langle x\rvert U\lvert y\rangle)\rvert\le1\) for the matrix \(U\) of any circuit over \(\mathcal{G}\).

\begin{proof}

The matrices of the gates X, Y, Z, S, T, S\(^{\dagger}\), T\(^{\dagger}\), CX, CY, CZ, SWAP, CCX, CCZ, CSWAP, RZ, P, CP, CRZ, RZZ are monomial: each row and each column has one non-zero entry, and it is a root of unity. The automorphism \(\gamma_k\) maps a root of unity to a root of unity, so the image of such a matrix is monomial with entries of absolute value 1, that is, unitary. For the other gates: \(\gamma_k(\sqrt2)=\pm\sqrt2\), so \(\gamma_k(\mathrm{H})=\pm\mathrm{H}\), and CH goes to a controlled \(\pm\)H; \(\gamma_k(e^{\mathrm{i}\theta/2})=e^{\mathrm{i}k\theta/2}\) and \(\gamma_k(\mathrm{i})=\pm\mathrm{i}\), so RX\((\theta)\) goes to RX\((k\theta)\), and RY\((\theta)\) to RY\((\pm k\theta)\). The matrix of a gate application is the tensor product of \(M_g\) and an identity matrix, and \(U\) is a product of such matrices; \(\gamma_k\) preserves sums and products, so \(\gamma_k(U)\) is a product of unitary matrices.

\end{proof}

\emph{Proposition 2} (Section 5.5 of the article; \(\kappa_<\), \(T'\) and the conditions for the six relations are given there): if for each \((n,a)\in\mathrm{dom}(R)\) with \(n<n_0\) we have \((n,a)\in\mathrm{dom}(G)\) and, modulo each of the first \(T'\) primes \(p\equiv1\pmod m\), \(p\ge p_1\), the matrices \(U_R(n,a)\) and \(U_G(n,a)\) satisfy the condition for the relation, then \(G\sim R\) at all \((n,a)\in\mathrm{dom}(R)\) with \(n<n_0\).

\begin{proof}

Let \(\xi\) be an entry of the difference, a minor or (for \(\equiv_{\mathrm{anc}}\), \(\simeq_{\mathrm{anc}}\)) an entry of \(U_R\) or \(U_G\) with \(z\ne0\) that is non-zero over \(\mathbb{K}\). By Lemma 5, \(\lvert\gamma_k(\xi)\rvert\le2\), and \(\sqrt2^{\,\kappa_<}\xi\in\mathcal{O}\). By Lemma 4 with \(t=p_1\), it vanishes modulo at most \(T'-1\) primes, so \(\xi\) is non-zero modulo one of the \(T'\) primes. Hence the conditions of the proposition imply that all differences of entries and all minors of the conditions are zero over \(\mathbb{K}\), and for \(\equiv_{\mathrm{anc}}\) and \(\simeq_{\mathrm{anc}}\) also that \(R\) and \(G\) return their ancillas at each \((n,a)\). For the relations up to phase, zero minors give \(M_G=\mu M_R\) for the compared matrices (vectors), and since both are unitary (the vectors are unit vectors; for \(\simeq_{\mathrm{anc}}\) these are \(V_G\) and \(V_R\)), \(\lvert\mu\rvert=1\). Hence \(G\sim R\) by Definition 2.

\end{proof}

%% file: supplement/S2_translator.tex
\section{The Python+Qiskit translator and the proof of Theorem 3}\label{sec:S2}

The translator takes a function \texttt{solve(n,\ a)} and returns a program \(P\) and a threshold \(n_{\rm tr}(P)\), or a refusal with a code from Table S2. The translator checks the conditions of Definition 3 as described in Sects. S2.3--S2.5. Theorem 3 concerns the returned programs that satisfy Definition 3. Assumptions (A1), (A2), (A4) and property (A3) are stated in Section 6.1.

\subsection{The supported fragment}\label{sec:S2.1}

The body of \texttt{solve(n,\ a)} is a sequence of statements:

\begin{itemize}
\tightlist
\item
  \texttt{qc\ =\ QuantumCircuit(n\ +\ e)} exactly once and \texttt{return\ qc} at the end;
\item
  an assignment of an expression of Sect. S2.2 to a name or a tuple of names (the reassignment rule is below);
\item
  a gate call \texttt{qc.g(...)} of fixed arity or \texttt{qc.g(range(...))} (expanded into a loop);
\item
  \texttt{for\ x\ in\ it:}, where \texttt{it} is a \texttt{range}, \texttt{list(range(...))}, \texttt{reversed(range(...))} or a list slice;
\item
  \texttt{if\ C:}, where \(C\) is evaluated at translation time (module constants only) or is a bit test (Sect. S2.3);
\item
  a call of a function defined in the same file or in a module from the same directory (import \texttt{from\ m\ import\ f}), without recursion; the arguments are evaluated in the scope of the caller. A call is a statement, and its value is discarded. Hence the last statement of the called function may be \texttt{return\ x}, where \texttt{x} is its parameter (usually \texttt{return\ qc} in a function that extends the circuit in place); for the translator this is an empty statement.
\end{itemize}

\emph{Reassignment.} A name may be assigned again if its previous assignment was made in the same body: at the top level of the function or in the same iteration of the same loop. An assignment in the body of a loop to a name assigned outside the loop gives the refusal \texttt{alias\_reassigned}: such a name accumulates its value across iterations. For example, \texttt{t\ =\ i\ +\ 1} followed by \texttt{t\ =\ t\ +\ 1} in the body of a loop is allowed, while \texttt{t\ =\ 0} before the loop and \texttt{t\ =\ t\ +\ 1} in its body is not. The operator \texttt{+=} and reassignment of the name of the circuit give the same refusal.

If the code has no loop with a bound of the form \(n+c\) but has a gate on the qubit \(n+c\), the program consists of fixed segments only. If it has neither, the circuit acts non-trivially only on qubits with constant indices. The translator does not accept such code (\texttt{no\_ladder}), although it can be written as a fixed segment.

\subsection{Values of expressions and proof obligations}\label{sec:S2.2}

The value of an expression has one of three forms:

\begin{itemize}
\tightlist
\item
  an affine form \(v(n,i)=c+k_0n+k_1i\) with integer coefficients, where \(i\) is the counter symbol of a ladder;
\item
  an interval of qubit indices \([\beta,\beta+\mu)\) (the value \texttt{list(range(b,\ b\ +\ l))}) with affine start \(\beta\) and length \(\mu\);
\item
  a \texttt{range} with affine bounds and a non-zero constant step.
\end{itemize}

Each rule generates \emph{proof obligations}, affine forms that must be non-negative for the value of the expression in Python to equal its abstract value.

\begin{table}[tbp]
\caption{Rules for expressions.}\label{tab:S1}
\small

\footnotesize\noindent
\begin{tabularx}{\textwidth}{@{}>{\raggedright\arraybackslash\hsize=1.43\hsize\hspace{0pt}}X>{\raggedright\arraybackslash\hsize=0.48\hsize\hspace{0pt}}X>{\raggedright\arraybackslash\hsize=0.69\hsize\hspace{0pt}}X>{\raggedright\arraybackslash\hsize=1.41\hsize\hspace{0pt}}X@{}}
\toprule
Expression & Value & Proof obligations & Reason \\
\midrule
$E_1\pm E_2$, $cE$ & sum, difference, multiple & --- & affine forms are closed under these operations \\
$Q[k]$, $Q=[\beta,\beta+\mu)$, $k$ an affine form & $\beta+k$ & $k\ge0$, $\mu-1-k\ge0$ & otherwise Python takes an element from the other end of the list or raises \texttt{IndexError} \\
$Q[k]$, $Q=[\beta,\beta+\mu)$, $k<0$ a constant & $\beta+\mu+k$ & $\mu+k\ge0$ & otherwise Python raises \texttt{IndexError} \\
$Q[lo:hi]$; $lo$, $hi$ affine forms other than negative constants & $[\beta+lo,\,\beta+hi)$ & $lo\ge0$, $hi-lo\ge0$, $\mu-hi\ge0$ & otherwise Python truncates the slice \\
\texttt{list(range(x, y))}, \texttt{len(range(x, y))} & $[x,y)$, $y-x$ & $y-x\ge0$ & otherwise the length is $\max(0,y-x)$ \\
\texttt{reversed(range(x, y, s))}, $s=\pm1$ & \texttt{range(y-s, x-s, -s)} & --- & the equality also holds for an empty \texttt{range} \\
\botrule
\end{tabularx}

\end{table}

A product of two variables, the operations \texttt{//}, \texttt{\%}, \texttt{**} and a conditional expression give a refusal.

\emph{Angles.} The translator evaluates the expression of a rotation angle separately from index expressions, in the form \(q\pi+r\) with rational \(q\) and \(r\): \texttt{math.pi} is the symbol \(\pi\), a numeric literal is its exact value in decimal notation (\texttt{0.1} is \(1/10\)), and a name assigned a constant angle or number has that value. Sum and difference, multiplication and division by a rational number, and an integer power of a rational number with exponent at most 64 in absolute value are computed exactly. All other expressions, including those depending on \(n\), the loop counter or the parameter, give the refusal \texttt{nonconst\_angle}.
If \(r\ne0\), the translator also refuses with the code \texttt{nonconst\_angle}; otherwise the gate of the program gets the angle \(q\pi\) (property (A3) of Theorem 3).

\renewcommand{\snthmname}{Lemma S.15}\begin{thmx}

If an expression \(E\) is accepted and its proof obligations hold for given \(n\) and \(i\), then the value of \(E\) in Python equals its abstract value at these \(n\) and \(i\).

\end{thmx}

\begin{proof}

Induction on the structure of \(E\). Each step is a row of Table S1; the proof obligations exclude exactly the cases in which Python computes another value or raises an exception.

\end{proof}

\renewcommand{\snthmname}{Lemma S.16}\begin{thmx}[names and calls]

The value of a name at the point of use equals the value of the last assignment to this name executed by Python.

\end{thmx}

\begin{proof}

Values are immutable: the translator does not accept list methods, \texttt{+=} or assignments to elements. Reassignment is allowed only in the same context, so the order of assignments in the translator agrees with the order of execution. Names assigned in the body of a loop cannot be used after the loop (in Python they are defined, but the translator refuses). An argument of a call is evaluated in the scope of the caller. Expressions have no side effects, and the called function cannot reassign names of the caller (there are no \texttt{global} and \texttt{nonlocal}). Hence substituting the value of the argument agrees with argument passing in Python.

\end{proof}

\subsection{Conditions}\label{sec:S2.3}

\begin{itemize}
\tightlist
\item
  A condition built from module constants is evaluated at translation time; by (A1), Python obtains the same value.
\item
  \emph{Bit test.} Let \texttt{b} be the expression \texttt{a{[}k{]}} or \texttt{(a\ \textgreater{}\textgreater{}\ k)\ \&\ 1}. By (A4) it takes the values 0 and 1, so \texttt{if\ b}, \texttt{b\ ==\ 1}, \texttt{b\ ==\ 0}, \texttt{b\ !=\ v} and \texttt{not\ b} are tests of the bit \(a_k\). The expression \texttt{t\ =\ a\ \&\ (1\ \textless{}\textless{}\ k)} takes the values \(0\) and \(2^{k}\); only \texttt{if\ t}, \texttt{t\ ==\ 0} and \texttt{t\ !=\ 0} count as bit tests. The comparison \texttt{t\ ==\ 1} gives a refusal, since for \(k\ne0\) it is false for every \(a\).
\item
  The bit index \(k\) is an affine form with the proof obligations \(k\ge0\) and \(n-1-k\ge0\). For \(k\ge n\), a list gives \texttt{IndexError} and an integer gives zero, so without this condition the value in Python would not agree with the semantics of \(\mathcal{L}\). A violation of the condition at an end of the range of the counter gives the refusal \texttt{bit\_index\_out\_of\_range}, and a negative index gives \texttt{bit\_index\_negative}.
\item
  In the body of a ladder, a condition may be only on a bit \(a[i+o]\) or, for step \(+s\), on a bit \(a[c]\) (Definition 3(b)). A condition on a bit of the form \(a[n+c]\) gives the refusal \texttt{bit\_on\_end\_fixed}. A condition on a bit \(a[c]\) in a ladder with step \(-s\) gives the refusal \texttt{bit\_const\_in\_down\_ladder}.
\item
  Assignments under a bit condition are forbidden (\texttt{alias\_under\_condition}). Hence a branch does not change the values of names and only adds the condition \(a_k=v\) to its gates. Nested conditions give a gate a set of conditions; the gate is applied if all of them hold.
\item
  Branching on \(n\), on the counter of a ladder or on the length of the parameter gives a refusal (\texttt{branch\_on\_size}, \texttt{branch\_on\_loop\_index}, \texttt{branch\_on\_param}).
\end{itemize}

\subsection{Loops}\label{sec:S2.4}

\renewcommand{\snthmname}{Lemma S.17}\begin{thmx}

(a) A loop whose bounds are both constant or both of the form \(n+c\) runs over a set of values that depends on \(n\) only by a shift by \(0\) or by \(n\). The translator fully unrolls such a loop, and the result agrees with execution in Python. (b) A loop over \texttt{range(c1,\ n\ +\ c2,\ s)} with \(s>0\) runs over the values \(i=c_1,c_1+s,\dots\) less than \(n+c_2\), that is, it agrees with the ladder \(\mathbf{for}\ i:=c_1\ \mathbf{to}\ n+c_2-1\ \mathbf{step}\ s\). A loop over \texttt{range(n\ +\ c1,\ c2,\ -s)} with \(s>0\) agrees with the ladder \(\mathbf{for}\ i:=n+c_1\ \mathbf{to}\ c_2+1\ \mathbf{step}\ {-s}\). A loop over \texttt{reversed(range(x,\ y,\ s))} with \(s=\pm1\) is, by Table S1, a loop over \texttt{range(y-s,\ x-s,\ -s)}. For a step \(\lvert s\rvert\ge2\), the first value of \texttt{reversed(range(...))} is not of the form \(n+c\), and the translator refuses (\texttt{range\_not\_ladder}). A loop over a list slice \(Q[lo:hi]\) is, by Table S1, a loop over the interval \([\beta+lo,\beta+hi)\) with step 1. The body is parsed once with the symbol \(i\). If the body is accepted, then for each value of \(i\) in the progression, the sequence of gates of the body in Python equals the substitution of this value into the symbolic sequence.

\end{thmx}

\begin{proof}[Proof of (b)]

An accepted body has no branches on \(i\) (\texttt{branch\_on\_loop\_index}), no nested ladders and no loops with bounds depending on \(i\) (\texttt{nested\_loop}). Hence the execution path of the body is the same for all \(i\), and the values of the expressions are affine forms that by Lemma S.15 equal the values in Python as long as the proof obligations hold.

\end{proof}

\subsection{The threshold $n_{\rm tr}(P)$ and the proof of Theorem 3}\label{sec:S2.5}

Each proof obligation is an inequality \(c+k_0n+k_1i\ge0\). The translator handles the proof obligations and the inequalities of Definition 4 as described after Definition 4: an inequality in the body of a ladder is checked at the ends of the range of the counter, an inequality violated for all sufficiently large \(n\) gives a refusal, and each of the others sets its own threshold. In addition, the translator checks two conditions of Definition 3:

\begin{itemize}
\tightlist
\item
  fixed positions (qubits and bits) of the form \(c\) lie to the left of the first window, and those of the form \(n+c\) to the right of the last (condition 3(c)). Like Definition 3, the translator counts the window, \(S\) and \(E\) over qubits and bits. If at some \(i\) a position of the window coincides with a fixed position (within one gate or between gates of the body; for example, the window bit \(a[i-1]\) at \(i=1\) and the fixed qubit 0), the refusal is \texttt{fixed\_overlaps\_window}. If a fixed position lies inside the range of the ladder but the windows do not touch it (for a step \(s>w\)), the refusal is \texttt{fixed\_inside\_ladder}.
\item
  A body whose window consists of bits only (for example, \texttt{for\ i\ in\ range(n):\ if\ a{[}i{]}:\ qc.z(n)}) satisfies Definition 3 and is accepted.
\end{itemize}

The translator checks the inequalities (T3) for each segment separately, over the positions of qubits and bits. In addition, the qubit expressions of one gate application must be pairwise distinct as expressions (Definition 3); otherwise the translator refuses (\texttt{qubit\_collision}). The threshold \(n_{\rm tr}(P)\) is the largest of the thresholds of all inequalities and one. The inequalities of Definition 4 are in this list, so \(n_{\rm tr}(P)\ge n_{\min}(P)\).

\emph{Proof of Theorem 3.} Let \(n\ge n_{\rm tr}(P)\) and \(a\in\{0,1\}^{n}\). All proof obligations hold. By Lemmas S.15--S.17 and the rules of Sect. S2.3, Python calls the same gates in the same order and with the same bit conditions as the program \(P\) at \((n,a)\). By (T1), the qubit indices of each call lie in the register; by Definition 3 and conditions 3(c) and (T3), they are pairwise distinct (Section 4.1). Hence by (A2) each call appends one gate application to the circuit and raises no exception. By the rule for angles (Sect. S2.2) and property (A3), the symbolic value of the angle expression of each call equals the angle \(q\pi\) of the gate of the program, and Python evaluates the same expression in floating-point arithmetic. Therefore \texttt{solve(n,\ a)} terminates and builds the circuit \(P(n,a)\) up to replacing each such angle with \(q\pi\). \(\square\)

\subsection{Refusal codes}\label{sec:S2.6}

The column ``Kind'' of Table S2 separates the two kinds of refusals (Section 6.1): \emph{class}, the code uses a construct absent from Definition 3 and the language \(\mathcal{L}\); \emph{fragment}, the translator does not parse the syntax of the code, although the same circuit may be given by a ladder program; \emph{both}, both cases occur under one code (for example, \texttt{2 ** j} with variable and with constant \(j\)).

\begin{table}[tbp]
\caption{Refusal codes of the translator and examples of code that causes them.}\label{tab:S2}
\small

\footnotesize\noindent
\begin{tabularx}{\textwidth}{@{}>{\raggedright\arraybackslash\hsize=1.10\hsize\hspace{0pt}}Xl>{\raggedright\arraybackslash\hsize=0.90\hsize\hspace{0pt}}X@{}}
\toprule
Code & Kind & Example \\
\midrule
\texttt{branch\_on\_size} & class & \texttt{if n > N0}, \texttt{if n \% 3 == 0} \\
\texttt{branch\_on\_loop\_index} & class & \texttt{if i == N0}, \texttt{if i \% 3 == 0} \\
\texttt{branch\_on\_param} & class & \texttt{if len(a) > 3} \\
\texttt{conditional\_expression} & both & \texttt{range(n - 1 if n \% 2 else n - 2)} \\
\texttt{nonaffine\_expression}, \texttt{nonaffine\_index} & both & \texttt{(i + 1) \% n}, \texttt{range(n // 2)}, \texttt{2 ** j} \\
\texttt{nested\_loop} & class & the double loop of the QFT, a ladder inside a ladder \\
\texttt{while\_loop}, \texttt{recursion} & both & doubling trees, recursive QFT \\
\texttt{growing\_arity} & class & \texttt{mcx}, \texttt{mcp} on a list of qubits whose length grows with $n$ or $i$ \\
\texttt{library\_gate} & fragment & \texttt{compose}, \texttt{append}, \texttt{.inverse()} \\
\texttt{nonconst\_angle} & class & $2\arccos\sqrt{1/(n-i)}$, $\pi/2^{i}$, an angle from the parameter \\
\texttt{two\_registers}, \texttt{fold\_index}, \texttt{stride\_index} & class & index $n+i$, $n-1-i$, $2i+1$ \\
\texttt{index\_out\_of\_range}, \texttt{qubit\_collision} & class & \texttt{cx(i, i + 1)} with \texttt{range(n)} on $n$ qubits; \texttt{cx(0, i)} at $i=0$ \\
\texttt{bit\_index\_negative}, \texttt{bit\_index\_out\_of\_range} & class & \texttt{a[i - 1]} at $i=0$; \texttt{(a \textgreater{}\textgreater{} (i + 1)) \& 1} at $i=n-1$ \\
\texttt{bit\_on\_end\_fixed} & class & a condition on a bit $a[n+c]$ in the body of a ladder \\
\texttt{bit\_const\_in\_down\_ladder} & class & \texttt{for i in range(n - 1, 0, -1): if a[0]: qc.x(i)} \\
\texttt{fixed\_overlaps\_window}, \texttt{fixed\_inside\_ladder} & class & the window position $i-1$ (qubit $i-1$ or bit \texttt{a[i-1]}, as in \texttt{for i in range(1, n): if a[i-1]: qc.cx(0, i)}) at $i=1$ coincides with the fixed qubit 0; \texttt{range(0, n, 3)} with the fixed qubit 5; \texttt{for i in range(n): if a[0]: qc.x(i)} \\
\texttt{alias\_reassigned}, \texttt{alias\_under\_condition}, \texttt{loop\_var\_after\_loop} & fragment & accumulator \texttt{t = t + 1}; assignment under \texttt{if a[i]}; counter after the loop \\
\texttt{bit\_param\_misuse} & both & \texttt{range(a)}, \texttt{a[i]} outside a condition \\
\texttt{range\_not\_ladder} & both & \texttt{reversed(range(0, n, 2))}, a loop over an explicit list, bounds not of the form $c$ or $n+c$ \\
\texttt{body\_without\_window} & class & \texttt{for \_ in range(n): qc.x(0)} (repetition without a window, not a ladder) \\
\texttt{no\_size\_dependence} & both & a fixed-size circuit \\
\texttt{no\_ladder} & fragment & gates only on qubits with constant indices (Sect. S2.1) \\
\texttt{unknown\_call}, \texttt{unknown\_name}, \texttt{unknown\_gate}, \texttt{unsupported\_statement}, \texttt{unsupported\_expression}, \texttt{extra\_params}, \texttt{shadowed\_builtin} & fragment & \texttt{getattr}, \texttt{try}, a default parameter, \texttt{QuantumRegister}, a redefined \texttt{range} \\
\botrule
\end{tabularx}

\end{table}

%% file: supplement/S3_plans.tex
\section{Exact and probabilistic plans on the benchmark}\label{sec:S3}

Figure S1 shows the numbers of primes and the times of the exact and probabilistic plans pair by pair. The experimental conditions and the summary numbers are in Section 7.4 of the article.

\begin{figure}[tbp]
\centering
\includegraphics{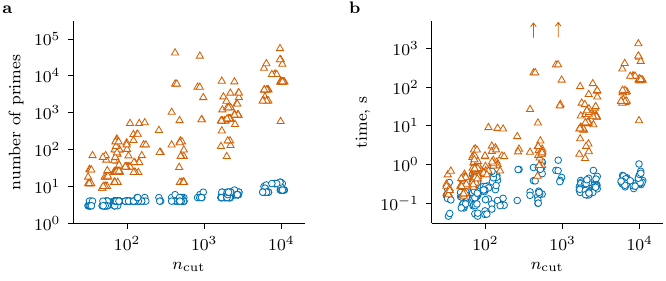}
\caption{The exact and probabilistic plans on the 224 equal pairs with the answer ``equal'' and \(\varepsilon\le2^{-40}\). \textbf{a} Number of primes; \textbf{b} time. Circles: the probabilistic plan; triangles: the exact plan; upward arrows (panel b): the exact plan not finished within 30~min (the time is at least the one shown). The \(n_{\rm cut}\) axis is the cutoff of the probabilistic plan; both axes are logarithmic, and points with the same \(n_{\rm cut}\) are spread horizontally by at most 10\%}\label{fig:S1}
\end{figure}

%% file: supplement/S4_bound.tex
\section{The bound of Theorem 1 on the ladders of the benchmark}\label{sec:S4}

Corollary 1 relies on the dimension bound of the automaton, so we checked the bound of Theorem 1 on all 1087 ladders that occur in the programs of the functions of the benchmark, including the functions of 132 pairs where the mutation does not change the circuit and which are not part of Section 7 (identical ladders counted once; 1054 of them occur in the 2200 pairs of Section 7). On each ladder, the bound is at least the dimension of the automaton built by the implementation (Section 6), and equals it on 688. The dimension after minimization is smaller than the bound by a factor of 1.56 in the median and of at most 31, and equals the bound on 190 ladders (Figure S2a). The median ratio of the bound to the least dimension grows with \(w+h\), from 1.0 at \(w+h=1\) to 4.7 at \(w+h=4\) (Figure S2b). The groups consist of different ladders, so this is the margin of the bound on the ladders of the benchmark, rather than the asymptotic scaling of a single ladder. The largest ratios (from 20 to 31) come from nine ladders with \(w=1\) and \(h=2\): for them the term \(D_0\) is from 129 to 1281, while the first term of the bound is 16 or 32.

\begin{figure}[tbp]
\centering
\includegraphics{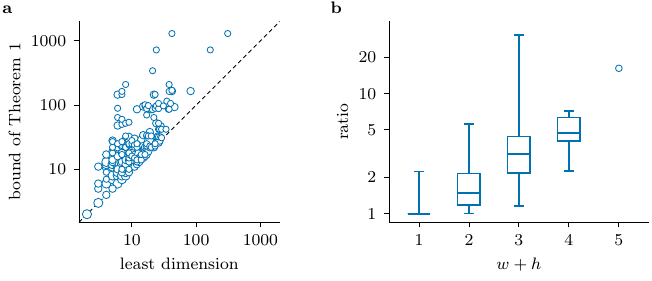}
\caption{The bound of Theorem 1 and the least dimension of the automaton on the 1087 ladders of the benchmark. \textbf{a} The bound versus the least dimension; both axes are logarithmic with the same decade scale, and the dashed line marks equality; a circle is one of 186 distinct pairs of values, and its size is the number of ladders with this pair: 1, 2 to 9, 10 or more; \textbf{b} the ratio of the bound to the least dimension (logarithmic axis) by the values of \(w+h\): the box spans the lower to the upper quartile, the line in the box is the median, and the whiskers are the minimum and maximum; at \(w+h=1\) the quartiles coincide, and at \(w+h=5\) there is one ladder (circle)}\label{fig:S2}
\end{figure}